\documentstyle[prl,aps,floats,epsf,epsfig]{revtex}

\begin{document}
\draft

\title{Duality in 2+1D Quantum Elasticity: superconductivity and
Quantum Nematic Order}

\author {J. Zaanen and Z. Nussinov$^*$}

\address{Instituut Lorentz for Theoretical Physics, 
Leiden University, P.O. Box 9506, NL-2300 RA Leiden, 
The Netherlands\footnote{$^*$ present address: Theoretical Division,
Los Alamos National Laboratory, Los Alamos, NM 87545, USA.}}

\author{S. I. Mukhin}

\address{Theoretical Physics Department, Moscow Institute for Steel \& Alloys,
119991 Moscow, Russia.}
\date{\today}

\maketitle

\begin{abstract}
Superfluidity and superconductivity are traditionally understood in
terms of an adiabatic continuation from the Bose-gas limit. Here we
demonstrate that at least in a 2+1D Bose system, 
superfluidity can arise in a strict
quantum field-theoretic setting. Taking the theory of quantum elasticity 
(describing phonons) as a literal quantum field theory with a bosonic
statistic, 
superfluidity and superconductivity (in the EM charged case) emerge 
automatically when the shear rigidity of the elastic state is destroyed 
by the proliferation of topological defects (quantum dislocations).
Off-diagonal long range order in terms of the 
field operators of the constituent particles is not required. 
This is one of the
outcomes of the broader pursuit presented in this paper. In essence, 
it amounts to the generalization of
the well known theory of crystal melting in two dimensions
by Nelson, Halperin and Young\cite{KTNHY}, to the
dynamical theory of bosonic states exhibiting quantum liquid-crystalline
orders in 2+1 dimensions. We strongly rest on the field-theoretic
formalism developed by Kleinert \cite{KleinertII}
for classical melting in 3D. Within
this framework, the disordered states correspond to Bose condensates
of the topological excitations, coupled to gauge fields describing
the capacity of the elastic medium to propagate stresses. Our focus
is primarily on the nematic states, corresponding with condensates
of dislocations, under the topological condition that disclinations
remain massive. The dislocations carry Burgers vectors as topological
charges.  Conventional nematic order, i.e. the breaking of
space-rotations, corresponds in this field-theoretic 
duality framework with an ordering of the Burgers vectors. However, 
we also demonstrate that the Burgers vectors can quantum disorder
despite the massive character of the disclinations. We identify
the physical nature of the `Coulomb nematic' suggested by
Lammert, Rokshar and Toner\cite{Lammert} on gauge-theoretical grounds.
The 2+1D  quantum liquid crystals differ in fundamental regards
from their 3D classical counterparts due to the presence of a
dynamical constraint. This constraint is the glide principle, well known from
metallurgy, which states that dislocations can only propagate in
the direction of their Burgers vector. In the present 
framework this principle plays a central role. This constraint is
necessary to decouple compression rigidity from the dislocation
condensate. The shear rigidity is not protected, and as a result
the shear modes acquire a Higgs mass in the dual condensate. This is 
the way the dictum that translational symmetry breaking goes hand in
hand with shear rigidity emerges in the field theory. 
However, because of the glide principle compression stays massless,
and the fluids are characterized by an isolated  massless compression
mode and are therefore superfluids. Glide also causes the shear Higgs mass
to vanish at orientations perpendicular to the director in the ordered
nematic, and the resulting state can be viewed as a quantum smectic of
a novel kind. Our most spectacular result is a new hydrodynamical way
of understanding the conventional electromagnetic Meissner state
(superconducting state). Generalizing to the electromagnetically
charged elastic medium (`Wigner Crystal') we find that the 
Higgs mass of the shear gauge fields,  becoming finite in 
the nematic quantum fluids, automatically causes a Higgs mass in the
electromagnetic sector by a novel mechanism.  
\end{abstract}


\maketitle


\section{Introduction}

The concept of duality has been around for a long time in high
energy- and statistical physics, but it seems that the condensed
matter community is still in the process of getting used to its
powers. As compared to its use in fundamental
field- and string theories, the concept acquires a quite vivid
and physical interpretation in the condensed matter arena. 
Quantum field theories
become of relevance as a description of the highly collective, long
wavelength properties of systems composed of strongly interacting
microscopic entities. Their emergence is tied to order. In their most
straightforward interpretation, field theories enumerate the physics
associated with spontaneous symmetry breaking. The fields are describing
the Goldstone modes. Implemented in this context, duality acquires the
meaning that states which appear as disordered are in fact also 
governed by order, although their order parameters might be beyond the
reach of the most brilliant experimentalists. In continuum field theories
this notion acquires a very precise meaning. Starting from an ordered
state, disorder is governed exclusively by the singular (in fact, multivalued)
configurations of the order field. These translate in `disorder operators' 
carrying 
quantized topological charges and the only admissible states of disorder
correspond with condensates of the topological excitations. Having 
identified the disorder fields, it is easy to deduce the physics of
the disordered states using the weaponry of order physics.

Duality is especially powerful in 2+1 dimensions. The reason is that
the topological excitations associated with conventional orders are
typically point particles in 2+1 dimensions. The disorder field
theories describing the collective properties of the `disorder matter'
are of the well understood Ginzburg-Landau-Wilson variety:
they are conventional Bose condensates albeit in terms of topological
matter. A popular example is vortex duality, describing in the quantum
context\cite{FisherLee,Fazio} the disordering of a superconductor in a 
Bose Mott-insulator driven by the disordering influence of the charging 
energies on the phase order\cite{banks,KleinertI,Sudbo,Senthil,Tesanovic}.
The disorder operators are the well known vortices carrying
quantized magnetic flux, corresponding with particle-like excitations
in 2+1D. At a critical charging energy these `vortex bosons' proliferate
in the vacuum and it can be shown (e.g., appendix A) that the field
theory describing this vortex matter is nothing else than the theory
of a neutral superfluid, with a massless Goldstone boson describing the
free propagation of the electromagnetic photon in the insulating state.

The most obvious form of order is, of course, crystalline order- the
breaking of the spatial Euclidean group down to a lattice group.
The first instance of a field theory describing a Goldstone
sector is of course the theory of elasticity which emerged in the
19-th century\cite{LandauLifel}. 
It is also the birthplace of topology in physics: 
the first topological defects which were appreciated as highly
relevant are the dislocations and disclinations with their Burgers
and Franck vectors as topological charges\cite{Nabarro}. 
In the context of 2D
classical phase transitions it formed the initial inspiration for
the work of Kosterlitz-Thouless\cite{KosterlitzThouless,Minnhagen}
placing the general duality motive
on center stage in statistical physics. Subsequently the
duality structure associated with the thermal melting of 2D solids
was uncovered by Nelson, Halperin and Young\cite{KTNHY} 
 (the KTNHY theory of 2D
melting).  In the 1980's the
theory for 3D crystal melting was further developed,
especially by Kleinert\cite{Kleinertpub}.
This strongly rests
on the similarities with vortex duality in 3D, although it
involves some non-trivial generalizations.
The most complete treatment of this subject
is found in the two volume textbook on gauge theories in
condensed matter physics by Kleinert\cite{KleinertI,KleinertII}, 
where the second volume is dedicated to elasticity.

Dynamics is not the same as statics and it is remarkable that the
elastic analogue of the quantum-mechanical vortex duality in 2+1D 
goes largely unexplored\cite{Stoofcom}. 
As the KTNHY theory already makes clear, 
liquid crystalline orders appear in a natural way within the duality
framework. As compared to the standard lore in terms of rod-like
molecules aligning their long axis etc., a considerable shift
of interpretation occurs. Nematic (and as we will see, also smectic-)        
orders appear as a consequence of the rich topological structure 
associated with spatial symmetries. These can be entirely classified
in topological terms. Nematics are those states of matter characterized
by a condensation of dislocations while disclinations are massive
excitations: the `hexatic' state predicted by KTNHY. The
isotropic fluid is a state where both dislocations and disclinations
are condensed and because these are interdependent, it is best termed
 a `defect condensate', see Kleinert \cite{KleinertII}.

Our interest was originally triggered by several suggestions regarding the
possible existence of zero temperature quantum manifestations of liquid 
crystalline order. These appeared at more or less the same time, although
aimed at different physical circumstances.
Mullen {\em et al}\cite{Stoof} suggested a possible super-hexatic phase 
in 2D Helium\cite{superhexatic}. Balents and Nelson\cite{Balents} 
investigated a smectic
phase in flux systems and its quantum analogue. Kivelson, Fradkin and 
Emery\cite{kivemfrad}
advanced the possible existence of smectic and nematic orders in
cuprate superconductors, shortly thereafter followed by suggestions
regarding smectic and nematic quantum Hall stripe phases\cite{quhallfr}
(see also \cite{Balents1,otherquha}). Viewed from a broader perspective,
these ideas are part of the current development in condensed 
matter physics to search for quantum fluids which are correlated to
a degree that the conventional quantum gas perspective (Fermi-liquids,
the Bogoliubov Bose gas, BCS theory) is no longer relevant. Instead,
it might be more appropriate to view such systems as on the verge
of becoming ordered (`fluctuating order', in the context of 
high Tc superconductivity see \cite{philmag,flucorder}).  
In a broader sense, one might want to equate this `fluctuating order' to
duality, as all that exists is order and the disorder derivatives of 
order. All degrees of freedom are of a collective kind and the system at long
distances has completely forgotten about the microscopic constituents
(like the electrons, Cooper pairs). 

As we will discuss in detail,  `fluctuating order' acquires a very precise
meaning in the context of the quantum melting of a crystal. We will in
this paper develop the theory of quantum liquids which are in a literal
way derivatives from the collective fields of the solid: the Goldstone
modes (phonons), and the topological defects (dislocations, disclinations).   
In a precise way, our construct exaggerates the orderly nature of matter
to an extent that the limiting case we describe cannot be
realized literally in any condensed matter system.  
The degrees of freedom of the constituent particles (the `interstitials';
the cooper pairs, He atoms, etc.) are not included in this theory. However,
the case can be made that the degrees of freedom of the interstitials
 are liberated at the moment that the solid undergoes quantum melting. 
These degrees of freedom are
of relevance for the long wavelength physics, while they are not an
intrinsic part of the field theoretic description. This is a main
short coming of our approach and the reader should view it as an
exposition of an unphysical limit which can nevertheless be
closely approached, at least in principle. It should also be regarded
as complementary to much of the existing work on quantum liquid 
crystals which has a (implicit) focus on the physics associated with
the interstitials\cite{Balents,Stoof,kivemfrad,slidingph,barci}. This
also includes D.-H. Lee's stripe-superconductivity 
duality\cite{dhleedual} which is also of the interstitial kind,
and obviously  the rich body of work explicitly dealing with
supersolid  physics\cite{Scalettar,Zaanensuso,Sachdevsuso,Zhangsuso}.

Besides the neglect of interstitial excitations, there are a number
of other limitations to this work. Most importantly, we limit ourselves
to bosonic matter, for the usual reason that we do not know how
to deal with the fermionic 
minus signs (see ref.\cite{flnematic} for some intriguing 
results on fermionic liquid crystals). The other limitations are
less essential and are motivated by technicalities. We limit ourselves to
2+1 dimensions because dislocations and disclinations are particle
like in this dimension, and the general form of the disorder
field theory is fully understood. In 3+1 dimensions the defects
are strings (dislocation, disclination loops in 3D space).
Although there is every reason to expect that  the outcomes will be 
qualitatively the same, it is not known how to construct string
condensates explicitly because of the lack of a second quantized 
formalism. Next, our focus will be entirely on the nematic states,
i.e. the dislocation condensates, for no other reason than to keep
the scope of this paper limited. We do not address therefore in
any detail the disorder condensates associated with the proliferation
of disclinations. In fact, because rotations are 
Abelian in two spatial dimensions, not much interesting is expected to
happen when disclinations proliferate, as 
associated with the melting of the quantum nematics into the isotropic 
state. This is different in three spatial dimensions where the non-abelian
nature of the rotational group comes into play leading to interesting
braiding and multiplicity of the disclinations structures.
Mathematical methods in the form of quantum double symmetries recently
emerged allowing a possible systematic investigation of such 
topological structures\cite{Muller}, and we defer this to a future study.
Finally, we limit ourselves to the theory of the isotropic quantum-elastic
medium (i.e., showing global rotational invariance, `no crystal faces') and
this is merely for convenience. Our findings are easily generalized
to any 2D space group. 

Within these limitations, there is much to be discovered. In
technical regards, we rest strongly on the mathematical methodology
developed by Kleinert\cite{KleinertII}. Our work can be 
viewed as an application of
his methods in a somewhat altered context, and we will review this
methodology in some detail in section II to make this paper
self-contained. At the core of this 
methodology is the realization that the universe of dislocations
and disclinations is governed by gauge fields. The defects act
 like sources in electromagnetism, exerting long range forces
on each other by the exchange of `photons'. It is, in itself, an
entertaining exercise to rewrite the familiar physics of dynamical
phonons in the language of `stress photons' residing in 2+1D
space-time as we discuss in sections IV and V. These stress
photons become quite meaningful when the solid quantum-melts due
to the proliferation- and Bose condensation of the dislocations.
The dislocations carry shear `stress' charge and the effect is that
together with the stress-gauge fields the analogue of an electromagnetically
charged superconductor emerges. This exhibits a Meissner effect (Higgs
phenomenon) causing the shear stress photons to become massive. This represents
the physical fact that in the fluid the shear rigidity associated with
the elastic state becomes short ranged.

The informed reader will recognize the similarity with vortex duality.
However, in two regards the dislocation condensate is radically different
and more interesting than the vortex condensate. First, vortices carry
scalar charges while dislocations carry vectorial charges, the
Burgers vectors. Under the condition that the disclinations are
massive, this vectorial nature of the charge does not pose a difficulty
because the governing symmetry stays Abelian (pure translations).
The Burgers vectors can be viewed as additional degrees of freedom,
responsible for the various forms of nematic order. 
In section III we will develop the basic formalism underlying the duality. 
We find that besides the conventional form of nematic order
(breaking spatial rotational symmetry, the 2+1D generalization of the hexatic
of 2D), a purely topological form of nematic order is to be expected:
although rotational symmetry is unbroken, disclinations are still massive. 
This `Coulomb nematic' was predicted on different grounds some time ago
by Lammert, Rokhsar and Toner\cite{Lammert}
and we find a natural physical interpretation
for it in our duality framework.

The other novelty refers to a very special phenomenon only occurring
on the dynamical level, involving the time axis. On the quantum level
this becomes most central, because space and time are entangled :
it is the `glide' principle, referring to the fact well known to
metallurgists\cite{Nabarro} 
that dislocations  can only propagate in the direction
of their Burgers vector. This glide constraint plays a
central role in the remainder and it renders our theory to be a
real quantum theory, having no classical analogy.
 
Glide has a remarkably deep meaning. We will
prove that the glide constraint is a necessary condition for the
decoupling of compressional stress from the dislocation currents.
This has far reaching consequences. It is a textbook wisdom that
translational symmetry breaking is associated with long range shear
rigidity. Upon melting the solid, the medium loses its capacity 
to transmit shear forces. However, liquids still carry sound, meaning
that they are characterized by a massless compression mode. On the 
quantum level, glide is needed to protect this compression mode against
the dislocation condensate!

A first direct consequence of glide is that the nematic breaking rotational
symmetry turns into a state which in a special sense is more like a
smectic, as will be discussed in section VIII. 
The director order correspond with a rotational order of the
Burgers vectors, and these in turn direct the `superfluid' dislocation
currents into one particular direction. This has the implication that
the shear Higgs gap vanishes in the direction exactly perpendicular to
the director where the phonons of the solid re-emerge.                   
  
We perceive the other consequence of glide as of a great general importance.
We will derive the excitation  spectrum of the dislocation condensate
explicitly and this will turn out to be of a quite universal form
(sections VI and VII).  Besides an overall scale vector (phonon velocity), 
it is completely  determined by the Poisson ratio 
and a single length, the `shear penetration depth'. It consists of
two massive shear modes and, in addition, a massless mode which is purely
compressional. 

In textbooks on quantum fluids, the emphasis is on the concept
of off-diagonal long range order (ODLRO) as introduced by Penrose 
to understand the fundamental nature of the superfluid state.
However, this way of thinking rests in last instance on the
continuation to the bose gas limit and the present field theory
is explicitly constructed so that this adiabatic connection is 
disrupted. However, Landau\cite{Landau} and Feynman\cite{Feynman}
constructed an alternative, purely {\em hydrodynamical} description
of superfluidity. As we will discuss in detail in section VIIC, this
rests on the assumption that the superfluid is characterized by
an isolated, propagating compression mode (`phonon') in the scaling
limit. This is consistent with the ODLRO in the Bose gas. However,
the present field theory shows that the Landau/Feynman hydrodynamic
theory is in a sense incomplete. Without referral to microscopics
it turns out to be possible to precisely specify in hydrodynamical terms
where the propagating compression is coming from: 

{\em it is sufficient condition for the existence of the  superfluid
state that a solid looses its rigidity to shear stresses due to
condensation of dislocations.}

This is just what is happening in the present field theory. The 
dislocation condensate is
characterized by an isolated massless compression mode, and according
to hydrodynamics it has to be a superfluid. However, it is a superfluid 
which is not based on `conventional' ODLRO. The ramification is that
the Landau/Feynman understanding of superfluidity is more general than
the one based on ODLRO, because we have identified a superfluid which
is disconnected from the Bose-gas limit!      
  
A next issue is what happens when electromagnetism is coupled in,
i.e. considering the duality for an
electromagnetically charged (`Wigner') bosonic crystal. If the
dual is indeed a superfluid, the charged case should exhibit      
an electromagnetic Meissner effect. General hydrodynamical
arguments are available demonstrating that `our' superfluid
should turn into an electromagnetic Meissner-Higgs state. 
These are of a much more recent origin. In 1989 Wen and Zee\cite{Wen}
demonstrated that the presence of an isolated massless compression
mode in the neutral case is sufficient condition for the appearance
of a Meissner-Higgs state when electromagnetism is coupled in. This
argument rests on duality: the compression mode can be dualized in
a pair of compression stress photons (see section IIIE) which are
coupled to EM photons. Integrating the former yields a Higgs mass
gap for the latter.

Although consistent with the Wen-Zee theorem, we find that in the
full elastic duality a EM Higgs mass is generated by yet a different
mechanism. We percieve this as of great general importance and this
counterintuitive affair is presented in section X. 
 By just coupling the electrical fields
to the displacement fields of the Wigner crystal, we find that a
miracle occurs upon dualizing to a stress-photon representation.
Automatically, {\em a Meissner term appears acting on the electromagnetic
vector potentials.} At first sight, it appears as if the crystal
is already exhibiting a Meissner effect yet this is not quite the case.
The electromagnetic photons are linearly coupled to the stress photons.
Upon integrating these out, a counter electromagnetic
 Meissner term is generated which
is exactly cancelling the `bare' Meissner. However, for this compensation
to work, the shear photon needs to be massless. When the shear-mass becomes
finite because of the dislocation condensate this compensation is no
longer complete and the system turns into a superconductor. In summary,
{\em in the stress-gauge field representation, the Meissner effect lies
in hide in the solid to get liberated in the superconductor because 
shear rigidity becomes short range.}

It appears that this mechanism is of a most general nature.
The specifics of the nematic states etc. do not enter the arguments
leading to these conclusions in any obvious way. In fact, we are 
under the impression that it offers a deep, hydrodynamical, insight 
into the nature of superconductivity, and the Higgs phenomenon in general. 
Giving matters a further thought, it appears that this mechanism is in not at
all conflicting with the usual understanding in terms of off-diagonal long
range order in terms of the constituent bosons. The latter is just a special
case of the former, derived from the gas limit where the shear length is 
vanishingly small.

To further emphasize this issue, we ask under what circumstance can one see
the difference between the conventional superconductor and our `order
superconductor'. The answer is that the superconductor is characterized
by {\em two length scales}: the magnetic penetration depth  and the
shear penetration depth. In the conventional description the latter is
just ignored. However, in our formalism the shear length takes a central
role and we can study explicitly what happens when the shear-length
exceeds the bare magnetic penetration depth. We find new physics
(section IX). In this regime the effective penetration depth turns into
the geometrical mean of the shear- and bare magnetic lengths. More
strikingly, by a novel mechanism, the magnetic propagator acquires poles
with real parts leading to oscillating magnetic screening currents. 
Upon entering the superconductor one finds a pattern of
screening and anti-screening  currents. 

Finally, we will conclude this paper with a discussion of potential
ramifications of our work in both the condensed matter- and 
cosmological context (Section X).

\section{General considerations: elasticity as a genuine field theory.}

The theory of elasticity is, of course,
overly well known. Here we will interpret it as a literal quantum 
field theory. In doing so, the theory acquires the status of a toy model which
might be good enough to reveal some most general features but it cannot be 
applied literally to any circumstance encountered in nature. 
The theory describes the
long wavelength collective behaviors of crystals: matter
spontaneously breaks spatial rotations and translations down to
a lattice group.  Historically, it is the first `emergent' field theory 
which was discovered.  In a much more modern setting, it is fascinating
that it can be reformulated in a differential geometric
language, with the outcome that at least the 3D isotropic theory turns
out to be Einstein-Cartan gravity in 2+1D, with the disclinations and
dislocations taking the role of curvature- and torsion sources, 
respectively\cite{KleinertII,kleinertgrav1,kleinertgrav2,kleinertgrav3}
In part motivated by the present work, interesting connections 
with quantum gravity have  been discussed in a
recent paper by Kleinert and one of the authors 
[JZ]\cite{KleinertZaanen}.  Except for a short
discussion in the conclusion section, we will ignore these aspects
and instead focus on 2+1D quantum elasticity. This theory remembers 
its origin  in the physics
of non-relativistic particles with the consequence that Lorentz invariance
is badly broken: the world-lines of non-relativistic particles are 
incompressible along the time direction and this fact 
is remembered by the long wavelength theory.

\subsection{Some basics and definitions.}

\begin{figure}[tb]
\begin{center}
\vspace{10mm}
\leavevmode
\epsfxsize=0.35\hsize
\epsfysize=5cm
\epsfbox{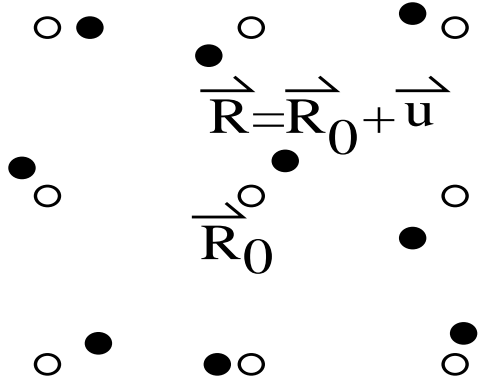}
\end{center}
\caption{In a crystal, the positions of
`atoms' (or electrons, Cooper pairs etc., the black dots)
can be uniquely  related to equilibrium positions  forming a
regular pattern (open circles) by displacements $\vec{u}$. 
A requirement for the crystal to exist is that the displacement
field $\vec{u} (\vec{x}, t)$ is non-singular.}
\label{displacements}
\end{figure}

Let us first summarize some basics of elasticity 
theory\cite{LandauLifel,KleinertII}, to remind the reader and
to introduce notations.
Starting with the crystal, the theory is derived by asserting that
 constituent particle $n$ has a equilibrium position $\vec{x}_n$,
and a real coordinate $\vec{x}'_n = \vec{x}_n + \vec{u}_n$ 
(Fig. \ref{displacements}).
Strictly speaking $\vec{u}_n$ should be finite in order to have a
meaningful continuum limit. At distances large compared to the lattice
constant $a$, one can define a displacement field $\vec{u} (\vec{x}, \tau)$
such that $\vec{x}' (\tau) = \vec{x} (\tau) + \vec{u} (\vec{x}, \tau)$
($\tau$ is imaginary time; $\vec{u} = (u^x, u^y)$; the
Latin labels  $\sim a, b, \cdots$ refer to space 
coordinates; Greek labels $\sim \mu, \nu, \cdots$ refer to space time;
Einstein summation conventions are used everywhere).
The distance
vector between two material points at $\vec{x}$ and $\vec{y}$ is changed
from $d\vec{x} = \vec{x} - \vec{y}$ to $dx'_a = dx_a + \partial_b u^a d x_a$,
and its length from $d r = \sqrt{d \vec{x}^2}$ to 
$d r' = \sqrt{ d \vec{x}^2 + 2 w_{ab} d x_a d x_b}$ where the 
strain tensor $w_{ab}$ is in linear approximation,

\begin{equation}
w_{ab}  = { 1 \over 2} \left(  \partial_{b} u^a  + \partial_{a} u^b 
\right).
\label{strains}
\end{equation}  

These strain fields are the physically meaningful entities entering
the theory, because the elastic action can only depend on the differences
$ d l - d l'$. For fundamental geometrical reasons,
this tensor is symmetric in the spatial indices. The part antisymmetric
in the space indices corresponding with the rotational field,
 in 2+1D, 

\begin{eqnarray}
\omega_{\tau} & = & { 1 \over 2} \epsilon_{\tau a b} \partial_a u^b
\nonumber \\
& = &  { 1 \over 2} \left( \partial_{x} u^y - \partial_{y} u^x \right) 
\label{asymconst}
\end{eqnarray}

cannot enter the action as $(\omega_{\tau})^2$ does not respect
rotational invariance. Nevertheless, upon considering higher orders in
the gradient expansion (`second gradient elasticity', see 
Kleinert\cite{KleinertII}) terms of the form  

\begin{equation}
S \sim \int d\Omega  \xi^2_R C_R ( \partial_a \omega_{\tau} )^2
\label{rotationalstiffness}
\end{equation}

will be encountered in the action. These invariants
express the `rotational stiffness' of the elastic medium. 
However, as compared to the leading order (`first gradient elasticity'), 
invariants $\sim w_{ a b}^2$, these involve two extra gradients meaning
that this rotational stiffness becomes only observable at length
scales smaller than the length $\xi_R$. Hence, it is of no 
consequence to the long wavelength dynamics, but as we will see later,
the finiteness of such terms is  sufficient condition for the existence
of nematic states.   

For notational purposes, we will adopt the following convention. 
Asymmetric tensor fields are written as

\begin{equation}
F^a_{\mu} = \partial_{\mu} u^a,
\label{defasym}
\end{equation}

such that, for instance, $ \dot{u}^x = \partial_{\tau} u^x = w^x_{\tau}$,
etc. The symmetrized fields are written as

\begin{eqnarray}
F_{a b} & = & { 1 \over 2} ( \partial_a u_b +  \partial_b u_a ) 
\nonumber \\
 & = & { 1 \over 2} ( w^b_a + w^a_b )
\label{defsym}
\end{eqnarray}

In the leading order gradient expansion, the (Euclidean) action should be 
$\sim w^2$. As no displacements in the time direction are allowed
($u^{\tau} = 0$) the kinetic energy density is simply $ - { {\rho} \over 2}
( \partial_{\tau} u^a )^2 = - { {\rho} \over 2} ( w_{\tau}^a )^2$, 
while the potential energy density
$C_{abcd} w_a^b w_c^d$ where the $C$'s are the elastic moduli,
and the theory of quantum elasticity becomes in 
Euclidean path integral form,

\begin{eqnarray}
Z & = & \int {\cal D} u^a e^{ - { 1 \over {\hbar}} \; S_{elas} } \nonumber \\
S_{elas} & = & \int d\Omega \left[ { 1 \over 2}  C_{abcd} w_a^b w_c^d
\; + \;  { {\rho} \over 2} ( w_{\tau}^a )^2 \right]
\label{quelasaction}
\end{eqnarray}

where $d\Omega = d^2x d\tau$ is the 2+1D Euclidean space time volume 
element. In the remainder, we will set $\hbar = 1$ and consider most
of the time just the form of the action $S$, leaving the path-integration
(including the measures) implicit. 

Let us now specialize to the theory of isotropic elasticity, describing
the elastic medium which is invariant under global spatial rotations.
This isotropic elasticity
is the standard theory used by metallurgists for the reason that metals like
steel are amorphous on macroscopic scales. For
single crystals this is not a valid assumption, except for the 2d 
closed packed triangular crystal where, for accidental reasons, 
the gradient expansion yields the isotropic theory. 
We focus on this particular case, because
it is the most basic and our findings are easily generalized to the
less symmetric cases.

The isotropic medium can be parameterized in terms of
two independent moduli: the compression- and shear moduli $\kappa$ and
$\mu$, respectively. $\kappa$ parameterizes 
the  response associated with uniform compression,
while shear ($\mu$) is associated with the response arising 
from moving opposite sides
of the medium in opposite directions. Compressional rigidity is universal
in interacting non-relativistic matter. Shear is the rigidity 
exclusively associated with the breaking of translational invariance.
The two respective compression
and shear moduli are related to the Poisson ratio $\nu$ via
\begin{equation}
\kappa = \mu \; { { 1 + \nu} \over { 1 - \nu} },
\label{poissonratio}
\end{equation}
and the potential energy density can be compactly written as 

\begin{equation}
{\cal L}_{pot} = \mu \left( \sum_{ab} w^2_{ab}
+ { \nu \over { 1 - \nu}} (\sum_a  w_{aa})^2 \right), 
\label{potendensity}
\end{equation}

using explicit 
summations for clarity. Combining this with the kinetic
energy, we obtain

\begin{equation}
S =  {\mu} \int d\Omega \left[
( w_x^x )^2 + ( w_x^y )^2  + ( w_y^x )^2 + ( w_y^y )^2 +
{ 1  \over {c^2_{ph}} }
( ( w_{\tau}^x )^2 + ( w_{\tau}^y )^2 ) + 
{ {\nu} \over { 1 - \nu} }
( w_x^x + w_y^y )^2 \right]
\label{strainact}
\end{equation}

where the `phonon' velocity 

\begin{equation}
c^2_{ph} = { { 2 \mu} \over {\rho} }.
\label{velocity}
\end{equation}

In the remainder we set this velocity to one, unless
stated otherwise. Time is measured in units of length,
with $c_{ph}$ being the conversion factor. 

Fourier transforming 
to momentum-(Matsubara) frequency space $(\vec{q}, \omega)$, this action
is diagonalized by a transversal-longitudinal ($T, L$) projection of the 
Fourier components $u^a$ of the displacement fields,

\begin{eqnarray}
u^x & = & \hat{q}_x u^L + \hat{q}_y u^T \nonumber \\
u^y & = & \hat{q}_y u^L - \hat{q}_x u^T,
\label{longtransproj}
\end{eqnarray}

where $\vec{\hat{q}} = ( \hat{q}_x, \hat{q}_y )$ is the unit vector in
momentum space. The end result is

\begin{equation}
S =  \mu \int d^2q d\omega \left[
(  { {q^2} \over 2} + \omega^2 ) |u^T|^2 +
(  { {q^2} \over {1- \nu} } + \omega^2 ) |u^L|^2 \right]
\label{strainact1}
\end{equation}

and one directly recognizes the transversal and longitudinal
acoustic phonons propagating with velocities 
$c_T =  c_{ph} / \sqrt{2} = \sqrt{ \mu / \rho}$ 
and $c_L = \sqrt { 2 \mu / ( 1 - \nu) \rho}$, respectively. 

Finally, for future use, let us consider the longitudinal and transversal
strain propagators, associated with the dynamical form factor as measured
in e.g. neutron scattering. The propagators are defined as
usual (see appendix D) and the longitudinal ($L$) and transversal
($T$) propagators are,

\begin{eqnarray}
G_L  & = & q^2 \langle \langle \hat{q}_x u^x  + \hat{q}_y u^y |
\hat{q}_x u^x  + \hat{q}_y u^y \rangle \rangle \nonumber \\
G_T  & = & q^2 \langle \langle \hat{q}_y u^x  - \hat{q}_x u^y |
\hat{q}_y u^x  - \hat{q}_x u^y \rangle \rangle \nonumber \\
 G & = &  G_L + G_T,
\label{propagatordef}
\end{eqnarray}

where $G$ is the total propagator. It immediately follows that

\begin{eqnarray}
G_L & = &  { 1 \over {\mu}} \; {  {q^2} \over { { {q^2} \over {1- \nu}}
+ \omega^2 } } \nonumber \\
G_T & = &  { 1 \over {\mu}} \; {  {(q^2 / 2)} \over { { {q^2} \over 2 }
+ \omega^2 } }
\label{eqnmotpropagator}
\end{eqnarray}
 
describing the phonon poles.

\subsection{The Singularities}

Thus far we reviewed the simple theory of phonons. Matters
become more interesting considering how this medium
can be destroyed. As long as the displacement field is finite,
we can uniquely associate a particular particle to a particular site
in the crystal such that translational invariance remains broken, and
quantum elasticity remains asymptotically exact because of Goldstone
protection. Hence, the question concerns the nature of the singularities
of the displacement fields. These come in two classes, the 
non-topological interstitial excitations and the topological 
dislocations and disclinations: \\

\begin{figure}[tb]
\begin{center}
\vspace{10mm}
\leavevmode
\epsfxsize=0.3\hsize
\epsfysize=5cm
\epsfbox{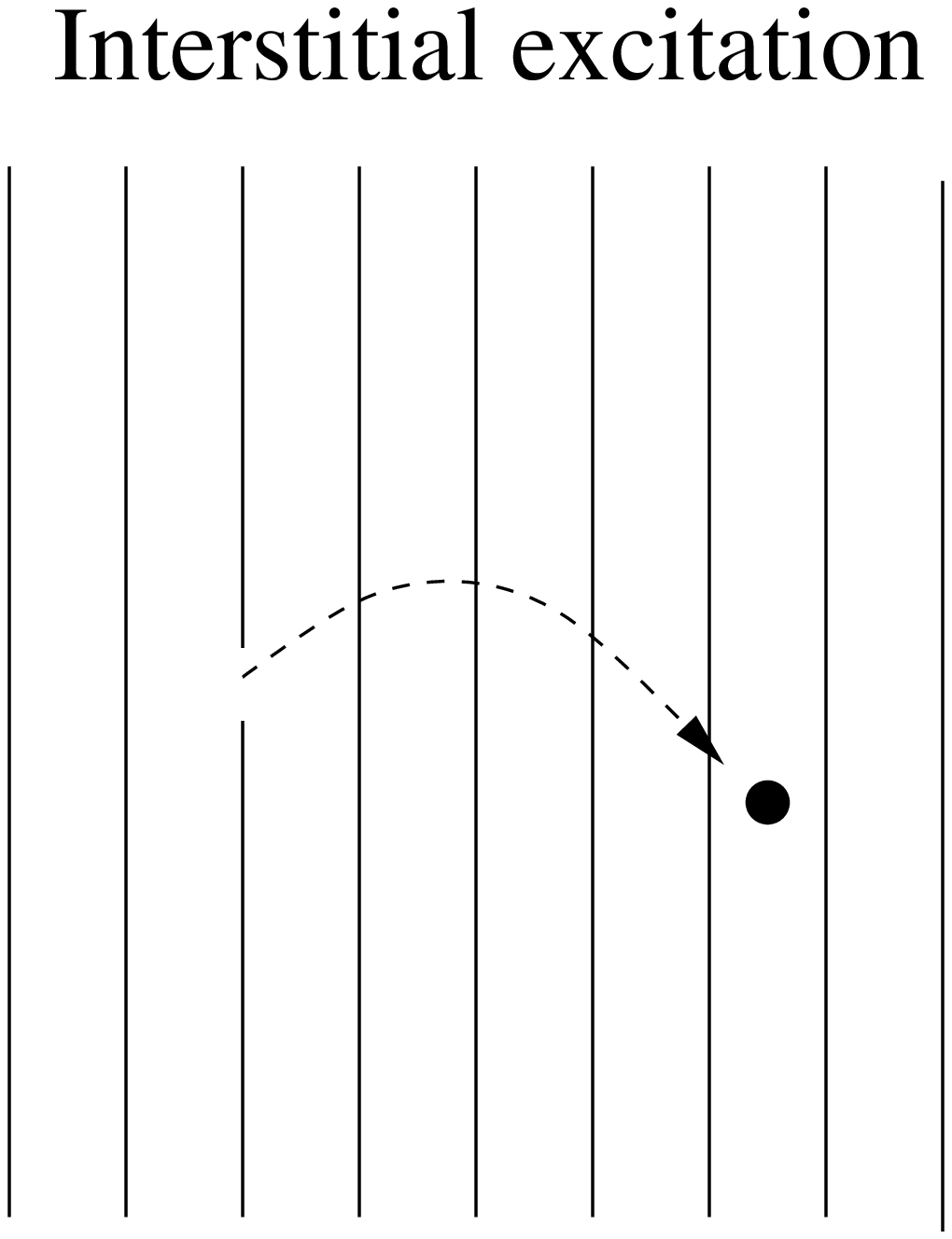}
\end{center}
\caption{Individual atoms can leave their equilibrium positions, leaving
behind a vacancy-interstitial pair, called in this paper `interstitial'.
These point-like defects cannot be avoided at any finite temperature 
while at zero temperature their condensation causes the supersolid. As
in other pictures, the lines just refer to `rows of atoms'.}
\label{interstitialpict}
\end{figure}

\begin{figure}[tb]
\begin{center}
\vspace{10mm}
\leavevmode
\epsfxsize=0.9\hsize
\epsfysize=10cm
\epsfbox{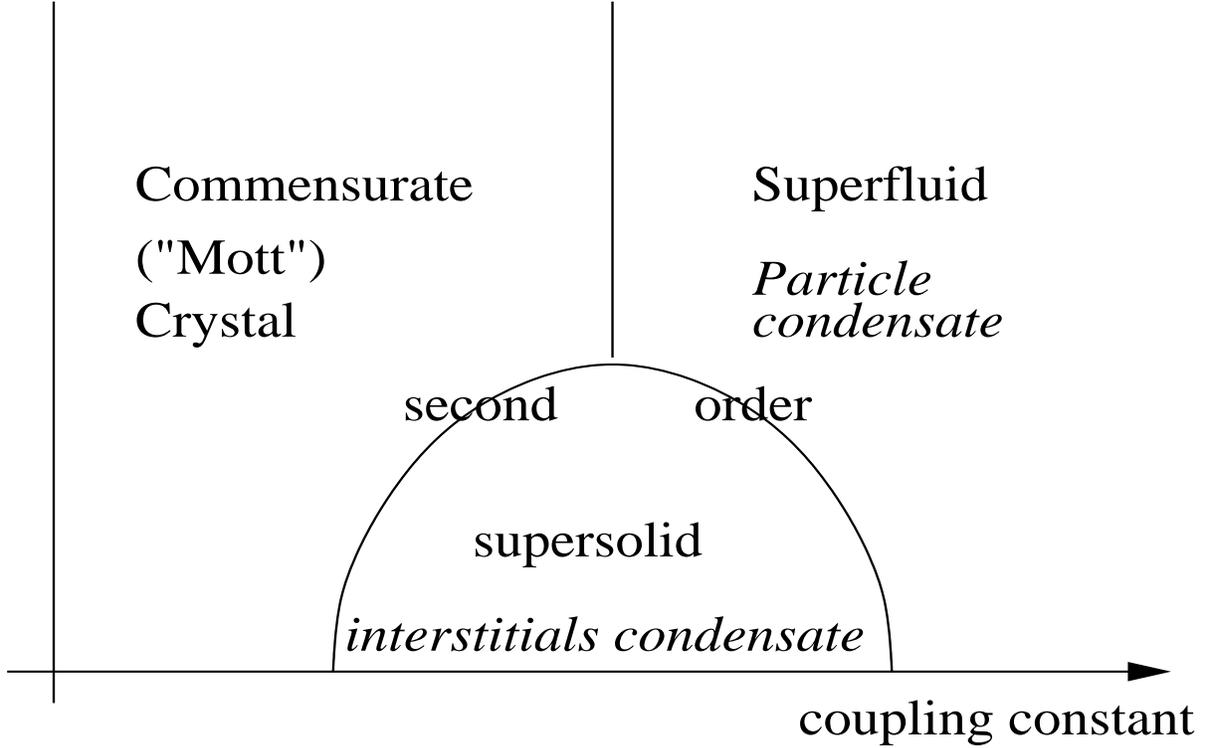}
\end{center}
\caption{The generic phase diagram of a system of Bosons subjected to
a lattice potential. When the kinetic energy is small (small coupling constant)
the Bosons will localize forming a Mott-insulating state. Upon increasing
the kinetic energy, a transition will follow into the superfluid state.
However, under the right conditions (finite thermodynamic potential, finite
range interactions) it might well happen that in between the Mott state and 
the superfluid a coexistence regime emerges: the supersolid, which can be 
viewed as a coexistence of a crystal and a Bose-gas of interstitials.}
\label{supersolid}
\end{figure}

(i) The non-topological defects- vacancies and interstitials. An
individual particle can meander away from its lattice position,
leaving behind a vacancy and moving via interstitial positions
infinitely far away from its original location 
(Fig. \ref{interstitialpict}). 
The displacement
$\vec{u}_n$ of this particular particle becomes infinite, but the
continuum field $\vec{u} (x)$ is by its very definition incapable
of keeping track of this type of singularity. Hence, interstitials
live `outside' the realm of field theory, and separate degrees of freedom 
have to be introduced to keep track of their physics. Since
interstitials and vacancies are point particles carrying a finite 
mass, they will always occur at a non-zero temperature. Therefore,
a real crystal at room temperature is in fact a gas of interstitials
coexisting with the ideal crystal. This coexistence is possible because
the interstitials just `dilute' the crystal, decreasing the amplitude
of the order parameter, and only at a density of order unity a 
transition will follow to a fluid state. At zero temperature, matters
are more sharply defined. At small coupling constant, interstitial-vacancy
pairs will only occur as virtual excitations, forming closed loops in
space time, and the crystal has a precise definition. At a critical coupling
constant these loops will blow out and interstitials will proliferate in
the vacuum. The system turns into a superposition of a crystal and a
quantum gas of interstitials. If these are bosons, the gas of interstitials
will Bose condense and a supersolid will form. Finally, a transition follows
where the amplitude of the crystal order parameter tends to zero and 
an isotropic superfluid will form (Fig. \ref{supersolid}). 

Although quite popular in condensed matter physics (e.g., ref.'s
\cite{FisherLee,dhleedual,Scalettar,Zaanensuso,Sachdevsuso,Zhangsuso}), 
this `supersolid' alley is not generic. In order for this to work, 
interstitials
should have a small mass relative to their topological competitors,
 and this is in practice only possible in the presence of strong 
Umklapp scattering.  Strong interactions between the bosons and an 
external potential are needed,
and these ingredients are wired in popular toy models like the
Bose-Hubbard model. \\ 

\begin{figure}[tb]
\begin{center}
\vspace{10mm}
\leavevmode
\epsfxsize=0.55\hsize
\epsfysize=5cm
\epsfbox{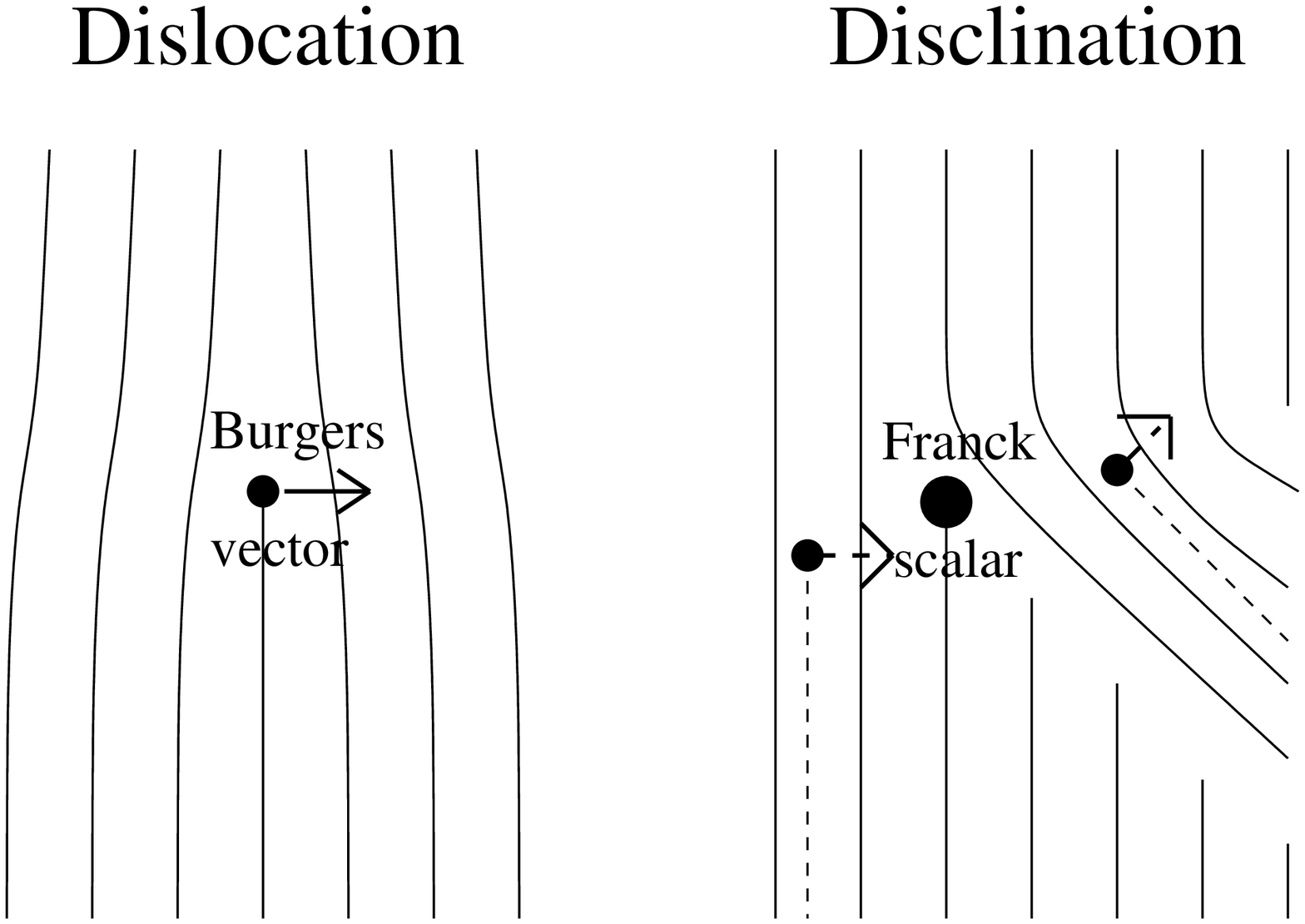}
\end{center}
\caption{As before, for convenience, the crystal lattice is encoded in lines
referring to `rows of atoms'. A dislocation corresponds with a 
half infinite row of atoms coming to and end somewhere in the solid.
It is the topological defect associated with translations and
its topological charge is clearly directional (the Burgers vector).
The disclination is associated with a rotation and it can be viewed
as a wedge of additional matter inserted by a Voltera process. 
Alternatively, it can be viewed as a bound state of dislocations
with parallel Burgers vector. By inserting test dislocations one
infers that their Burgers vector is transported as if a curvature
source resides at the disclination core. The opening angle corresponds 
with its topological charge, the Franck scalar (in 2+1D).} 
\label{defectpict}
\end{figure}

(ii) The topological excitations occurring in the elastic medium 
are the dislocations and disclinations, associated with the
restoration of translational- and rotational invariance, respectively. 
These are both point-particle like in 2+1D suggesting that their 
associated disorder field theories are relatively simple -- in fact
our main reason to specialize to this dimension. However,
as discussed in great detail by Kleinert\cite{KleinertII}, 
the tensorial nature of
the theory complicates matters greatly: for instance, the
observation that the full disorder theory has to do with 
Einstein-Cartan gravity. Contrary to the interstitials, the topological
singularities are a natural part of the full field theory, and they
enumerate the singularities of the continuum displacement fields $\vec{u}$.
The topological invariants in 2+1D of the dislocation (the
Burger two component vector $\vec{n} = \; (n^x, n^y) \; = \; |n|\; 
( \hat{n}^x, \hat{n}^y)$,
$\vec{\hat{n}}$ being the unit vector) and 
the disclination 
(the Franck `scalar' $\Omega$) can be defined through circuit 
integrals (Fig. \ref{defectpict}),

\begin{eqnarray}
\oint d u^a & = & n^a \nonumber \\
\oint d\omega_R & = & \Omega
\label{circuitint}
\end{eqnarray}

where $\omega_R = {1 \over 2} (\vec{\nabla} \times \vec{u})$, 
the rotation of $\vec{u}$ on the time slice (Eq. \ref{asymconst}). 
Using Stokes theorem, these can be written as (tensorial) dislocation- 
($J^a_{\mu}$) and disclination currents ($I_{\mu}$) as

\begin{eqnarray}
J^a_{\mu} & = & \epsilon_{\mu \nu \lambda} \partial_{\nu} 
\partial_{\lambda} u^a,
\nonumber \\
I_{\mu} & = &  \epsilon_{\mu \nu \lambda} \partial_{\nu} 
\partial_{\lambda} \omega_R.  
\label{topcurrents}
\end{eqnarray}

These topological currents are not independent (the 
proper current is the `defect density', see section IIIC). 
Amongst others,  the disclination can be seen as a bound state 
of dislocations with parallel
Burger vectors, while at the same time dislocations can be regarded as 
bound disclination- anti-disclination pairs.
Disclinations act as sources for dislocations,

\begin{equation}
\partial_{\mu} J^a_{\mu} = -\varepsilon_{a \nu} I_{\nu},
\label{dislconservation1}
\end{equation}
with $\varepsilon_{a \nu}$ the two dimensional
antisymmetric tensor.

Disclinations are the bad players, greatly 
complicating the theory. Two simple statements can be made directly.
When dislocations and/or disclinations proliferate in the vacuum, the
crystal is immediately destroyed because of their topological nature.
Further, when they proliferate together the phase transition will, under
all circumstances, be first order. One might want to view conventional
solid-liquid transitions in this way yet this notion is not very useful 
as the transition is strongly first order and the physics stays
near the lattice constant, rendering field theory meaningless.  

Our central assumption is that a regime exists where disclinations
continue to be massive excitations, even when dislocations proliferate.
If this is the case, it follows
from Eq. (\ref{dislconservation1}) that the dislocation currents are
conserved,

\begin{equation}
\partial_{\mu} J^a_{\mu} = 0.
\label{dislconservation}
\end{equation}

This Bianchi identity simplifies the theory to such an extent that it
becomes completely tractable. 
The theory describing the proliferation
of dislocations becomes a straightforward extension of the well known 
Abelian Higgs duality, reviewed in appendix A. By definition, we call
disordered states satisfying this topological condition nematic
(quantum) fluids.

The `first gradient' theory Eq.(\ref{strainact}) 
governing the long wavelengths 
regime cannot supply the ingredients needed to distinguish the dislocation
and disclination masses. The reason is discussed in section IIA. The
lowest order theory does not allow an invariant expressing rigidity
associated with rotations. According to Eq. (\ref{rotationalstiffness}),
such a rigidity shows up first in second gradient elasticity and starts to
exert its influences at length scales less than the length $\xi_R$. 
Since disclinations are rotational defects one anticipates that their
energetics is influenced by the presence of such terms. As Kleinert
\cite{KleinertII} shows, this is indeed the case and he demonstrates 
that a sufficiently large rotational stiffness $C_R$ is sufficient condition 
for the existence of a nematic regime where Eq. (\ref{dislconservation}) is
obeyed. This shows that nematic states are entities which can be addressed
within continuum field theory. In condensed matter physics, however, the issue
will be decided by the specifics of the physics at the lattice constant.
It is questionable if the right conditions are ever met starting with a
closed pack lattice of spherical particles interacting via van der Waals
forces - the existence of a (super) hexatic state in 2+1D Helium is still
subject of controversy\cite{KleinertII}. On the other hand, as Kivelson
{\em et al.} pointed out, this might be
quite different starting out with the strongly anisotropic stripe-crystals
as found in high Tc cuprates and quantum-Hall systems\cite{kivemfrad}.

\begin{figure}[tb]
\begin{center}
\vspace{10mm}
\leavevmode
\epsfxsize=0.9\hsize
\epsfysize=10cm
\epsfbox{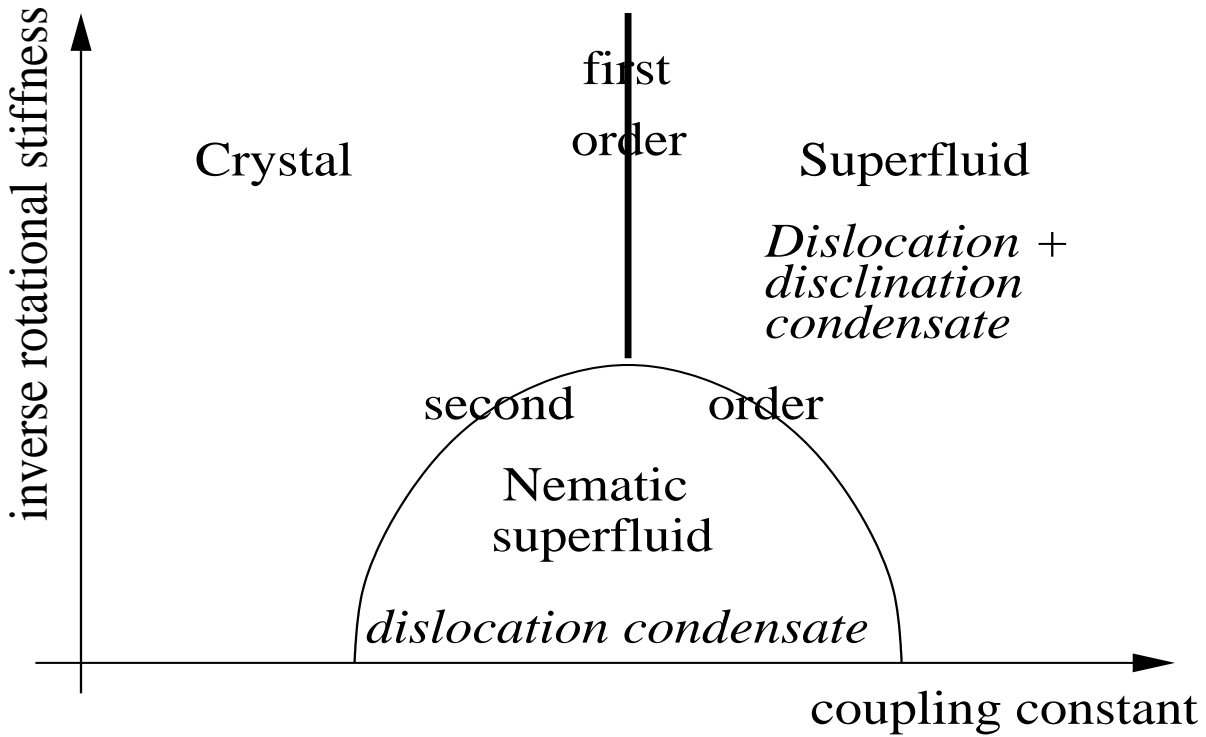}
\end{center}
\caption{The generic phase diagram associated with 2+1D Bosonic matter,
assuming that the field theoretic description based on quantum elasticity
is of relevance. When the rotational stiffness is small a first order
transition from the crystal directly into the superfluid is expected,
corresponding with a simultaneous Bose condensation of dislocations and
disclinations (or better condensation of `defects'). Upon increasing
the rotational stiffness, disclinations become expensive and at intermediate
coupling constants a dislocation-only condensate will emerge. These 
are the nematic superfluids which are at the focus of this paper.}
\label{nematicphase}
\end{figure}

Let us assume that the right microscopic conditions are present. 
In addition,
it appears to be necessary to explicitly forbid interstitials in order to
keep full control of the theory. Under these circumstances, one expects a
phase diagram with a topology as indicated in Fig. (\ref{nematicphase}). 
Whenthe rotational stiffness is to small, a first order transition 
from the crystal to the isotropic quantum fluid should occur 
when the coupling constant
increases. In the field theory, this isotropic fluid corresponds with a
combined dislocation/disclination (`defect') condensate. 
Upon increasing the
rotational stiffness, a tricritical point will occur where the first order
line bifurcates in two second order lines. The crystal first melts into a
nematic fluid, corresponding with a dislocation condensate, 
and at a larger
coupling constant a transition follows to the isotropic fluid. This is 
of course nothing else than the straightforward extension of the KTNHY
theory of 2D classical melting\cite{KTNHY} to the zero-temperature, 
2+1D quantum melting context.

This phase diagram is not surprising, and it should be taken as an input
for what follows. At this stage, it should come as a surprise to the
reader that we call both the nematic- and the isotropic fluids 
{\em superfluids}.
In the existing literature it is an automatic reflex to associate 
the superfluid
to the presence of interstitials coexisting with the topological 
condensates (e.g., ref.[\cite{Stoof}]), and we are considering the situation
where interstitials are explicitly forbidden. A main aim of this work will
be to prove that the pure dislocation condensate is at the same time a
conventional superfluid (sections VI, IX). The other novelty is associated
with the precise nature of the nematic regime. Again under the condition
that interstitials are excluded, we will find two distinct nematic phases 
(Fig. \ref{totalphase}): a phase displaying topological nematic order which
does not break rotational invariance, and a phase which does break rotational
invariance. The latter is, in a sense which will be explained, 
a `quantum smectic'. However, it is distinctly different from 
the quantum smectic
or `sliding phases' discovered by Kivelson et al.\cite{kivemfrad,slidingph}.
The latter occur under special strong Umklapp conditions and their fluid
characteristics reflect the motions of interstitials. Our quantum smectic 
originates in topology and the special dynamical condition of glide, introduced
in the next section.

\section{Duality in elasticity: the fundamentals}

After these preliminaries, we now arrive at the core of our
paper. In this section we will introduce the basic ingredients.
To address the quantum dynamics in 2+1D we need a
systematic mathematical formalism and, in this
regard, we
rely heavily on Kleinert's treatise 
\cite{KleinertII}. The first, and most crucial step is that we need
to dualize the familiar dynamical phonons (Goldstone modes associated
with crystalline order) into a gauge-theoretical `stress-photon'
representation. Instead of the familiar phonon fields expressed in
infinitesimal displacements one finds instead that the motions
are parameterized in terms of non-compact $U(1)$ gauge fields 
carrying a number of distinct flavors referring to Burgers vector
components. We are actually not aware of an explicit treatment
of the theory of stress-photons in the context of 2+1D quantum
elasticity and we will dedicate sections IV-V to a detailed
analysis of this counterintuitive affair. The advantage of this
unfamiliar representation is that the dislocations act as sources
for the stress photons, and the description of interacting dislocation
matter turns into a straightforward extension of electromagnetism
(section IIIB). 
Accordingly, the fluid state realized as the dislocations spontaneously
proliferate turns into a Bose-condensate of particles carrying stress
charge and this will turn out to be a close analogy of the electromagnetic
Meissner-Higgs state (section IIIC). A novelty is that the dislocation
currents and stress gauge fields carry the `Burgers flavors'. In section
IIIC we present a first main result: using general symmetry principles 
we derive the disorder field theory describing the collective behaviors of
dislocation matter. 
 Conventional nematic order (breaking of space-rotational symmetry
by a director order parameter) follows straightforwardly.  However,
depending on microscopic conditions, also a nematic state is possible 
which is 
characterized by a mere topological order. This is discussed in section
IIID where we establish the connection with the Ising gauge-theory
of Lammert, Rokshar and Toner\cite{Lammert}. Last but not least, 
in section IIIE we introduce the final 
ingredient which will play a central role
in the quantum theory: the glide principle. 
We will present in this section
the proof for the surprising fact that the glide constraint is equivalent 
to the requirement that translational symmetry breaking 
leads to the emergence 
of shear rigidity, leaving compression rigidity unaffected. After the
technical Sections IV-V, these various ingredients will
be brought together in Section VI where the true nature of the
nematic duals will be exposed.

\subsection{General considerations}

We will demonstrate that under the assumptions discussed
in the previous section, the transition between the elastic medium
and the nematic superfluids is governed in the scaling limit by
a straightforward extension of
the well known Abelian Higgs duality(e.g., see 
ref.\cite{KleinertI} and Appendix A,B).
 This turns out to be a feat
of symmetry. Abelian Higgs duality is associated with the 
quantum phase transition associated with a simple phase degree of
freedom (global $O(2)$ symmetry), and it demonstrates that the
$U(1)$ Meissner phase in 2+1D is dual to XY order. Among others,
it describes the quantum phase transition between the superfluid
and the Bose Mott-insulator. The theory of quantum elasticity 
is, because of its tensorial nature, far more complicated. However,
these complications are not essential so long as disclinations
are avoided. Dislocations are just about translations and their
Abelian nature makes possible that the theory
describing the proliferation of dislocations 
simplifies dramatically, to turn into a 
Abelian Higgs duality in disguise.

Goldstone modes are the excitations implied by order, and the
topological excitations are associated with the destruction
of this order. The Abelian Higgs duality (see Appendix A) 
rests on a simple transformation showing that the vortices associated with 
$O(2)$ (XY) long range order in 2+1D can as well be viewed as particles 
carrying electrical charge, interacting with each other via 
$U(1)$ gauge bosons (`photons') parameterizing  the long range 
interactions between the vortices mediated by the spin waves 
(Goldstone bosons). 
Hence, non-compact electromagnetism in 2+1D
can be seen literally as just a reparametrization of the 
physics of global $O(2)$ (dis)order. Starting out with a small
coupling constant in the global $O(2)$ `universe', vortices are
massive excitations appearing as vortex-antivortex pairs (closed
loops in space time). Upon increasing the coupling constant 
these loops grow until they become as large as the
size of the system. At this point, the vortices become real excitations 
proliferating in the vacuum. This corresponds with the quantum 
phase transition to the disordered, symmetrical state of the
$O(2)$ system. However, because in the dual `universe' vortices
are just bosonic particles carrying an `electrical' charge mediated
by photons, nothing prohibits the dual particles to Bose condense,
and this corresponds with a Meissner state because the dual system
is $U(1)$ gauged. The general lesson is that order and
disorder are just a matter of viewpoint. An observer having machinery
allowing him  to measure the XY degrees of freedom will insist that
his/her spins break symmetry spontaneously at low coupling constant with
symmetry restored at large coupling constant. Alternatively,
an experimentalist not knowing better that electromagnetism exists,
will insist that in his/her universe superconducting order sets in at
large coupling constant, getting destroyed upon decreasing the coupling
constant. We refer the reader unfamiliar with these notions to
Appendix A and B where we present a synopsis of this 
duality transformation.

Superficially, the theory of (isotropic) quantum elasticity, 
Eq. (\ref{strainact}), is quite similar 
to the $O(2)$ quantum non-linear sigma model of appendix A
(Eq. \ref{phasedynL}), with the displacement
fields $u^a$ of the former taking the role of the phase fields $\phi$
of the latter. A main difference is in the `upper' labels $a$, 
expressing that particles can move both transversally and longitudinally
relative to the propagation direction of the phonon, 
while XY spins can only precess.
However, considering matters more closely, the theory is much richer. 
The reason is symmetry. The symmetry principle behind the theory of 
elasticity is the breaking of the Euclidean group down to a lattice group
(in 2+1D elasticity: $E(2) \times O(2)$, $E(2)$ is the 2D Euclidean group,
$O(2)$ imaginary time dimension). 
The latter correspond with an infinite group formed from
discrete lattice translations- and rotations. In contrast to the XY
problem, we are dealing here with the complications of non-abelian
symmetry. 

The disclinations are the topological defects exclusively associated
with translations, and translations commute. Disclinations carry the
`magnetic' (dual) charges associated with the rotations and the 
difficulties coming from the non-abelian nature of the underlying 
symmetry become central when disclinations start to play a dynamical
role. Assuming that the disclination stay massive, the vacuum sector
can be described entirely in terms of the dislocations, and since
these involve only translations the theory is of an abelian nature and
tractable in principle. The theory is still more complicated 
than phase dynamics because dislocations carry vectorial
charges (the Burgers vectors) but these do not cause essential
complications. In essence, under dualization Burgers vectors turn into 
labels `flavoring' an abelian gauge theory with a structure
similar to the phase dynamics theory. Disclinations correspond with massive
particles appearing as excitations relative to the nematic vacua, and
their physics is in principle remarkably complex. The richness associated
with non-abelian duality structures becomes manifest on this 
level\cite{Muller} and we will leave this for future study.

\subsection{From phonons to stress photons}

Here we will present the first step of the duality: 
re-parameterizing the phonons in stress-photons and identifying 
the dislocations as sources. This is discussed at great length
in \cite{KleinertII}, and we just summarize here the main steps.
These follow the same 
pattern as the Abelian Higgs duality summarized in Appendix A.

Starting with the action Eq. (\ref{strainact}), we should 
take into account the defects that might be present. The 
presence of defects renders the 
displacement field configurations to become multivalued. In analogy 
with vortices, these can be made explicit. To this
end we introduce {\em plastic strain tensors},

\begin{equation}
w_{\mu, P}^a = \partial_{\mu} u^a_{MV}
\label{plasticstrains}
\end{equation}

where $u^a_{MV}$ singles out the multivalued (or `singular') configurations.
The elastic energy can only depend on the difference
between the elastic strains and the plastic strains. Hence, we should
insert for the strains in Eq. (\ref{strainact}) the total strain,

\begin{equation}
w_{\mu, tot}^a = w_{\mu}^a - w_{\mu, P}^a.
\label{elasticstrains}
\end{equation}

We now rewrite Eq. (\ref{strainact}) in terms of the
asymmetric strains $w_{\mu, tot}^a$,  keeping implicit the condition that
only the spatially symmetric strains enter the action. 
Next, we introduce auxiliary fields $\sigma^a_{\mu}$ and apply the
Hubbard-Stratanovich transformation to Eq. (\ref{strainact}),

 \begin{eqnarray}
Z & = & \int D u_{a} \int D \sigma_{\mu}^{a} \exp[-\int \tilde{\cal{L}}
 [u_{a},\sigma^{a}_{\mu}]]~ d\Omega] \nonumber \\
\tilde{\cal{L}} & = & -\frac{1}{2} \sigma^{a}_{\mu} C_{\mu \nu ab}^{-1}
 \sigma^{b}_{\nu} + i \sigma^{a}_{\mu} ( w^a_{\mu} - w_{\mu, P}^a ).
\label{stressact0}
\end{eqnarray}

The elastic strains $w^a_{\mu}$ correspond with the single valued 
displacement field configurations. Accordingly, 

\begin{eqnarray}
i \sigma^a_{\mu} w^a_{\mu} & = &  i \sigma^a_{\mu} \partial_{\mu} u^a,
\nonumber \\
  & = & - i u^a \partial_{\mu} \sigma^a_{\mu}.
\label{shiftder}
\end{eqnarray}

The derivative can be shifted as the fields are integrable. 
We observe that the smooth fields $u^a$ just enter as a Lagrange multiplier.
After an integration, a Bianchi identity for the  $\sigma$ fields follows,

\begin{equation}
\partial_{\mu} \sigma^a_{\mu} = 0.
\label{stressconser}
\end{equation}

Thus these fields are conserved, independently, for every
upper label $a$. The $\sigma$ fields are none other than the familiar 
stress fields, and the above operation merely corresponds with the
standard stress-strain duality. Eq. (\ref{stressconser}) encapsulates
the conservation of stress. By including the time axis ($\mu = \tau$),
Eq. (\ref{stressconser}) automatically represents the equations of motion. 

Alternatively, the stress action can be obtained by varying the action
with respect to the strains,
\begin{equation}
\delta S = C_{\mu \nu a b} w^a_{\mu} \delta w^b_{\nu},
\label{varyS}
\end{equation}
and the stress fields are, by definition,
\begin{eqnarray}
\sigma^b_{\nu} & = & {  {\delta S} \over {\delta w^b_{\nu}} }
\nonumber \\
 & = &   C_{\mu \nu a b} w^a_{\mu}.
\label{strestra}
\end{eqnarray}

Hence, stress can be directly expressed in terms of strain and this will turn
out to be quite useful on several occasions. Specializing to the
isotropic 2+1 dimensional case ($\hbar= c_{ph}=1$),

\begin{eqnarray}
\sigma^a_b & = & 2 \mu w^a_b + ( \kappa - \mu) \delta_{a b} ( w^x_x + w^y_y)
\nonumber \\
\sigma^a_{\tau} & = & \mu \partial_{\tau} u^a.
\label{stressstrainiso}
\end{eqnarray}

A second constraint follows from the fact that only the strains 
symmetrical in the space-indices enter the strain action.
One has to add this as a constraint in the 
above duality transformation and it follows immediately that the
stress fields have to be symmetric in the space-indices,

\begin{equation}
\sigma^x_y = \sigma^y_x.
\label{stresssym}
\end{equation} 

This condition is known in the elasticity literature as
the Ehrenfest theorem\cite{LandauLifel}.

Following Kleinert, let us now depart from the classic treatise
of elasticity by realizing that, like in the Abelian Higgs
problem,  the conservation law Eq. 
(\ref{stressconser}) implies that the dynamics can be parameterized
in terms of $U(1)$ gauge fields $B_{\mu}^a$. As compared 
to the XY case, the  difference is
that these gauge fields now carry an additional `Burgers flavor' $a = x, y$,

\begin{equation}
\sigma^a_{\mu} = \epsilon_{\mu \nu \lambda} \partial_{\nu} B^a_{\lambda}.
\label{stressgauge}
\end{equation}

We stress that these fields are {\em not} two-forms, but instead one-forms
with additional `internal' degrees of freedom $\sim a$ which suffice to
keep track of the translations. In
terms of the `stress photons'  $B^a_{\mu}$, the theory of elasticity turns 
into a form of electromagnetism `flavored' by the Burgers labels,

 \begin{eqnarray}
Z & = & \int {\cal D} B_{\mu}^{a} \; \delta ( \partial_{\mu} B^a_{\mu})
\exp{\left[  -\int d\Omega \;
{\cal L}^{dual}_0 ( B^a_{\mu})\; \right] }
\nonumber \\
{\cal{L}}_0^{dual} & = &  \frac{1}{2} \;
\epsilon_{\mu \nu \lambda} \partial_{\nu} B^a_{\lambda} \; 
C_{\mu \mu' ab}^{-1} \;
\epsilon_{\mu' \nu' \lambda'} \partial_{\nu'} B^b_{\lambda'}
\label{stressmaxwell}
\end{eqnarray}

excluding the gauge volume in the measure. The
Ehrenfest condition Eq. (\ref{stresssym}) should also be imposed on
the gauge fields,

\begin{equation}
\partial_y B^y_{\tau} - \partial_{\tau} B^y_{y} =
- \partial_x B^x_{\tau} + \partial_{\tau} B^x_{x}.
\label{ehrengauge}
\end{equation}

Up to this point we have just re-parameterized the overly familiar 
theory of acoustic phonons in a theory describing force carrying
gauge bosons. Although it is still a free Abelian gauge theory,
one already anticipates that it looks quite different from the 
familiar phonon representation. The next two sections are devoted to 
technicalities needed to uncover the physics in this mathematical
construction. The advantage becomes obvious considering the singularities.

We have yet to deal with the plastic strain tensors appearing   
in Eq. (\ref{stressact0}). These are associated with multivalued
displacement field configurations $u^a_P$ which may not be 
integrated out,

\begin{eqnarray}
{\cal{L}}^1_{dual} & = & i \sigma^a_{\mu} w^a_{\mu, P} \nonumber \\ 
& \equiv & i \sigma^a_{\mu} \partial_{\mu} u^a_{P} \nonumber \\
  & = & i \epsilon_{\mu \nu \lambda} \partial_{\nu} B^a_{\lambda}
\partial_{\mu} u^a_{P}
\nonumber \\
  & = & i B^a_{\mu} J^a_{\mu}.
\label{dislosource}
\end{eqnarray}

The dislocation currents $J^a_{\mu} = \epsilon_{\mu \nu \lambda} 
\partial_{\nu} \partial_{\lambda} u^a$ (Eq. \ref{topcurrents}) are recognized
as sources for the stress gauge fields. In `stress electromagnetism',
the dislocations have the same role as charge particles have in 
normal electromagnetism, at least in 2+1D. Elasticity,  on this level,
is sufficiently similar to the XY problem in 2+1D that the essence
of Abelian Higgs duality is still applicable. The Goldstone modes
dualize in gauge fields, while the topological defects acquire the
role of sources. Elasticity is richer in the regard that
the theory is flavored by the Burgers labels, reflecting the fact
that translations are more interesting than an internal 
$O(2)$ symmetry. However, although these make the 
problem richer, they do not pose 
a problem of principle.  

Another matter is that the above duality is suffering from the
shortcoming that the disclination currents are implicit; they just
enter as sources for the dislocation currents, Eq. (\ref{dislconservation1}). 
The disclinations can be made explicit in the formalism 
by using the double curl
gauge fields discovered by Kleinert\cite{KleinertII}. 
Schematically, the stress fields are parameterized by,

\begin{equation}
\sigma_{\mu}^a = \epsilon_{\mu \nu \lambda} \epsilon_{a \nu' \lambda'} 
\partial_{\nu} \partial_{\nu'} h_{\lambda \lambda'},
\label{doublecurl}
\end{equation}  

where $h_{\mu \nu}$ are genuine two-forms.  It is readily found that these
fields are minimally coupled to a source $\sim i h_{\mu \nu} \eta_{\mu \nu}$ 
where $\eta_{\mu \nu}$ 
corresponds to a two-form called `defect density'. This can be
decomposed as

\begin{equation}
\eta_{\mu \nu} = \eta^{\Omega}_{\mu \nu} + { 1 \over 2} 
\left[ \epsilon_{\lambda \mu a} \partial_{\lambda} J^a_{\nu}
+ \epsilon_{\lambda \nu a} \partial_{\lambda} J^a_{\mu}
- \epsilon_{\mu \nu a} \partial_{\lambda} J^a_{\lambda}
\right],
\label{defectdensity}
\end{equation}
where the $J_{\mu}^a$'s are the dislocation currents while 
$\eta^{\Omega}_{\mu \nu}$ corresponds with the `full' disclination
currents. In terms of these double curl fields the dynamics of
disclinations in the dislocation condensate  can be directly addressed --
see ref.\cite{KleinertZaanen} for the Lorentz-invariant case. 
One infers that at the moment that disclinations start to play a dynamical
role there is a need for a different kind of theory. However, imposing
the nematicity condition just means that disclinations can be neglected
and under these circumstances dislocations are just like flavored 
vortices.

\subsection{The structure of the dual disorder field theory.}

In the previous paragraphs we introduced a useful formalism
for the description of the physics of isolated (`classical')
dislocations. Dislocations are sources of kinetic energy. These
will occur in any crystal as virtual excitations in the form of
closed dislocation-antidislocation loops in space time. Upon
increasing the coupling constant, the characteristic dimension
of these loops will grow, until a loop blow out will occur,
signaling the transition to the quantum fluid state. On the
disordered side of the phase transition, one is dealing with quantum
matter composed from strongly interacting defects,
exhibiting a dual order. In the case of the Abelian-Higgs
problem this is just a Bose condensate of vortices exhibiting a
dual Meissner effect because of the coupling to the dual photons.
Resting on universality, this state is just described by a dual
Ginzburg-Landau-Wilson theory describing the `disorder parameter'
dynamics corresponding with the superconducting order parameter
expressing the off-diagonal long range order of the vortices. 
In this section we will derive the
form of the universal disorder field theory describing dislocation
matter. To the best of our knowledge our treatment is novel; from
this point onward we depart from the established wisdoms.  

As compared to the vortices, disclination condensates are 
more complicated than vortex condensates because both the stress
gauge fields and the dislocation currents are flavored by the
Burgers labels. However, these don't pose a fundamental problem
as we will now show.
As a consequence of the abelian nature of the translations, the
dislocation currents can be factorized,

\begin{equation}
J^a_{\mu} (\vec{r} ) = \delta^{(2)}_{\mu} (L, \vec{r}) \; n_a, 
\label{linedelta}
\end{equation}

where $\delta^{(2)}_{\mu} (L, \vec{r})$ is the line delta function
specifying the locus of the dislocation world-line in space-time $\vec{r}$,
and $n^a$ is the $a$-th component of the Burgers vector.
 This factorization makes possible to follow the standard
strategy to obtain the effective Ginzburg-Landau disorder field-theory
describing the tangle of dislocation world-lines in terms of a complex
scalar G-L field $\Psi = |\Psi| e^{i \phi} $, with the interpretation 
that $|\Psi|^2$ corresponds the density of dislocations in the condensate 
while the phase field $\phi$ parameterizes the entanglement (see e.g.
ref. \cite{KleinertI}, appendix A,B). 

The minimal coupling between the dislocation currents and the stress
gauge fields can be written as ($d\Omega$ space-time volume element),
\begin{eqnarray}
S_{BJ} & = & \int J^a_{\mu} B^a_{\mu} d\Omega \nonumber \\
       & = & \int  \delta^{(2)}_{\mu} (L, \vec{r}) n_{a}  
              B^a_{\mu} d\Omega \nonumber \\
       & = & \int n_a B^a_{\mu} d r_{\mu}.
\label{mincoupexp}
\end{eqnarray}
The canonical momentum follows immediately,
\begin{equation}
P_{\mu} = p_{\mu} + n_a B^a_{\mu}.
\label{canonicalmomentum}
\end{equation}

We observe that the effective gauge field in the Hamiltonian
formulation is just $\vec{n} \cdot \vec{B}_{\mu}$. This simple
result is very important. As we are dealing with a line-entanglement
problem, the disorder field theory should have the form of a 
Ginzburg-Landau-Wilson action. The kinetic part of this action should
be $\sim P^2$. Upon the substitution with $p_{\mu} = i \partial_{\mu}$
this acts on some disorder field $\Psi$. In case one is dealing with a
vector order parameter theory based on an internal symmetry, the disorder
field carries itself the vector labels $\sim \Psi^a$ 
(e.g., the `soft spin' parameterization of non-linear sigma models). 
Here the situation is different.
The Burgers vectors themselves refer to space and a-priori one cannot 
exclude that the covariant derivatives would explicitly depend on the
Burgers labels as well, such that these would act in a non-trivial way
on the order parameter field $\Psi^a$. Eq. (\ref{canonicalmomentum}),
however, demonstrates that the canonical momentum does not carry an
explicit dependence on the Burgers vectors because of the product
with the stress gauge fields. It just depends on the space-time
direction $\mu$. 

This will become crucial when we develop further
the disorder field theory. We will find that the glide constraint
(section IIIE) has the unusual consequences that the dislocation
worldline loops are oriented in planes spend by the Burgers vector
and the time direction. One could be tempted to think that this
would have the effect that the dual order parameter field could 
become one dimensional. However, the above argument shows that 
this is never the case: the ordering field knows about the full
embedding space. In appendix B these matters are analyzed from the
loop gas perspective and the outcome is fully consistent with
the Landau style derivation of the previous paragraph. 

Let us start out assuming that 
the field acted on by the covariant derivatives  
is just  a simple complex scalar field $\Psi$.  We find 
for the piece of the disorder field theory involving the gauge fields,

\begin{equation}
S_{LR} = \int d\Omega \left[ | ( \partial_{\mu} - i n_a B^a_{\mu} )
\Psi |^2 \; + \;   \frac{1}{2} \sigma^{a}_{\mu} c_{\mu \nu ab}^{-1}
 \sigma^{b}_{\nu} \right].
\label{Slongrange}
\end{equation}

In addition, a mass term $\sim m^2 |\Psi|^2$ is present. The
mass controls the size of the
dislocation loops. These have a finite length for $m^2 >0$,  while
the loops `blow out' when $m^2 < 0$. Finally, a term $\sim w \Psi^4$ has to
be added describing the short range interactions\cite{KleinertI}. We 
observe that $|J^a_{\mu}| \sim |\Psi| n_a$ with the implications
that the Burgers vectors enter the disorder field theory like
$m^2 |\Psi|^2 n^2 \hat{n_a} \hat{n_a} + w |\Psi|^4 n^4 ( \hat{n_a} )^2
( \hat{n_b} )^2$. In the absence of disclinations one has however to
impose the condition
that in a finite volume of space no `ferromagnetic
polarization' of Burgers vectors can occur: Burgers vectors
have to be locally anti-parallel. The reason for this {\em local}
charge neutrality condition is topological. A finite uniform
`Burgers polarization' corresponds with a disclinations and
these we assumed to be massive. 
The local `Burgers vector neutrality'  implies that the only allowed
invariants are scalars and  traceless tensors,

\begin{equation}
Q_{ a b} = |n|^2 \left( \hat{n}_a \hat{n}_b - {1 \over 2} \delta_{a b} \right)
\label{directors}
\end{equation}

These describe directors in two dimensional space,  `$O(2)$ vectors 
with head and tails identified'. Because $Q_{ab}$ is traceless, the
mass term cannot depend on it. It follows that the 
disorder field theory has  the form,

\begin{equation}
S_{\Psi} = S_{LR} +  \int d\Omega \left[ m^2_{\Psi} | \Psi |^2 +
w_{\Psi} |\Psi|^4 - r |\Psi|^4 Q_{a b} Q_{b a} \right].
\label{disorderpsi}
\end{equation}

The `director sector' associated with the $Q_{a b}$'s should have its own 
dynamics. When dislocations with anti-parallel Burgers vectors collide
they might annihilate pairwise and re-emerge with a different overall
director orientation. The director does not commute with the Hamiltonian
and is therefore subjected to fluctuations. These can be incorporated
by adding,

\begin{equation}
S_Q = \int d\Omega \left[ \partial_{\mu} Q_{ a b} \partial_{\mu} Q_{b a}
+ m_Q^2 Q_{ab} Q_{ba} + w_Q Q_{a b} Q_{b c} Q_{c d} Q_{d a} \right]
\label{directoraction}
\end{equation}
 
corresponding with a non-linear sigma model for the traceless tensors $Q_{a b}$
in a soft spin representation.
Since directors can only exist when the dislocations have proliferated,
one should impose that
$m_Q^2 > 0$ although the total mass $m_Q^2 - r |\Psi|^4$ can become negative
( $r > 0$ should be imposed). Notice that for the `semi-circle directors' of
relevance to 2D space cubic invariants are forbidden while they are allowed
for the projective plane director order parameters of relevance to 3D space.

Using just the duality notion and symmetry principles, we have
derived a description of nematic liquid crystalline order which
is radically different from the standard interpretations found in
textbooks on liquid crystals. In spirit it is of course quite similar to
the KTNHY-theory of melting in
two classical dimensions\cite{KTNHY}. It can be viewed as a generalization of KTNHY
to 2+1 quantum- or 3 classical dimensions although in some essential
regards our formulation is more complete. Most importantly, we arrive
at a physical interpretation of the
{\em `topological nematic
order'} identified by Toner, Lammert and Rokshar on basis of abstract
gauge theoretical arguments. This has been overlooked by KTNHY,
and the possible existence of such a state implies that the standard
order parameter theory of nematic states is in essential regards 
incomplete and potentially misleading. Let us explain this in more
detail.    

\subsection{The topological- or `Coulomb' nematic.}

The order parameter  theory for nematics was established in the 
early 1970's by de Gennes and others\cite{deGennes}, 
and it is 
of precisely the form Eq.(\ref{directoraction}). However, the
physical interpretation is entirely different. The physical
perspective is that of kinetic gas theory. A gas of rod-like
molecules is considered  and it is assumed that potential energy
is gained at collisions when the long axis of the molecules
line up. The $Q$'s parameterize this size anisotropy and  
Eq.(\ref{directoraction}) follows after averaging. The macroscopic
director order parameter reflects, in a direct way, the properties
of the microscopic constituents and the director is viewed as an
internal symmetry (the orientations of the molecules) detached
from the symmetries of space-time. In the language of this paper,
it is a theory addressing the degrees of freedom of the interstitials.

Here we take the opposite perspective. The starting point is
the state breaking the space symmetries maximally
(the crystal, elastic medium) and disordered states are derivatives of
this maximally ordered state.  These correspond with condensates of the
topological excitations associated with the maximal order. Since 
crystalline order involves the breaking of the Euclidean group down
to a lattice group, the charges of the topological excitations 
(Burgers-, Franck vectors) relate to the symmetries of space itself,
and the order parameters characterizing the partial orders refer to
the breaking of the  space symmetries, and not to some internal
symmetry. In this perspective liquid crystalline orders correspond
with dual
orders in terms of the disorder parameters and these allow for
a richer structure. KTNHY\cite{KTNHY} follow, in essence, 
the same path as we do but their attention is limited to the breaking
of  manifest symmetries (algebraic
translational and algebraic rotational order in the `solid' and
`hexatic' phases, respectively). There is, however, yet another phase
which can only be understood in terms of the disorder parameter
structure and is best called a state characterized by topological
nematic order.

As we already repeatedly emphasized,
a state should be called nematic when it is a liquid characterized 
by massive disclinations carrying quantized, sharply defined, Franck
vectors. This topological definition is more general than the usual
definition stating  that a nematic state is one which is translationally
invariant while it breaks spontaneously rotational symmetry. It
follows immediately from Eqs. (\ref{disorderpsi},\ref{directoraction})
that a state exists which is {\em not} 
breaking rotational invariance while
it is a nematic in the topological sense.

\begin{figure}[tb]
\begin{center}
\vspace{10mm}
\leavevmode
\epsfxsize=0.9\hsize
\epsfysize=10cm
\epsfbox{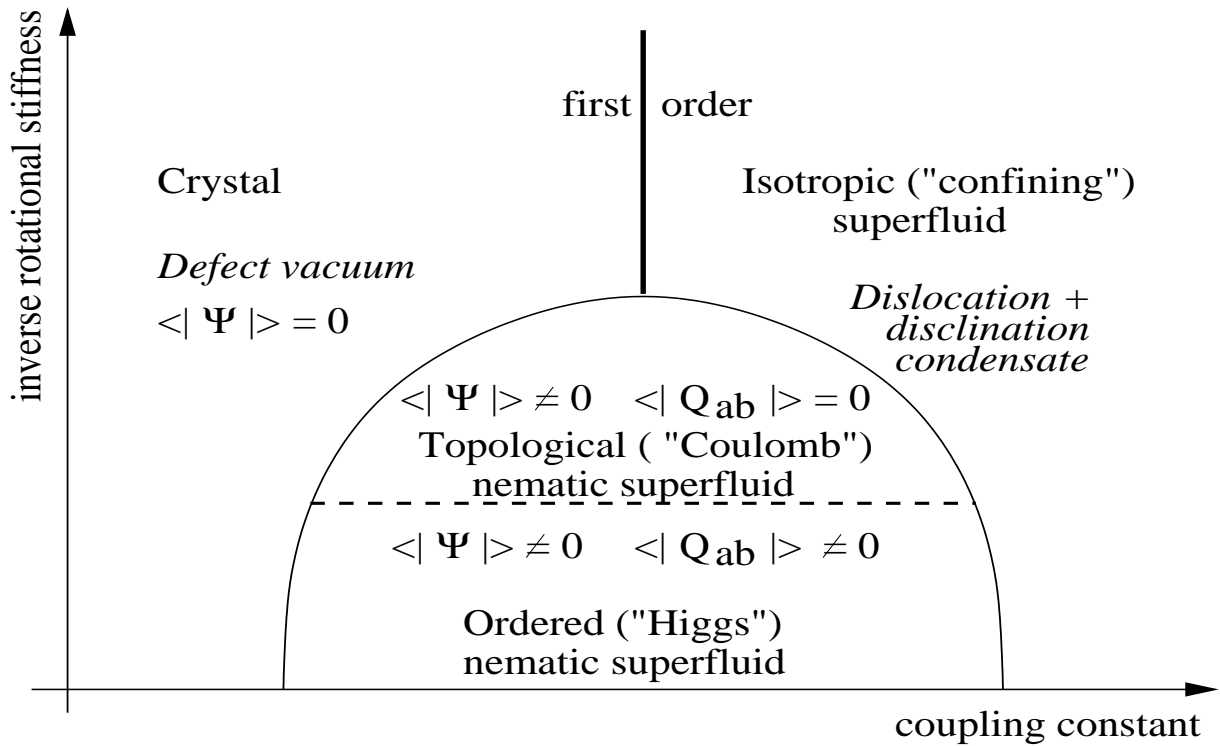}
\end{center}
\caption{A refinement of the phase diagram Fig. 5 based
on the dual disorder field theory described in this section. The nematic
states are associated with Bose condensate of dislocations having a 
primary order parameter $\langle | \Psi | \rangle$. In addition, the
states are characterized by a secondary director order parameter
$\langle | Q_{ab} | \rangle$ which is referring to the breaking of
space-rotational invariance. In the dislocation condensate the
director might (ordered nematic) or might not (topological nematic)
acquire an expectation value. In both cases the disclinations are
massive, condensing only at the phase boundary with the isotropic
superfluid. This offers an explanation of the states first identified
by Lammert, Toner and Rokhsar [3] using abstract gauge
theoretical arguments.}  
\label{totalphase}
\end{figure}

Let us consider Eqs.(\ref{disorderpsi},\ref{directoraction}) in 
more detail. This theory only makes sense in the parameter regime where
$\Psi$ is the primary order parameter.
The dislocation loops have first to blow out, and only when
free dislocations  
occur at a finite density it is meaningful to consider the order 
of their Burgers vectors. This implies that $w_{\Psi}  - r |Q|^2 > 0$
to assure stability, while $m^2_Q > 0$ in order to prevent the $Q$'s
to condense before the $\Psi$'s have condensed. Under these conditions,
the directors will always follow parasitically the Bose-condensation
of the dislocations. In this regime, the effect of the mode-coupling 
will be to renormalize the mass of the director field into
$m^2_{Q, eff} = m^2_Q - r |\Psi|^4$. The squared amplitude 
$|\Psi|^2$ will grow like
$ \left[ ( g - g_c ) / g_c \right]^{2 \beta}$ as function of the deviation
from the critical coupling constant $g_c$ with $\beta$ 
being the order parameter
exponent, and it follows that $m^2_{Q, eff} = m^2_Q - r       
\left[ ( g - g_c ) / g_c \right]^{4  \beta}$. If $m^2_Q = 0$, the director
order parameter switches on parasitically, directly  upon entering the
the dislocation Bose condensate. On this level, the transition
appears as a 3D XY transition from the solid to a conventional nematic.
However, when  $m^2_Q > 0$, the dimensionless coupling constant
 $(g-g_c)/g_c$ has to exceed
a critical value before the effective director mass turns negative at
a coupling constant $g_c'$. Hence, in between $g_c$ and $g_c'$, a state 
exists which does
not exhibit director order (rotational symmetry breaking). At the same time,
it is not a normal isotropic fluid because disclinations are massive. 
We conclude that
a nematic state exists with an order which can only be measured
by non-local means, by the insertion of disclinations, and it therefore
corresponds with a truly topological order. One expects such a state
to become stable at relatively small rotational stiffness and the topology
of the phase diagram should be as indicated in Fig. (\ref{totalphase}).

Obviously, such a state cannot be imagined starting from a gas of rods.
In order to make it work, one needs that the director is a composite 
of vectors, where the vectors can have a life of their own, while at the
same time it is necessary that one can identify disclinations without
referral to  director fields. This is unimaginable, starting
from the gas limit. Nevertheless, this topologically ordered state was
identified before, using abstract but elegant arguments based on gauge
invariance. Lammert, Rokshar and Toner\cite{Lammert} realized that 
the theory   
Eq.(\ref{directoraction}) is characterized by an Ising gauge 
symmetry: the action is invariant under $\vec{n} \rightarrow - \vec{n}$
because the physical meaningful entity is $Q \sim n^2$. They argued
that the theory can be generalized by making this gauge symmetry explicit
in terms of the theory of rotors minimally coupled to $Z_2$
gauge fields. This can be regularized on a lattice as,

\begin{equation}
S = \int d\Omega \left[ - J \sum_{<ij>} \sigma_{ij} \vec{n}_i \cdot
\vec{n}_j - K \sum_{plaq} \sigma \sigma \sigma \sigma \right]
\label{Z2Odgauge}
\end{equation}

describing rotors $\vec{n}$ living on the sites of the square lattice
coupled through links where Ising valued ($\pm 1$) fields $\sigma_{ij}$
live, determining the sign of the `exchange' interactions. The second term
is the Ising Wilson action corresponding with the product of the Ising 
gauge variables encircling the plaquettes. This theory is invariant under
the gauge transformations corresponding with flipping all $\sigma$'s 
departing from a given site and simultaneously multiplying the vector
on the same site by $-1$. It is well known that theories 
like Eq. (\ref{Z2Odgauge})
have three phases: (1) The Higgs phase corresponding with ordered rotors
turning into directors under gauge transformations, i.e. just the
ordered nematic, (2) The confining phase, where the rotors are disordered,
while also the gauge fluxes have proliferated. This is in one to one
correspondence with the state where conventional disclinations have
condensed, i.e. the isotropic fluid. (3) The Coulomb phase, which is the 
surprise. The rotors are disordered but the gauge fluxes (or `visons') are
still massive. This state carries the topological order.

Obviously, such a Coulomb phase cannot be imagined starting from the
conventional `gas of rods' perspective. However, in terms 
of the elastic duality, it  acquires a simple physical interpretation.
 The Ising gauge fluxes are just the $\pi$ disclinations associated
with the solid, and the vectors are in one to one  
correspondence with the  Burgers vectors. The gauge symmetry is 
associated with the very physical local constraint coming from
the Burgers vector neutrality condition. Surely, besides the
`conventional'  ordered nematic/Higgs- and isotropic 
fluid/confining phase there is nothing against a phase where 
Burgers vectors refuse to order while the 'crystal' 
disclinations are still massive. 

This Coulomb phase will be at center stage in much of the remainder 
of this paper for technical reasons:
it is a simpler state than the ordered nematic, and the properties 
of the latter are easily deduced from those of the former.

\subsection{Dynamics and the glide principle.}

The next novelty has everything to do with the `quantum' in the title
of this paper. Up to this point it could have been as well a treatise
on 3D classical elastic matter, except for some symmetry adjustments. Quantum
physics implies that space- and time have to be considered simultaneously and
starting from non-relativistic matter (see ref. \cite{KleinertZaanen}
for the relativistic extension) dislocations are subjected to a fundamental,
yet independent {\em dynamical} constraint.  
 In classic elasticity theory\cite{Nabarro} this is known as the
{\em glide principle}: dislocations can only move in the direction of
their Burgers vector. In the present  field-theoretic formulation, 
this dynamical principle has remarkably far reaching implications.

We already encountered twice the condition that the fields relevant
to elasticity are symmetric tensors in the space indices. This
is true for the strain- and stress fields, Eqs. (\ref{asymconst}, 
\ref{stresssym}),
but it should also be imposed  on  the 
dislocation currents,

\begin{equation}
J^x_y = J^y_x
\label{glidecondition}
\end{equation}

This is less obvious than one might think at first. 
It is an independent constraint and it is readily seen that
it cannot be deduced from the symmetry of stresses and
strains. In the dynamical (quantum) context, it can be
viewed as an additional condition descending from the ultraviolet,
needed to keep the theory recognizable as a theory derived from
the microscopic breaking of translational invariance. Let us present the proof
that the glide condition is needed to ensure that the dislocations are the
disorder operators associated with the restoration of translational
symmetry.

\begin{figure}[tb]
\begin{center}
\vspace{10mm}
\leavevmode
\epsfxsize=0.6\hsize
\epsfysize=5cm
\epsfbox{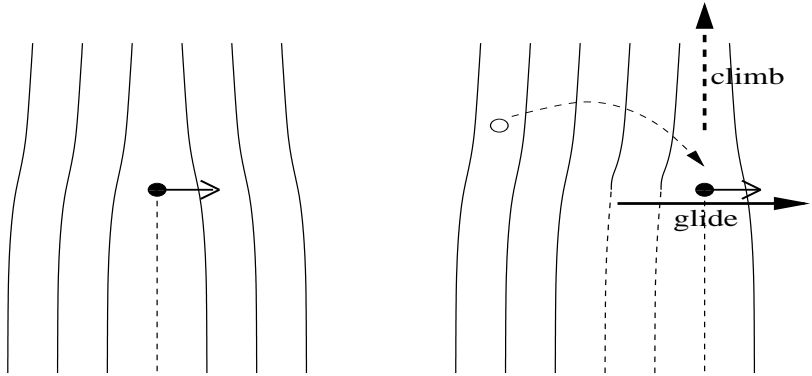}
\end{center}
\caption{Even in the complete absence of interstitials (a requirement
for the validity of the field theory), dislocations can propagate
easily in the direction of their Burgers vector (`glide motion'): cut
the neighboring row of atoms, move over the `tail' and invoking only
local atomic motions the dislocation has been transported. In order 
to move in a direction perpendicular to the Burgers vector, the `row
of atoms' has to become longer and this is only possible in the
presence of delocalized interstitials. In the field theory, the glide
constraint acquires a central role.}   
\label{glidepict}
\end{figure}

Eq. (\ref{glidecondition}) has the implication that dislocations
move {\em exclusively} in the direction of their Burgers vector. This can
be easily seen by rewriting the dislocation currents in terms of
$\delta$ functions acting on the time slice, $\vec{n}$ is the Burgers vector, 
$\vec{\bar{r}} = (x, y)$ the position of the dislocation on the time slice, 
\begin{eqnarray}
J^{a}_{b} & = & \epsilon_{b\nu\lambda}\partial_{\nu}\partial_{\lambda}
u^{a}_{P}
\nonumber\\
& = &\delta^{(2)}_{b}(L)n_{a}\equiv n_{a}\int_{L}d\tau
'\frac{d\bar{r}_{b}}{d\tau '} \delta^{(3)}(\vec{r}-\vec{\bar{r}})
\nonumber\\
& = & n_{a}\int_{L}  d\tau '\dot{\bar{r}}_{b}(\tau ')
\delta^{(2)}(\vec{r} - \vec{\bar{r}})\delta^{(1)}(\tau-\tau
')
\nonumber\\
& = & n_{a}\dot{\bar{x}}_{b}(\tau)\delta^{(2)}(\vec{r}-\vec{\bar{r}}(\tau)))
\nonumber \\
& \equiv & n_{a}v_{b}\delta^{(2)}(\vec{x}_{1,2}-\vec{\bar{x}}_{1,2})
\label{vj's}
\end{eqnarray}

where $\vec{v} = ( v_x, v_y)$ is the dislocation velocity and the
subscripts in $\vec{x}_{1,2}$ refer to
the spatial planar projection of the general 
space-time coordinate. It immediately follows that

\begin{equation}
 J^x_y - J^y_x  =  \left( n_x v_y -  n_y v_x 
\right) \delta^{(2)}(\vec{x}_{1,2}-\vec{\bar{x}}_{1,2}) = 0,
\label{glidedyn}
\end{equation}

implying that $ \vec{v} \times \vec{n} = 0$, meaning
that the dislocation velocity is finite only in the direction 
parallel to the Burgers vector $\vec{n}$.

Starting from the crystal lattice it is easily seen why this condition
has to be imposed. This is just the usual argument found
in the classic elasticity textbooks\cite{Nabarro}, becoming precise in the
zero temperature limit. As discussed, to have a meaningful
field theory, we have to insist that the interstitial degrees of 
freedom are
virtual. This implies that the `elementary' particles constituting
the solid cannot be transported. Consider the dislocation motions
on a microscopic scale. The dislocation with unit Burgers vector 
corresponds with the endpoint of a row of `atoms' (in 2+1D) of 
half-infinite length. The `glide' motion in the direction of 
the Burgers vector (Fig. \ref{glidepict}) is still possible, 
even under the condition that the elementary matter cannot flow: 
cut the row of `atoms' adjacent to the dislocation into a 
dislocation-antidislocation pair and annihilate
the original dislocation with the newly created anti-dislocation.
The effect is
that the dislocation has moved over one lattice spacing. This 
requires only  microscopic motions of individual particles and is therefore
not violating the no-interstitial constraint. The `climb' motion
in the direction 
perpendicular to the Burgers vector is a different matter. In this
case, the length of the half-infinite row of atoms has to increase or
decrease and this necessarily involves the presence of delocalized
interstitial particles. As these are not present, climb motion
is impossible. At finite temperatures, interstitials are always present.
The glide condition is thereby no longer rigorous, although it
is still invariably true that in real solids the conservative glide
motions are much easier than the climb motions, driven by the diffusion
of interstitials.  
               
The theory of elasticity describes two distinct rigidities: shear-
and compression rigidity. Assuming that the theory
describes the long wavelength physics associated with crystalline
order, the capacity to carry shear forces is a {\em consequence}
of the breaking of translational invariance. At the same time,
compression rigidity is a much more general property which does
not require the breaking of translational invariance. Also fluids
and gases carry pressure.

Dislocations are the topological excitations which are exclusively
associated with the restoration of translational symmetry. Accordingly,
dislocations can only interact with the {\em shear} components of
the stresses carried by the medium. By principle, dislocation currents 
cannot couple to the components of the stress gauge fields $B$ 
responsible for compressional stresses and the relevant charge
in the stress-gauge field formalism should vanish. 
If this charge would be finite, it would lead 
to the absurdity that the fluid only carries
short range compression forces. It is straightforward to prove
that the  vanishing of the couplings between
dislocations and compressive stresses is identical
to the requirement that the space-like dislocation currents are 
symmetric tensors, Eq. (\ref{glidecondition}).

In full generality, the action of a medium carrying exclusively 
compression rigidity is,
\begin{equation}
S = { 1 \over \hbar} \int d^2x d \tau \left[ 
{\kappa \over 2} ( w_x^x + w_y^y )^2 + {\rho \over 2} ( 
(w^x_{\tau})^2 + (w^y_{\tau})^2 ) \right].
\label{Scom}
\end{equation}

Using Eq.(\ref{stressstrainiso}), we find for the stress fields

\begin{eqnarray}
\sigma^x_x & = & \sigma^y_y = \kappa ( w^x_x + w^y_y ) \nonumber \\
\sigma^x_y & = & \sigma^y_x = 0 \nonumber \\
\sigma^a_{\tau} & = & ( \rho / 2) \partial_{\tau} u^a
\label{dualcompr}
\end{eqnarray}

As compared to the general case, this involves a number of extra
constraints. Expressing the stress fields in terms of the
stress gauge fields via $\sigma^a_{\mu} = \epsilon_{\mu \nu \lambda}
\partial_{\nu} B^a_{\lambda}$ the constraints Eq. (\ref{dualcompr})
 are uniquely resolved by the
`compression gauge': all $B$'s are zero except for $B^x_y = - B^y_x =
\Phi$, i.e. compression is governed by a single scalar field $\Phi$.
The stress fields become
$\sigma^x_{\tau} = \partial_x \Phi,  \sigma^y_{\tau} = \partial_y \Phi, 
\sigma^x_x = \sigma^y_y = - \partial_{\tau} \Phi$. Notice that
compression  involves a single space-like gauge field ($\Phi$) which
is time-like (i.e. like the scalar field in electromagnetism).

It follows immediately that the dual action becomes, keeping all units
explicit,

\begin{eqnarray}
S & = & \hbar \int d\Omega 
\left[ { 2 \over \kappa} (\partial_{\tau} \Phi)^2
+ { 1 \over {\rho} } ( (\partial_x \Phi)^2 + (\partial_y \Phi)^2 ) \right]
\nonumber \\
 & = & { {2\hbar} \over {\kappa} } \int d^q d \omega 
( \omega^2 + ({ {\kappa} \over {\rho}} ) q^2 ) | \Phi|^2.
\label{Scom1}
\end{eqnarray}

This clearly describes a compression mode. Consider now 
the coupling to the dislocation currents
in the compression gauge,

\begin{eqnarray}
i J^a_{\mu} B^a_{\mu} & = & i \left( J^x_y B^x_y + J^y_x B^y_x \right)
\nonumber \\
& = & i \left(  J^x_y - J^y_x \right) \Phi.
\label{pressureglide}
\end{eqnarray}

This proves that the the dislocation current tensor has to be symmetric
in the space indices to satisfy the fundamental requirement
that compression does not a carry a gauge coupling to the 
dislocation currents!  Remarkably, in this field theoretic
formulation the kinematical glide condition follows from symmetry
principle: dislocations should not climb in order to satisfy
the condition that translational symmetry breaking is associated
exclusively with shear rigidity.

\begin{figure}[tb]
\begin{center}
\vspace{10mm}
\leavevmode
\epsfxsize=0.3\hsize
\epsfysize=5cm
\epsfbox{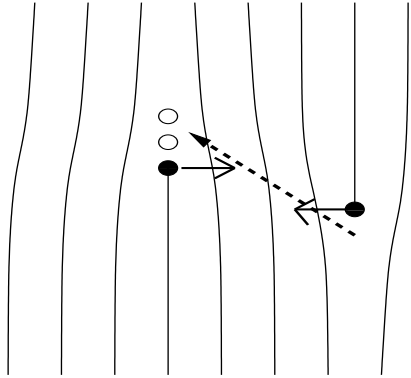}
\end{center}
\caption{In condensed matter systems, interstitials cannot be
avoided at the moment one enters a fluid state. The reason is
that dislocations have a finite probability to collide. At a
collision particles can tunnel from one dislocation to the other,
thereby liberating climb motions. These processes are of relevance 
for the long wavelength physics, with the effect that the field-theoretic
fluids discussed in this paper cannot be literally realized.}
\label{collisionpict}
\end{figure}

Anticipating that glide will play a central role in the remainder,
one can already infer the sense in which the field-theoretic
dual can never become literal in condensed matter systems.
The culprit is the incapacity of the field theory to keep track
of the interstitials. The interstitials interfere in a catastrophic
way, because these destroy the glide condition. In real condensed
matter systems interstitials cannot be avoided in the nematic quantum
fluids. When dislocations are present at a finite density, the
probability that dislocations `collide', i.e. approach each other
to microscopic distances, becomes finite. At collisions, the 
constituent particles can tunnel between dislocations causing the
`rows of atoms' to become shorter and longer (Fig.\ref{collisionpict}), 
thereby liberating climb motions. Climb is the same as interstitial transport.
Obviously, just releasing climb in the field theory would lead
to the absurdity that compression rigidity would vanish, and 
one has somehow to dress up the theory with separate fields describing
the interstitials. At present it is unclear how to construct such a
theory.

\section{The physical fields: helical projection and linear
polarization.}

In the previous section we have collected all necessary 
ingredients to further develop the theory. However, at first sight
it appears as a quite complicated affair. In total we are
dealing with 6 stress gauge fields $B^a_{\mu}$, while 
gauge invariance implies 6 dislocation currents $J^a_{\mu}$
as well. In terms of these gauge fields the dual action acquires
a quite complicated form, further complicated by the various
constraints. The gauge transversality condition implies that
2 out of the six fields are unphysical anyhow. In addition,
the Ehrenfest symmetry constraint has to be satisfied by the
$B$'s, while the glide condition acts as an extra constraint 
on the $J's$. In total, the theory is about three {\em physical}
stress gauge fields. Since the dislocation currents describe
the singularities in the physical fields, there are three
dislocation currents associated with the physical fields, while
one of these three dislocation currents is unphysical because
of the glide constraint. 

In order to isolate the physical content of the theory it is
convenient to employ helical projections. This is discussed
at length by Kleinert for the 3D isotropic theory\cite{KleinertII}.  
Helical projections are well known from e.g. quantum electrodynamics.
Transform the theory to momentum space and consider Dreibeins (in 3D)
$e^{(\alpha)}$, $\alpha= 0, 1, -1$, with their $(0)$ component 
parallel to the propagation direction, while the $(\pm 1)$ 
components represent  left- and right circularly polarized 
transversal `photons'. By neglecting
the $(0)$ components of the projected gauge fields transversality is 
imposed (`transversal gauge'). As discussed by Kleinert, the
tensorial stress gauge fields in the 3D isotropic theory
can be decomposed into $S=0$ (compression)
and $S=2$ (shear) helical components. In the present 2+1D case,
Lorentz invariance is badly broken due to the `anomalous' time 
axis and the helical projections have to be modified accordingly. 
This space-time anisotropy is reflected in the `upper' Burgers 
indices $a$ of the stress-gauge field tensors $B^a_{\mu}$ referring 
exclusively to spatial directions but also in the `lower' space-time
indices $\mu$. As we will now demonstrate, there appears to be a
unique projection satisfying all constraints, putting the
stress photons on `normal coordinates'.

To put matters in perspective, let us first shortly review the
standard helical projection. Consider a $U(1)$ gauge field $A_{\mu}$
in 2+1D with an Euclidean  Maxwell action 
$S \sim  F_{\mu \nu} F^{\mu \nu}$, 
$F_{\mu \nu} = \partial_{\mu} A_{\nu} -  \partial_{\nu} A_{\mu}$
and the gauge constraint $\partial_{\mu} A_{\mu} = 0$.
Fourier transform to (Matsubara) frequency-momentum space
${\bf p} = \left( q_x, q_y, \omega \right)$ and introduce the unit
vector $  \hat{p} = {\bf p} / | p |$. The action
becomes $S \sim |  p |^2 \left( \hat{p}_{\mu} A_{\nu} -
\hat{p}_{\nu} A_{\mu} \right)   \left( \hat{p}_{\mu} A_{\nu}^* -
\hat{p}_{\nu} A_{\mu}^* \right)$ with gauge constraint
$\hat{p}_{\mu} A_{\mu} = 0$. The helicity basis is constructed
as follows: choose an orthogonal set of three basis vectors (Dreibein)
${\bf e}^1, {\bf e}^{-1},  {\bf e}^0$ in the subspace of a 
fixed momentum $\hat{\bf p}$ such that ${\bf e}^0 = \hat{\bf p}$, 
coinciding with the propagation direction of the photon.
Lorentz invariance turns into the Euclidean group $E(3)$ after analytic
continuation and this is wired in by constructing a spin-one 
representation,

\begin{eqnarray}
{\bf e}^{(+1)} (\hat{p})  & = & { 1 \over {\sqrt{2}} } 
\left( {\bf e}^1 + i {\bf e}^{-1} \right) \nonumber \\
{\bf e}^{(-1)} (\hat{p})  & = &- { 1 \over {\sqrt{2}} } 
\left( {\bf e}^1 - i {\bf e}^{-1} \right)   \nonumber \\  
{\bf e}^{(0)}  (\hat{p}) & = & {\bf e}^0 = \hat{\bf p}.
\label{helidreibein}
\end{eqnarray}  

Define the projection matrices $P^{(h)}$

\begin{eqnarray}
P^{(h)}_{\mu \nu}  (\hat{\bf p}) & = & e^{(h)}_{\mu} (\hat{\bf p})
e^{(h)}_{\nu} (\hat{\bf p})
\nonumber \\
\sum_{ h}  P^{(h)}_{\mu \nu}(\hat{\bf p}) & = &  
\delta_{\mu \nu}  \nonumber \\
P^{(h)}_{\mu \lambda}(\hat{\bf p}) P^{(h)}_{\lambda \kappa} (\hat{\bf p}) & = & 
P^{(h)}_{\mu \kappa}(\hat{\bf p}) \delta_{\mu \kappa} (\hat{\bf p}),
\label{projdreibein}
\end{eqnarray} 

showing that a vector function like the 
gauge field ${\bf A}$ can be expanded in the helical basis,

\begin{eqnarray}
A_{\mu} ({\bf p})& = & \sum_{h} P^{(h)}_{\mu \nu} ( \hat{\bf p} )
A_{\nu} ({\bf p}) = \sum_{h} e^{(h)}_{\mu} (\hat{\bf p}) A^{(h)} ({\bf p})
\nonumber \\
A^{(h)} & = & \sum_{\mu} e^{(h)}_{\mu} (\hat{\bf p})  A_{\mu}({\bf p})
\label{Aproj}
\end{eqnarray}

and the  $A^{(h)} ({\bf p})$ are the helicity components of $A_{\mu}$.
One infers immediately that  the gauge constraint,

\begin{equation} 
\sum_{\mu} p_{\mu} A_{\mu} = |p| \sum_{\mu} e^{(0)}_{\mu}  
\sum_{h} e^{(h)}_{\mu} 
 A^{(h)} = |p| A^{(0)} = 0
\label{heltransvers}
\end{equation}

It follows that the longitudinal helical component $A^{(0)}$ 
corresponds with the unphysical content of the gauge field, while
the $A^{(\pm 1)}$ fields are the physical components satisfying
the gauge-transversality requirement. Using Eq.'s 
(\ref{Aproj}, \ref{heltransvers}) and the orthonormality of the 
Dreibeins it is easily shown that the Maxwell action becomes

\begin{equation}
{\cal L} \sim |p|^2 \left( | A^{(+1)} |^2 \; + \;  | A^{(-1)} |^2 \right),
\label{EMMaxwell}
\label{helmaxwell}
\end{equation}

expressing the fact that the physical excitations of
the electromagnetic field correspond with left- and right circularly
polarized photons.

The above has to be modified in the quantum-elasticity context 
as the Euclidean invariance $E(3)$ is broken in
2+1D space-time to $E(2) \times O(2)$ due to the special role 
of the imaginary
time axis. However, rotational symmetry is maintained on the
time slice, suggesting that a decomposition in purely
spatial longitudinal- and transversal components should be useful.
This is surely the case for the `upper' Burgers indices $a$ of
the stress gauge fields $B^a_{\mu}$. Parameterize the spatial momentum
vector in terms of a phase $\phi_{\bf q}$ as  $\vec{q} = |q| \;
( \hat{q}_x +  \hat{q}_y ) = |q| \; \left( \hat{e}_{x} \cos(\phi_{\bf q}) +
\hat{e}_{y} \sin(\phi_{\bf q}) \right)$ 
and decompose the stress gauge fields
in longitudinal- and transversal spatial components of their 
Burgers flavors ($L,T$, respectively),

\begin{eqnarray}
B^L_{\mu} ({\bf p}) & = &  \cos{\phi_{\bf q}} B^x_{\mu} +
\sin{\phi_{\bf q}}  B^y_{\mu} \nonumber \\
B^T_{\mu} ({\bf p}) & = &  \cos{\phi_{\bf q}} B^y_{\mu} -
\sin{\phi_{\bf q}}  B^x_{\mu}.
\label{burgersLT}
\end{eqnarray}

\begin{figure}[tb]
\begin{center}
\vspace{10mm}
\leavevmode
\epsfxsize=0.7\hsize
\epsfysize=10cm
\epsfbox{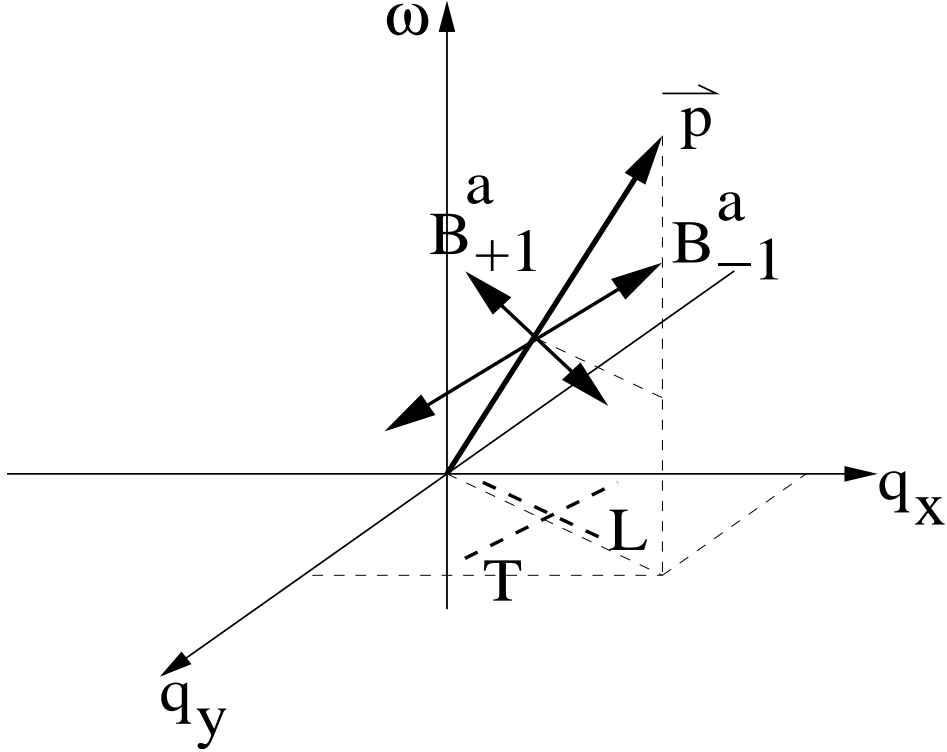}
\end{center}
\caption{Explanation of the helical projection used to evaluate
the stress gauge fields in terms of their physical content (`stress
photons'). The propagation direction of the photon in frequency-momentum
space is $\vec{p}$. The Burgers components of the stress-gauge field
are decomposed in longitudinal (L) and transversal (T) spatial momentum
components, where L is parallel to $\vec{p}$. In order to satisfy the
gauge transversality condition, the gauge fields are decomposed in
a magnetic component -1 which is parallel to the T spatial component
and an electrical component which is orthogonal to both $\vec{p}$ and
the magnetic component.}
\label{dreibeinexpl}
\end{figure}

This requirement imposes a preferred reference frame on the
Dreibeins living in space-time, see Fig. (\ref{dreibeinexpl}). 
We keep the velocity convention
$c^2_{ph} = 2 \mu / \rho = 1 $ as introduced in section IIa,
to define  space-time `momentum' as
${\bf p} = ( {\bf q}, \omega ) = (q_x, q_y, \omega) =
|p| \; ( \hat{q}_x, \hat{q}_y, \hat{\omega} ) = |p| \; e^{0} ( {\bf p} )$.
The space-time longitudinal unit vector $e^{0} ( {\bf p} )$
can be written in polar coordinates as,

\begin{equation}
e^{0} ( {\bf p} )=\left(\begin{array}{c}
  \sin{\theta_ {\bf p}}\cos{\phi_{\bf p}}\\
  \sin{\theta_{\bf p}}\sin{\phi_{\bf p}}\\
 \cos{\theta_ {\bf p}}
\end{array}\right)
 \label{elin0}
\end{equation}

In non-relativistic electromagnetism problems it is natural to
single out the spatial components of the 
vector potentials $A_{x,y}$ as responsible for
magnetic phenomena. Since Lorentz invariance is badly broken
in the present case, one should look for some superposition
of the $B^a_{x,y}$ fields representing the `magnetic' sector
of the stress gauge fields. 
These live on the time-slice as
the physical fields have to be orthogonal to  Eq. (\ref{elin0})
the direction of this transverse component has to coincide with
the transverse spatial direction which we call `$-1$' or
`magnetically'  polarized,

\begin{equation}
 e^{-1} ( {\bf p} ) =\left(\begin{array}{c}
   \sin{\phi_{\bf p}}\\
    -\cos{\phi_{\bf p}} \\
0
\end{array}\right).
\label{elin-1}
\end{equation}

The other transverse component has to be orthogonal to both 
Eq. (\ref{elin0}) and Eq. (\ref{elin-1}) and this fixes the
direction of the remaining $ +1$ or `electrical'  component,

\begin{equation}
e^{1} ( {\bf p} ) =\left(\begin{array}{c}
  -\cos{\theta_{\bf p}}\cos{\phi_{\bf p}}\\
  -\cos{\theta_{\bf p}}\sin{\phi_{\bf p}}  \\
  \sin{\theta_{\bf p}}
\end{array}\right).
\label{elin+1}
\end{equation}

Introducing separate `zweibeins' for the space-like transverse
and longitudinal projections,

\begin{eqnarray}
\tilde{e}^L ( {\bf p} ) & = &  \left(\begin{array}{c} 
 \cos{\phi_{\bf p}} \\ \sin{\phi_{\bf p}}
\end{array}\right), \nonumber \\
\tilde{e}^T ( {\bf p} ) & = &  \left(\begin{array}{c} 
-\sin{\phi_{\bf p}} \\ \cos{\phi_{\bf p}}
\end{array}\right).
\label{tildeelin}
\end{eqnarray}

Now, we have to define rules of change of the basis vectors 
introduced above in the case of the sign change of the momentum 
$\vec{p}\rightarrow -\vec{p}$ in the frequency-momentum space. Namely,
we associate the inversion of momentum $\vec{p}\rightarrow -\vec{p}$ with
the following change of the angles:

\begin{eqnarray}
\theta_{\bf p}\rightarrow \pi-\theta_{\bf p};\;\;\phi_{\bf p}\rightarrow
\phi_{\bf p}+\pi
\label{inversion}
\end{eqnarray}

Thus, we arrive at the following relations between basis vectors sets 
corresponding to the momenta $\vec{p}$, $-\vec{p}$:

\begin{eqnarray}
e^{0} (-{\bf p} ) & = &-e^{0} ({\bf p} )\nonumber\\
e^{-1} (-{\bf p} ) & = &-e^{-1} ({\bf p} )\nonumber\\
e^{1} (-{\bf p} ) & = &e^{1} ({\bf p} )\nonumber\\
\tilde{e}^L (-{\bf p} ) & = &-\tilde{e}^L ({\bf p} )\nonumber\\
\tilde{e}^T (-{\bf p} ) & = &-\tilde{e}^T ({\bf p} ).
\label{ibasis}
\end{eqnarray}

These relations have a straightforward implication for 
our definition of the projected fields in analogy with  Eq.(\ref{Aproj}). 
Namely, we want to preserve 
the usual relation $B^a_{\mu}(-\vec{p})=B^a_{\mu}(\vec{p})^*$ between 
the Fourier components of the real stress gauge fields also in the 
$\vec{p}$-dependent basis vectors set  defined above in Eq.'s 
(\ref{elin0},\ref{elin-1},\ref{elin+1},\ref{tildeelin}). For this purpose,
we define projected stress gauge fields $B^{L,T}_{h}$ in the following way:

\begin{equation}
B^a_{\mu} = \sum_{E=L,T}i\tilde{e}^{E}_a (e^{1}_{\mu} B^E_{1}+ie^{-1}_{\mu}
B^E_{-1})\label{projfields}\equiv\sum_{E=L,T}\tilde{e}^{E}_a (ie^{1}_{\mu} 
B^E_{1}-e^{-1}_{\mu}B^E_{-1}).\label{linear}
\end{equation}

The multiplication by the imaginary $i$ in 
Eq. (\ref{linear}) for the basis 
vectors leads to a change in
sign when the momentum $\vec{p}$ is reversed, 
precisely serving the purpose explained above.

To guide the intuition it is useful to consider what the projections
Eqs. (\ref{elin0},\ref{elin-1},\ref{elin+1}) mean in the
familiar electromagnetism context. Consider the magnetic-
($H^2$) and electrical ($E^2 = E^2_x + E^2_y$) energy densities
in energy-momentum space,

\begin{eqnarray}
H^2 & = & | q_x A_y - q_y A_x|^2 \nonumber \\
    & = & q^2 | A_{-1}|^2 \nonumber \\
E^2 & = & | q_y A_{\tau} - (\omega / c ) A_y|^2
+ | - q_x A_{\tau} + (\omega /c) A_y|^2 \nonumber \\
    & = & q^2 |A_{+1}|^2 + ( \omega / c)^2 ( |A_{+1}|^2 + |A_{-1}|^2 ).
\label{emlinear}
\end{eqnarray}

This demonstrates that this has to do with a linear polarization,
such that in the non-relativistic regime $q >> \omega / c$,
the $h = -1$ components are responsible for magnetic phenomena
while the $h = +1$ components are in charge of the electrical fields.
Although not to be taken literal, we will in the remainder
repeatedly refer to `magnetic-like' and `electrical-like' stress fields.  
 
The projected fields in Eq.(\ref{projfields}) are very convenient. 
First, by neglecting the `$0$' components of the $B$ fields, 
the gauge transversality is imposed. Secondly, the space-like 
transversal-longitudinal projection has the effect that the
Ehrenfest symmetry condition $\sigma^x_y = \sigma^y_x$
(Eq. \ref{stresssym}) acquires a simple form,

\begin{eqnarray}
iB^{L}_{+1} & = & -\cos{(\theta)}B^{T}_{-1} 
\nonumber \\
& = & -\hat{\omega} B^{T}_{-1}. 
\label{constr}
\end{eqnarray}

This demonstrates that the Ehrenfest condition renders the
$B^{L}_{+1}$ field redundant. Together with the gauge
transversality this implies that three physical `stress photons'
exist: $B^T_{-1}$,  $B^L_{-1}$, and $B^T_{+1}$. As we will
see in the next section, in terms of the projected fields  
Eq.(\ref{projfields}) the stress-gauge field action for the
isotropic medium Eq. (\ref{stressmaxwell}) acquires a quite simple form
which is easily diagonalized. 

\section{2+1D quantum elasticity in gauge fields}

Before we turn to the more interesting physics associated with the
dislocations, let us first consider how the theory looks like in
the absence of these defects. This is just the simple phonon problem
discussed in section II, now rewritten in terms of the dual degrees
of freedom: the stress gauge fields. Although the technology introduced
in the previous section simplifies matters to a great extent, the
stress-gauge field formulation remains remarkably complicated, at least
as compared to the couple of lines of simple algebra needed to
identify the phonons in  the strain field formalism. We will first
derive the action of isotropic elasticity in terms of the stress-photons
living on normal coordinates. Although the result has a simple form, it
poses a considerable physical interpretation problem. Among others,
it carries information in the $\omega \rightarrow 0$ limit which
seems absent in the strain formulation. In addition, we find stress-photons
resembling longitudinal- and transversal phonons, but also 
a third photon is present which has no counterpart in the strain 
formalism. At first sight this is quite confusing, but on closer
inspection it gives away some interesting insights.  Phonons are living
in the `universe' derived from order, while the stress photons
belong to the dual `universe' governed by the disorder fields,
the matter associated with the dislocations.  The richer, 
more complex structure associated with the stress photons is needed
to understand the physics of the disordered state, as we will see
in the next sections. In the ordered state, the stress photons are
rather clumsy entities. This will be illustrated in  subsection 
B where we will calculate the elastic propagators  starting with the 
stress photons. This will turn out to be a remarkably long and tedious 
detour, but this technology will become most useful when dealing
with the disordered states.

\subsection{Derivation of the stress photon action.}

In terms of the stress fields $\sigma^a_{\mu}$, 
the dual `stress-Maxwell' action of isotropic quantum
elasticity becomes,

\begin{eqnarray}
S(\nu) & = & { 1 \over {4\mu} } \int d\Omega 
\left[ {\sigma_{x}^x}^2 + {\sigma_{x}^y}^2 + {\sigma_{y}^x}^2 + 
{\sigma_{y}^y}^2 \right.
\nonumber \\
  &  & \left.-\frac{\nu}{1+\nu}(\sigma_{x}^x+\sigma_{y}^y)^2 +
 {\sigma_{\tau}^x}^2 + {\sigma_{\tau}^y}^2 \right]
\label{fullaction}
\end{eqnarray}

with the constraint $\sigma_x^y = \sigma_y^x$ to be imposed.

Let us evaluate this action using the projections
introduced in the previous section. Substitute $\sigma^a_{\mu} =
\epsilon_{\mu \nu \lambda} \partial_{\nu} B^a_{\lambda}$, transform to 
momentum-frequency space, and insert the inverse of Eq. (\ref{projfields}).
A major simplification occurs in the case $\nu = 0$, such that $\kappa = \mu$.
The action simplifies to a Maxwell action, separately for the $x$ and $y$
Burgers flavors,

\begin{eqnarray}
S (\nu = 0) & = & { 1 \over {4\mu} }  \sum_{a=x,y} \int d\Omega \;
\left[ (\sigma^a_x)^2 + (\sigma^a_y)^2 +  (\sigma_{\tau}^a)^2 \right]
\nonumber \\
  & = & { 1 \over {4\mu} } \sum_{a=x,y} \int d\Omega  \; F^a_{\mu \nu}
F^a_{\mu \nu},
\label{nu=0action}
\end{eqnarray}

with $F^a_{\mu \nu} = \partial_{\mu} B^a_{\nu} - \partial_{\nu} B^a_{\mu}$.
In terms of the projected fields in frequency-momentum space this simply 
becomes (compare with Eq. \ref{helmaxwell}),

\begin{equation}
S (\nu = 0) = {  1 \over {4\mu} } \int d^{2} q d\omega |p|^2  
\left( |B^T_1|^2 +  |B^T_{-1}|^2 + |B^L_1|^2 +  |B^L_{-1}|^2 \right).
\label{acnu01}
\end{equation}

For $\nu \neq 0$ we have to evaluate the compression-only part
$\sim ( \sigma^x_x + \sigma^y_y)^2$.
A straightforward calculation yields

\begin{equation}
\sigma^x_x + \sigma^y_y = i |p| (i B^T_1 - \hat{\omega} B^L_{-1} ).
\label{comprfields}
\end{equation}

Finally, we have to impose the symmetry condition 
$\sigma^y_x = \sigma^x_y$,
and as we already discussed this translates into the condition
$iB^L_1 = -\hat{\omega} B^T_{-1}$ 
(Eq. \ref{constr}), 
so that the $B^L_1$ field can be eliminated. 
Collecting, we find for the full 
action,

\begin{eqnarray}
S & = & { 1 \over {4 \mu} } \int d^2 q d\omega \; |p|^2 \;  \left[
|B^T_1|^2 +  ( 1  + \hat{\omega}^2 ) |B^T_{-1}|^2 + |B^L_{-1}|^2  \right. 
\nonumber \\
&  &  \left. - { \nu \over { 1 + \nu} } 
|i B^T_1 - \hat{\omega} B^L_{-1} |^2 \right].
\label{elaction0}
\end{eqnarray}

The compression-only term $\sim \nu$ causes a linear coupling 
of the $B^T_1$ and $B^L_{-1}$ modes which is proportional to frequency, 
while it vanishes in the static limit. This coupling is diagonalized by

\begin{eqnarray}
{\cal B}^T_{1} = { 1 \over {\sqrt{ 1 + \hat{\omega}^2}}}
\left(i B^T_1 - \hat{\omega} B^L_{-1} \right) \nonumber \\
{\cal B}^L_{-1} = { 1 \over {\sqrt{ 1 + \hat{\omega}^2}}}
\left( -i\hat{\omega} B^T_1 - B^L_{-1} \right).
\label{eltransform}
\end{eqnarray}

The labels of the ${\cal B}$ fields are chosen to reflect  
the $\omega \rightarrow 0$ origin of these fields. At
finite frequencies these are of course linear superpositions
of the `bare' $B$ fields. Using $|p|^2 = \omega^2 + q^2$ and
$\hat{\omega} = \omega / |p|$ we find the diagonal action describing
the stress photons $B^T_{-1}$, ${\cal B}^T_{1}$ and ${\cal B}^L_{-1}$,

\begin{equation}
S  =  { 1 \over {4 \mu} } \int d^2 q d\omega \left[
( 2 \omega^2 + q^2 ) | B^T_{-1} |^2 +
{ 1 \over {1 + \nu} } ( ( 1 - \nu) \omega^2 + q^2 ) |{\cal B}^T_{1}|^2
+ ( \omega^2 + q^2 ) |{\cal B}^L_{-1}|^2 \right].
\label{elasticaction}
\end{equation}

This is an important result: it is the theory of quantum elasticity in
the dual, gauge field, representation.
In a magical way, it seems, familiar information has resurfaced.
In the terms multiplying the  $B^T_{-1}$ and ${\cal B}^T_{1}$
stress photons the dispersion relations of the 
transverse and longitudinal phonons are recognized. 
Halfway the calculation
this information was completely scrambled. Consider, for instance, 
the $B^T_{-1}$ stress photon. This mode appears as an eigenmode
already in step Eq. (\ref{acnu01}). However, the unit of velocity
is $c^2_{ph} = 2 \mu / \rho$ while the transverse phonon velocity
is $c^2_{T} = \mu / \rho$ (compare Eq. \ref{strainact1}), 
a factor of two mismatch.
In the final result, the correct velocity has re-emerged thanks to 
the Ehrenfest constraint. By eliminating the $B^L_1$ field, the
propagator of the $B^T_{-1}$ field picks up an additional $\hat{\omega}^2$
factor and this factor takes care of halving the `natural' 
velocity $c^2_{ph}$,

\begin{equation}
p^2 ( 1 + \hat{\omega}^2) = ( \omega^2 + c^2_{ph} q^2 ) ( 1 
+ { {\omega^2} \over { \omega^2 + c^2_{ph} q^2 } } ) = \frac{1}{2}
( \omega^2 + { {c^2_{ph} q^2} \over 2} ).
\label{halving}
\end{equation}

It is easy to verify that also in the strain formalism the halving
of the transversal velocity finds its origin in the symmetry 
of the strain tensor.

The ${\cal B}^T_{1}$  clearly propagates like the longitudinal phonon.
However, at the same time it is solely responsible for the propagation
of stresses through the medium in the high temperature, classical 2D
limit. This is easily seen: the $-1$ fields describe time-like phenomena
and disappear with the time axis. The action reduces to

\begin{equation} 
S = \int d^2q  { {q^2} \over {4\mu (1+\nu)} } |B^T_{+1}|^2,
\label{static}
\end{equation}

which indeed
corresponds with the action in the 2D limit.
The reader is referred to appendix C for a derivation
of the static 2D action.  

The real surprise is the presence of a second time-like field,
${\cal B}^L_{-1}$, propagating with the natural velocity $c_{ph}$. This
mode has no counterpart in the strain formulation! A priori there is
no reason that the mode content in the stress-gauge field formulation
should be the same as in the strain formulation. In the latter one
counts the Goldstone mode content, while the meaning of the gauge fields
is fundamentally different, since they parameterize the capacity of the
medium to mediate forces. For instance, in the Abelian Higgs duality
a single spin wave translates into two physical photons in the dual
representation. The specialty in the present context is that the theory
of elasticity involves different velocities, and it is just curious that
stress photons exist propagating differently from any phonon. The resolution
of this puzzle is that these photons do exist in their dual world, but
they cannot be made visible using order (Goldstone mode) means. At the same
time, when disorder gets in charge this extra photon acquires a meaning
on the disorderly side of the duality transformation: as we will see
the third photon has everything to do with the fact that the superfluid
carries {\em three} modes. To shed further light on these matters, 
let us consider the elastic propagators.

\subsection{Gauge fields and elastic propagators.}

In order to appreciate the physical meaning of the stress
photons, it is quite helpful to interrogate them with physical means: 
the elastic propagators.  These were already introduced in section II
(Eqs. \ref{propagatordef}, \ref{eqnmotpropagator}) and their 
spectral functions correspond with the dynamical form factors as
measured by neutron- and X-ray scattering. From the strain formalism
of section II, it is obvious that the poles of the spectral functions
correspond with the Goldstone modes, the phonons. 
 
How to compute these propagators in a stress gauge fields language?
As we discuss in Appendix C, the relations between the strain 
propagators ($S_0$ is  the strain action),

\begin{equation}
\langle \langle \partial_{\mu} u^a (\vec{r}_1)| 
\partial_{\nu} u^b (\vec{r}_1) \rangle \rangle
\equiv
{ 1 \over Z} \int {\cal D} u^a  \; (\partial_{\mu} u^a (\vec{r}_1))\;
(\partial_{\nu} u^b (\vec{r}_2)) \; \exp{\left( - S_0 \right)}
\label{strainpropdef}
\end{equation}

and the stress propagators ($S_{dual}$ is  the stress action)

\begin{equation}
\langle \langle \sigma^c_{\lambda} (\vec{r}_1) | \sigma^d_{\kappa} (\vec{r}_2) 
\rangle \rangle
\; \equiv  \; 
{ 1 \over Z} \int {\cal D} \sigma^a_{\mu} \delta 
( \partial_{\mu} \sigma^a_{\mu} ) \; \sigma^c_{\lambda} (\vec{r}_1) \; 
\sigma^d_{\kappa} (\vec{r}_2) \; \exp{\left( - S_{dual} ( \sigma^a_{\mu})
\right)}
\label{stresspropdef}
\end{equation}

are simple, but non-trivial,

\begin{equation}
\langle \langle \partial_{\mu} u^a (\vec{r}_2)| 
\partial_{\nu} u^b (\vec{r}_1) \rangle \rangle
\; =  \;  \delta (\vec{r}_1 - \vec{r}_2) \delta_{\mu, \nu} 
\delta_{a, b} C^{-1}_{\mu \mu a a} \; - \; C^{-1}_{\mu \lambda a c} 
C^{-1}_{\nu \kappa b d } \langle \langle \sigma^c_{\lambda} (\vec{r}_1) |
\sigma^d_{\kappa} (\vec{r}_2) \rangle \rangle.
\label{genpropag}
\end{equation}

Hence, in momentum space these have the form 
$\langle \langle p_{\mu} u^a | p_{\nu} u^b \rangle \rangle 
\sim C^{-1}_{\mu \mu a a} -
 C^{-1}_{\mu \lambda a c} C^{-1}_{\nu \kappa b d } 
\langle \langle \sigma^c_{\lambda} | \sigma^d_{\kappa}  \rangle \rangle$. The
algorithm is therefore as follows: (a) use the stress-strain relations
to determine the correspondence $p_{\mu} u^a =  C^{-1}_{\mu \lambda a c}
\sigma^c_{\lambda}$, (b) calculate the stress propagators 
$\langle \langle \sigma^c_{\lambda} | \sigma^d_{\kappa}  \rangle \rangle$ 
exploiting the
stress-gauge fields; these become proportional to the transversal
stress-photon propagators, (c) subtract the stress propagator from
the constant $\sim C^{-1}_{\mu \mu a a}$ and the elastic propagator
follows. 

We are particularly interested in the elastic (or phonon) propagators
Eqs. (\ref{propagatordef},\ref{eqnmotpropagator}) referring only to 
spatial momenta, corresponding with the observables in condensed matter
experimentation. The relevant stress strain relations are in isotropic
elasticity,

\begin{eqnarray}
i q_x u^x & = & { 1 \over {2\mu (1+\nu)} } 
\left[ \sigma_x^x - \nu \sigma_y^y \right] \nonumber \\
i q_y u^y & = & { 1 \over {2\mu (1+\nu)} } 
\left[ \sigma_x^x - \nu \sigma_y^y \right] \nonumber \\
i q_x u^y & = &  { 1 \over {2\mu} } \sigma_x^y \nonumber \\
i q_y u^x & = &  { 1 \over {2\mu} } \sigma_y^x \nonumber \\
\label{stressstrain}
\end{eqnarray}

where the two last relations are true under the constraint 
that $\sigma_x^y=\sigma_y^x$. Using these relations, the
calculation of the elastic propagators becomes a remarkably
lengthy but straightforward algebraic exercise.  

Let us first consider a simple example to illustrate 
matters: the compression-only case (Eqs. (\ref{Scom}, \ref{Scom1}) 
in section III). Using the strain fields,
the propagator follows directly from Eq. (\ref{Scom}),

\begin{equation}
G = G_L  = { 2 \over {\kappa} } \; { {q^2} \over { q^2 + {\rho \over \kappa}
\omega^2}}
\label{comprprop0}
\end{equation}

keeping the velocity explicit for clarity.

Since $\partial_x u^x + \partial_y u^y = - ( 2 / \kappa) \partial_{\tau}
\phi$ in compression gauge, the stress-strain relation becomes 
 $q_x u_x + q_y u_y = - ( 2 / \kappa)
\omega \phi$. The propagator of the `compression photon' $\phi$ follows
immediately from the action Eq. (\ref{Scom1}),

\begin{equation} 
\langle \langle \phi | \phi \rangle \rangle = {\kappa \over 2} \; { 1 \over { \omega^2 +
{ {\kappa} \over {\rho} } q^2 }}
\label{compphotonprop}
\end{equation}

The elastic propagator is given in terms of the compression photon
propagator as,
\begin{eqnarray}
G & = & { 2 \over {\kappa} } - { { 4 \omega^2} \over { \kappa^2}} \;
\langle \langle \phi | \phi \rangle \rangle \nonumber \\
  & = & { 2 \over {\kappa} } - { { 4 \omega^2} \over { \kappa^2}} \; 
{\kappa \over 2} \; { 1 \over { \omega^2 +
{ {\kappa} \over {\rho} } q^2 }} \nonumber \\
& = & { 2 \over {\kappa} } \; { {q^2} \over { q^2 + {\rho \over \kappa}
\omega^2}}
\label{strainstresscomprprop}
\end{eqnarray}

and this is the desired result.

Let us now consider the computation of the elastic propagators 
of the full  problem.  Consider first
the longitudinal propagator. A useful relation is, 

\begin{equation}
q_x u^x + q_y u^y = { { 1 - \nu} \over { 2\mu ( 1 + \nu)}}
{p} \left(i B^T_1 - \hat{\omega} B^L_{-1} \right)
\label{longaux}
\end{equation}

and it follows immediately that

\begin{equation}
G_L = Const. -
({ { 1 - \nu} \over { 2\mu ( 1 + \nu)}})^2  
\; p^2 \;\langle  \langle iB^T_1 - \hat{\omega} B^L_{-1} |
i B^T_1 - \hat{\omega} B^L_{-1} \rangle \rangle.
\label{longstressprop}
\end{equation}

The transverse propagator is more cumbersome and after
some lengthy but straightforward algebra we find

\begin{eqnarray}
&&G_T = Const. - { 1 \over {4 \mu^2} } \; p^2 \; 
\displaystyle\left(2 \hat{\omega}^2 \langle \langle B^T_{-1} | B^T_{-1} \rangle \rangle
-i\hat{\omega}[ \langle  \langle B^L_{-1} | B^T_{+1} \rangle \rangle-
\langle  \langle B^T_{+1} | B^L_{-1} \rangle \rangle]\right.\nonumber\\
&&\left.
+ { {2\nu} \over { (1 + \nu)^2 } } \langle \langle iB^T_1 - \hat{\omega} B^L_{-1} |
 iB^T_1 - \hat{\omega} B^L_{-1} \rangle \rangle \right).
\label{transstressprop} 
\end{eqnarray}

Adding these up, we find for the total propagator
\begin{eqnarray}
&&G =  { {2 (\nu + 2)} \over {\mu ( \nu +1)} } \; - \;
{ 1 \over {4 \mu^2} }  \; p^2 \;
\displaystyle\left( 2\hat{\omega}^2 \langle B^T_{-1} | B^T_{-1} \rangle-
 i \hat{\omega}[  \langle B^L_{-1} | B^T_{+1} \rangle-
\langle B^T_{+1}| B^L_{-1} \rangle]\right.
\nonumber\\
&&\left.
+ { { 1 + \nu^2} \over {(1+\nu)^2}} 
\langle B^T_1 + \hat{\omega} B^L_{-1} |
 B^T_1 + \hat{\omega} B^L_{-1} \rangle \right).
\label{totstressprop}
\end{eqnarray}

The general results Eqs.(\ref{longstressprop},
\ref{transstressprop},\ref{totstressprop}) 
are an expression of the remarkably convoluted way 
in which the stress photons relate to observables associated with the 
orderly side of the duality. Considering the intact elastic medium,
these expressions should eventually simplify to the simple phonon
propagators Eqs.(\ref{propagatordef}), and this is indeed what happens!

Transforming to the diagonal photon representation using Eq. 
(\ref{eltransform}), and using the photon propagators following
from (\ref{elasticaction}), the propagator Eq. (\ref{totstressprop})
can be computed. Remarkable cancelations occur, and eventually

\begin{eqnarray}
G  & = &  {1  \over {\mu}} \left(  { {2 (\nu + 2)} \over { \nu +1} }
\; - \; { {\omega^2} \over { \omega^2 + q^2 /2 }} \; - \; { 
{ 2 \omega^2 (1 - \nu) + q^2 ( 1 + \nu^2)} \over
{ ( 1 + \nu)  ( \omega^2 ( 1 - \nu) + q^2 )} } \right)
\nonumber \\
      & = & { 1 \over {\mu} } \left[
 { {(q^2 / 2)} \over { \omega^2 +  (q^2 / 2) }}
\; + \; { {q^2} \over { ( q^2/ ( 1 - \nu) ) + \omega^2 }} \right].        \label{totpropelas}
\end{eqnarray}

We indeed recover the simple phonon propagator 
Eq.(\ref{eqnmotpropagator}).

Alltogether, this is a remarkably tedious calculation, especially
as compared to the elementary derivation of the propagators using
the strain fields. Is this just an un-necessary detour,
or is there a deeper meaning to it? It is actually so that the 
stress photons parameterize the physics of the medium in a more complete
way than the strain phonons do. We already pointed at the ${\cal B}^{-1}_L$
stress photon which has no image in the phonon sector. Nevertheless, this
photon is physical. In order to parameterize all the forces which can
be transmitted in the elastic medium one needs a physical photon
for every dimension, meaning that three stress-photons have to exist
given that space-time is 2+1 dimensional. At the same time, in order
to explain any outcome of an experiment on the medium using 
machines constructed from orderly matter
(like neutron sources) it is clear that one can get away
just knowing about the Goldstone bosons which are counted according
to the number of space dimensions (the two phonons). How can this be?
The above calculation gives the answer. A first major complication 
occurs due to the transformation to the ${\cal B}$ photons: the
${\cal B}^{-1}_L$ propagators acquire zero weight and the neutrons,
etcetera, only couple to the longitudinal phonon-like   ${\cal B}^{1}_T$
stress photons. In other words, the  ${\cal B}^{-1}_L$ are genuine modes
but experimental physics does not posses the machines to measure them!

This becomes more manifest when dislocations come into play. When these
proliferate, the disorderly side of the duality gets in charge, imposing
its physics also on the orderly side of the universe where neutron sources
are found. As we will see in the next sections, this has the
consequence that the third stress-photon will become visible: the
order derived superfluid is characterized by three collective modes.     

\section{Completing the duality: Coulomb nematics as superfluids.}

In the last two sections we acquired the tools required to
unveil the physics hidden in the formal construction introduced
in section III. We learned how to incorporate all the
constraints to isolate the physical fields and what remains to
be done is to see what happens in the presence of the dislocation
condensate of section III: the stress photons get coupled to
the dislocation condensate and these will exhibit a dual 
Higgs mechanism.     
 In section VI.A we will derive the precise form of the coupling
between the dislocation condensate order parameter field and the
stress gauge fields. In section VI.B we will derive the effective
Gauge-field action describing the nematic states.

\subsection{The charges in stress-gauge theory.}

Let us start using the lessons of the previous sections to find
out what the physical dislocation currents are and how they
interact with the projected stress-gauge fields.
In the duality framework, minimal coupling acquires the 
meaning that it just enumerates the number of ways in which order 
fields can become singular. As we already discussed, dislocations
enter via the minimum coupling action Eq. (\ref{dislosource}),

\begin{equation}
{\cal L}^{dual}_1 = i B_{\mu}^a J_{\mu}^a.
\label{ dislosource1}
\end{equation}

The singularities can of course only occur in physical field
configurations, and because the stress gauge fields $B_{\mu}^a$
have unphysical components, the dislocation currents $J_{\mu}^a$
also correspond with a redundant set: a gauge volume has to
be divided out. In the previous sections we learned how to isolate
the physical content in the stress-gauge sector, and gauge invariance
implies that a physical source is uniquely associated with 
every physical stress photon,
        
\begin{equation}
S_{BJ} = i B^T_{-1} {J^T_{-1}}^* + i B^T_{+1} {J^T_{+1}}^* + i B^L_{-1} 
{J^L_{-1}}^{*},
\label{dislcoup0}
\end{equation}

\noindent where the 
Fourier components of the corresponding fields are assumed.
As only three physical stress photons exist, no more than three
components ($J^T_{-1},  J^T_{+1}, J^L_{-1}$) of the six components of the
tensor $J^a_{\mu}$ are physical.

At this stage, the theory still remembers the fact that dislocations
carry vectorial charges. As we discussed, the  `lower index' $-1$ 
implies that the currents $J_{-1}^{L,T}$ are space like, meaning that 
these refer to literal currents in this non-relativistic problem. 
Using Eq. (\ref{tildeelin}) the $L, T$ components can be transformed
into $x, y$ components and $(J^x_{-1}, J^y_{-1})$ refers to a current
of dislocations with a Burgers vector having components $\sim (n_x, n_y)$.

We have now to recall the glide principle, which we discussed at length
in section III. This is an independent constraint which has
to be imposed in order to keep the theory meaningful. It amounts
to a simple symmetry condition on the spatial components of the
dislocation currents $J_x^y = J_y^x$. After the projection, it becomes
evident why this is an independent constraint: 

\begin{equation}
J^L_{-1} = i \hat{\omega} J^T_{+1}.
\label{glidecond}
\end{equation}

The set  $\{J^T_{-1},  J^T_{+1}, J^L_{-1}\}$ is 
still redundant and because of the glide 
constraint either $J^L_{-1}$ or $J^T_{+1}$ may be 
eliminated. It is convenient to
eliminate $J^L_{-1}$ in favor of $J^T_{+1}$. We end up with 
$(J^T_{+1}, J^T_{-1})$ as the physical currents.  These are space-time
filling because both polarizations $ \pm 1$ are present (i.e. they
correspond with $J_{\mu}^T, \mu = x, y, \tau$ before the projection).
It is surprising that only a single, transversal Burgers flavor $T$
is left. It means that when the dislocations would only communicate 
via the long range elastic forces they would at least in 2+1D
become indistinguishable
from either vortices or just electromagnetically charged scalar 
matter in the static limit. A difference is that the stress gauge fields 
have a richer structure than electromagnetic fields, and this renders
the dynamics richer. By inserting the glide
condition Eq. (\ref{glidecond}) into  Eq. (\ref{dislcoup0}) it follows
that

\begin{equation}
S_{BJ}  =  i B^T_{-1} {J^T_{-1}}^* + 
i ( B^T_{1} -i \hat{\omega} B^L_{-1} ) {J^T_{+1}}^{*}.
\label{charges0}
\end{equation}

The coupling to the $B^T_{\pm 1}$ fields is as usual, and the novelty 
is that the $J^T_{+1}$ current couples to the $B^L_{-1}$ field with
a charge $g \sim \omega$. Matters further clarify by transforming to
normal coordinates, Eq. (\ref{eltransform}),

\begin{equation}
S_{{\cal B}J} =  i B^T_{-1} J^T_{-1} 
- { i \over {\sqrt{ 1 + \hat{\omega}^2 }}}
\left[ ( 1 - \hat{\omega}^2 ) {\cal B}^T_1 
- 2 \hat{\omega} {\cal B}^L_{-1} 
\right]  { J^T_{+1}}^*
\label{charges1}
\end{equation}

The gauge field action is given by Eq. (\ref{elasticaction}). 
In the scaling 
limit, approaching the vacuum, the higher time derivatives $\sim 
\hat{\omega}^2$ are clearly irrelevant. The ${\cal B}^L_{-1}$ 
mode decouples,
and one ends up with a structure which is except for a velocity anisotropy
just like electromagnetism coupled to a scalar Higgs field.  

\subsection{The dual dislocation-Higgs mechanism.}

Let us now recall the discussion in section III. When the dislocation 
loops blow out the dislocation world-lines entangle, forming a 
Bose condensate described by the order parameter field $\Psi$. This
is in turn minimally coupled to an effective gauge field $n_a B^a_{\mu}$.
The condensate itself is living in the static $\omega \rightarrow 0$
limit and the relevant coupling becomes,

\begin{eqnarray}
S_{{\cal B}J} (\omega \rightarrow 0) & = & 
i \left( B^T_{-1} {J^T_{-1}}^* + B^T_{+1} {J^T_{+1}}^* \right) \nonumber \\
              & = & i B^T_{\mu}{ J^T_{\mu}}^* 
\label{chargesstatic}
\end{eqnarray}

Interestingly, due to the intervention of the glide principle, the
Burgers vectors are in a sense `eaten' and turned into a scalar,
purely transversal label. This has the interesting consequence 
that if dislocations would only interact by the long range deformations
as parameterized by the stress gauge fields {\em true nematic order
would be impossible} at least in the isotropic medium. As we explained
in section III, the breaking of rotational invariance involves an
ordering of the Burgers vectors and according to Eq. (\ref{chargesstatic})
the long range fields do not carry information  regarding the vectorial
nature of the topological charges. However, starting from a real crystal
this information will be available at short distances and this translates
into the coupling $|\Psi|^4 Q_{a b} Q_{b a}$. 

Recalling the static Ginzburg-Landau-Wilson action for the dislocation
disorder field theory Eq. (\ref{disorderpsi}) and taking into account
that only the transversal components of the stress Gauge fields carry charge,
it follows from Eq. (\ref{mincoupexp}),

\begin{equation}
S_{disorder} = \int d^2x \left[ | (\partial_a - i n_T B^T_a) \Psi |^2
+ m^2_{\Psi} | \Psi^2| + w_{\Psi}  |\Psi^4| - r |\Psi|^4 Q_{a b} Q_{b a} 
\right]
\label{Coulombnematic}
\end{equation}

This action shows that when $m^2$ turns negative, a Higgs mass will be 
generated in a conventional way: $\Psi \rightarrow |\Psi| \exp{(i \phi)},
\; |\Psi| > 0$ and the covariant derivative term $\sim |\partial_a \phi
- n_T B^T_a| \rightarrow |n_T B^T_a|^2$.

In addition, we
also must to take into account the {\em dynamical} coupling to the 
${\cal B}^L_{-1}$ photon. Although power counting demonstrates
that this coupling is clearly irrelevant for the vacuum, it plays a
crucial role in the dynamics. This full dynamics is easily derived. 
 Writing Eq. (\ref{charges1}) as

\begin{equation}
{\cal L} _{BJ}  =  i g^a_{k, l} (\hat{\omega}) B^{a *}_k J^T_l + h.c.,
\label{charges2}
\end{equation}

where the charges $g$ are zero except for

\begin{eqnarray}
g^{T}_{-1, -1} & = & 1 \nonumber \\
g^{T}_{+1, +1} & = & { { ( 1 - \hat{\omega}^2 ) } 
\over {\sqrt{ 1 + \hat{\omega}^2 }}}  \nonumber \\
g^{L}_{-1, +1} & = & { { ( - 2 \hat{\omega} ) } 
\over {\sqrt{ 1 + \hat{\omega}^2 }}}.
\label{chargeex}
\end{eqnarray}
  
It follows directly that the Meissner term acquired by the stress
gauge fields in the dual condensate has to be

\begin{equation}
{\cal L}_{Meissner} = { {q^2_0} \over {2 \mu} } \sum_{k k' a a'} \; |n_T|^2 \; 
g^a_{k, l} g^{a'}_{k', l} \left( B^{a *}_{k} B^{a'}_{k'} + h.c. \right).
\label{Meissner0}
\end{equation}

The square root of the Higgs mass ($q_0$) is an inverse `magnetic' 
penetration depth which we will soon learn to appreciate as a shear
penetration depth. Written explicitly,

\begin{equation}
{\cal L}_{Meissner} = { {q^2_0 \; |n_T|^2} \over {2 \mu} } 
\left[ |B^T_{-1}|^2 + 4 { { \hat{\omega}^2 } \over { 1 + \hat{\omega}^2}} \;
|{\cal B}^L_{-1}|^2  - 
{ { 2\hat{\omega}{( 1 - \hat{\omega}^2) }} \over { 1 + \hat{\omega}^2}}
( {\cal B}^{T *}_1  {\cal B}^L_{-1} + h.c. ) \right].
\label{DuMeissner}
\end{equation}

In combination with the `Maxwell' action Eq. (\ref{elasticaction}),
we arrive at a next central result, describing the effective total 
stress-gauge field action characterizing the nematic dual 
of elasticity,

\begin{eqnarray}
{\cal L}^{eff}_{BB} & = & { 1 \over {4\mu } }
\left[ ( 2 \omega^2 + q^2 + 2 |n_T|^2 q^2_0 ) |B^T_{-1}|^2
+ ( \omega^2 + q^2 + 8 |n_T|^2 q^2_0 { { \hat{\omega}^2 } \over { 1 + \hat{\omega}^2}} )
|{\cal B}^L_{-1}|^2 \right. \nonumber \\
& & \left. + ( { {( 1 - \nu) \omega^2 + q^2 } \over { 1 + \nu} } 
+  2 |n_T|^2 q^2_0 { { ( 1 - \hat{\omega}^2)^2 } 
\over { 1 + \hat{\omega}^2}} ) 
|{\cal B}^T_{+1}|^2
  - 4 |n_T|^2 q^2_0 { \hat{\omega}{( 1 - \hat{\omega}^2) } 
\over { 1 + \hat{\omega}^2}}
( {\cal B}^{T *}_1  {\cal B}^L_{-1} + h.c. ) \right].
\label{dualeffaction}
\end{eqnarray}

This expression characterizes completely the nematic duals of the crystal,
in the same way that the Meissner action is fully representative for
the universal characteristics of the superconducting state. A remaining
issue is how this action knows about the presence or absence of 
rotational symmetry breaking. This information enters via the factors 
$|n_T|^2$. Neglecting the frequency dependency of the  charges for 
clarity, 
terms of the kind  $|n_T  B^T_{\mu}|^2$ can be rewritten as

\begin{eqnarray}
|n_T B^T_{\mu}|^2 & = & ( \hat{q}_y n_x -  \hat{q}_x n_y )^2 |B^T_{\mu}|^2
\nonumber \\
 & = & Q^2 \left( {1 \over 2} + \hat{q}^2_y Q_{xx} + \hat{q}^2_x Q_{yy}
- 2  \hat{q}_x \hat{q}_y Q_{xy} \right) |B^T_{\mu}|^2
\label{qdepmeissner}
\end{eqnarray}

in terms of the unit directors $Q_{ab}$ introduced in section III. One
infers that the effect of the director order $\langle | Q_{ab} | \rangle
\neq  0$ is in causing {\em the shear-Higgs mass to become orientation
dependent}. The glide principle lies at the origin. The glide constraint
is responsible  for the coupling being entirely in the transversal 
channel. 
The Higgs mass has to do with the kinetic energy of the dislocations and,
as we explained in section III, glide does imply that the
dislocation motions are `directed' by their Burgers vectors, which in 
turn order along the director. A surprising feature is that the order
parameter field $\Psi$ is  still fully two (space) dimensional, and the
condensate is just conventional. The rational is found in section IIIC
and Appendix B. The dimensional reduction 
(`dynamical compactification')
implied by Eq. (\ref{qdepmeissner}) in the presence of director order
as will be discussed in section VIII becomes only manifest on the level
of the stress-photons.

As we discussed
in section III at length, the other possible state of nematic quantum
matter is the Coulomb nematic which is not breaking rotational invariance,
such that $\langle | Q_{ab} | \rangle  = 0$. This is the more basic problem
which we will analyze first. In the absence of director order, it follows
directly from Eq. (\ref{qdepmeissner}) that $|n_T|^2 = Q^2 /2$, a simple
scalar which can be absorbed in the Higgs mass $Q^2 q^2_0 /2 \rightarrow
q^2_0$. Therefore, in the Coulomb nematic one is dealing with an
isotropic Higgs mass and this will be further analyzed in the next section.
As we will see in section VIII, the ordered nematic is just a spatially
anisotropic extension of this theme.

\section{The Coulomb nematic as a topological superfluid.}

The hard work is done and it becomes now a matter of computing
observables to establish what it all means physically. The
Coulomb nematic is the most basic example, and in this section we will
use it to extract the most important results of this paper. We will
demonstrate  that the dual dislocation-Higgs condensate is
at the same time a superfluid which is indistinguishable from
a conventional superfluid except for the presence of the
topological nematic order. The mechanism driving the superfluidity
is novel. The order parameter in a conventional
superfluid is governed by off-diagonal long range order in terms
of the field operators associated with the constituent particles
(`interstitials' in our jargon). We are dealing with a state having
off-diagonal long range order in terms of the dual dislocation fields, 
objects
corresponding with an {\em infinity} of constituent particles.
The dislocation condensate is minimally coupled to the stress photons,
and these acquire a Higgs mass. However, they do so in a special way.
By imposing the glide condition, only the shear components of
the stress photons acquire a mass: as we discussed in section III,
compression is decoupled. The effect is  that the dual dislocation
condensate is characterized by two massive shear `photons' and
a massless purely compressional mode. According to a theorem by
Feynman a bosonic state carrying an isolated massless compression
mode has to be a superfluid. Hence, we have discovered a way to
understand superfluidity without invoking Bose condensation of
particles: a superfluid is a Bose-solid having lost its rigidity
against shear stresses!

\subsection{The stress photons in the Coulomb nematic state.}

The starting point is the action Eq. (\ref{dualeffaction}) and
still some work has to be done. The `magnetic'
transversal phonon-like  stress photon $B^T_{-1}$ is easy: its
spectrum becomes in real frequency $\omega = \sqrt{ q^2 + q^2_0}$, implying
that this photon just acquires a Higgs mass $\omega_0 = q_0$. One observes that
its coupling constant is just $1 /\mu$, the inverse of the shear modulus. 
This is therefore a pure shear mode and this sheds already some light on
the nature of this dual Higgs condensate. The elastic medium is
associated with translational symmetry breaking with the consequence 
that shear forces can propagate over infinite distances. 
Dislocations restore the translational invariance with the consequence
that shear rigidity vanishes in the dual condensate. The physical
interpretation of the dual Higgs mechanism is now obvious: the
dislocation condensate just takes care of the requirement that shear
forces can only propagate over finite distances in the liquid state,
and $q_0$ (carrying the dimension of inverse length) just parameterizes
the inverse length scale over which shear can penetrate into the fluid:
the `shear penetartion depth'.

Eq. (\ref{dualeffaction}) tells a more complicated story 
regarding the fate of the longitudinal phonon-like 
stress photon ${\cal B}^L_{-1}$ and the `third photon'  ${\cal B}^T_{+1}$
in the presence of the dislocation condensate. The Meissner effect causes
a complicated, frequency dependent mode coupling between these two 
stress photons. With some effort, the action be diagonalized

\begin{eqnarray}
{\cal L}^{eff} ( {\cal B}) & = &  { 1 \over {4\mu} } \left[
( 2 \omega^2 + q^2 + 2q^2_0 ) |B^T_{-1}|^2 \right. \nonumber \\
& & \left. \; + \; { 1 \over {(1 + \nu) } }
\sum_{\pm} \left( \omega^2 + ( 1 + { {\nu} \over 2} ) q^2 +
( 1 + \hat{\omega}^2) (1 + \nu ) q^2_0  \right. \right. \nonumber \\
&  & \left. \left. \pm
\sqrt{ ( \nu ( \omega^2 + { {q^2} \over 2} ) + ( 1 + \hat{\omega}^2)
(1 + \nu ) q^2_0 ) ^2 - 2 \nu ( 1 + \nu ) ( 1 - \hat{\omega}^2) q^2 q^2_0 }
\right)  | {\cal B}_{\pm}|^2 \right]
\label{dualeffaction1}
\end{eqnarray}

and we will analyze this result in more detail in the remainder of 
this section.

By considering the zero-frequency limit one can already arrive
at a crucial insight,

\begin{equation}
\lim_{\omega \rightarrow 0} \;  {\cal L}^{eff} 
=  { 1 \over {4\mu} } \left[ ( q^2 + 2q^2_0 ) |B^T_{-1}|^2
\; + \;  q^2 | {\cal B}_{+}|^2 \; + \; ( { {q^2} \over {1 + \nu}} +2q_0^2 ) 
| {\cal B}_{-}|^2 \right]
\label{dualeffactionq0}
\end{equation}

Although the penetration depths are different, both the $B^T_{-1}$
and ${\cal B}_{-}$ stress photons have acquired a mass. However,
the ${\cal B}_{+}$ photon has remained massless. 
 The elastic medium carries both shear and
compression rigidity. Different from shear, compression should be
left unaffected by the dislocation condensate: as we discussed in
section III the theory is explicitly constructed to satisfy this 
requirement (the
glide principle). The longitudinal phonon is  of a
mixed shear- and compressional nature, and so is the ${\cal B}^L_{-1}$
stress photon which is traveling at the same velocity. As a 
consequence, the longitudinal phonon is destroyed by the
dislocation condensate. However, the dislocations do not
communicate with compression and therefore the dual state has to
be characterized by a pure massless compression mode, characterized
by a  velocity $c \sim \sqrt {\kappa /\rho}$. As we will see, this
is precisely what is happening. Clearly, in order to turn the
longitudinal phonon into a compression mode, another actor is
needed and this  the ${\cal B}^T_{+1}$ 
stress photon! The bottom line is that the mysterious, invisible
`third photon' of the elastic state turns into an entity with
a straightforward physical interpretation in the dual state.
It is not only needed to `repair'
compression, but by combining itself with the `longitudinal' photon
it gives rise to yet another physical  massive `electrical shear' mode
having a different gap and velocity than the `magnetic shear' $B^T_{-1}$
mode. The reader might already have inferred that these are 
statements regarding the generic modes of the superfluid in the
asymptote of strong correlations.

\subsection{The long wavelength limit: isolation of compression.}

To derive the elastic propagator using Eq. (\ref{totstressprop}) for
arbitrary $q, \omega$ is quite 
tedious, and barely worth the effort. It reduces to the phonon propagator
Eq. (\ref{totpropelas}) in the regime $q^2 + \omega^2 >> q^2_0$ and 
the physical interest is in first instance in the long wavelength
limit $q << q_0$ where the dislocation condensate is in charge of 
the physics. 

In the long wavelength limit $q << q_0$ matters simplify considerably.
Let us focus on the `electrical' shear- and massless compression photons.
 To get a handle on the $ q << q_0$ limit, let us
consider the dislocation-Meissner action in terms of the projected
fields $B$ itself, instead of the transformed ${\cal B}$ fields,

\begin{equation}
{\cal L}_{Meissner} = { {q^2_0} \over { 2 \mu} } | B^T_1 -i 
\hat{\omega} B^L_{-1}|^2
\label{Dumeiss1}
\end{equation}

It is directly seen that this is diagonalized by,

\begin{eqnarray}
{\cal A}^T_{+1} = { i \over {\sqrt{ 1 + \hat{\omega}^2}}}
\left( B^T_1 -i \hat{\omega} B^L_{-1} \right) \nonumber \\
{\cal A}^L_{-1} = { i \over {\sqrt{ 1 + \hat{\omega}^2}}}
\left( \hat{\omega} B^T_1 +i B^L_{-1} \right)
\label{distransform}
\end{eqnarray}

such that

\begin{equation}
{\cal L}_{Meissner} = { {q^2_0} \over { 2 \mu} } ( 1 +  \hat{\omega}^2 )
|{\cal A}^T_{+1}|^2 
\label{meissner1}
\end{equation}
 
showing that the ${\cal A}^L_{-1}$ mode decouples completely from
the dislocation condensate. The problem is, however, that this
transformation Eq. (\ref{distransform}) does not remove the mode
couplings associated with $\nu \neq 0$: $\hat{\omega}$ has reversed
its sign as compared to the transformation Eq. (\ref{eltransform}).
Let us see what happens when we transform Eq. (\ref{elaction0}) to
the ${\cal A}$ basis,

\begin{equation}
{\cal L}_0 = { {|p|^2} \over { 4 \mu} } \left[
| {\cal A}^T_{+1} |^2 + | {\cal A}^L_{-1} |^2 -
{ {\nu} \over { ( 1 + \nu) ( 1 + \hat{\omega}^2 )}}
| ( 1 - \hat{\omega}^2 ) {\cal A}^T_{+1} + 2 \hat{\omega} 
{\cal A}^L_{-1} |^2 \right]
\label{elasacA}
\end{equation}

and it is immediately inferred that the prefactor of the mode 
coupling $ {\cal A}^T_{+1 *} {\cal A}^L_{-1} + h.c.$ is proportional
to $ \hat{\omega} ( 1 - \hat{\omega}^2)|p|^2/ ( 1 + \hat{\omega}^2)
\sim q^2$, while the mode-splitting $\sim q^2_0$, implying that
in the long wavelength limit $q^2 << q^2_0$ this mode coupling
can be neglected, and the  ${\cal A}$ stress photons become diagonal.

The total action can be written at long-wavelength as

\begin{eqnarray}
{\cal L}_{q << q_0} & = & { {( 1 - \nu )} \over { 4 \mu( 1 + \nu )
( 1 + \hat{\omega}^2) }} \left[ 2 \omega^2 + 
{ {( 1 + \nu )} \over {( 1 - \nu )}} q^2 \right]\; |{\cal A}^L_{-1}|^2
\nonumber \\
& & + { 1 \over { 4 \mu} } \left[ ( 2 \omega^2 + q^2 + 2q^2_0 ) | B^T_{-1}|^2
+ ( \omega^2 + q^2 + 2 q^2_0 ( 1 + \hat{\omega}^2 ) ) | {\cal A}^T_{+1} |^2
\right]
\label{smallqact}
\end{eqnarray}

omitting terms  $O( q^2 / q^2_0)$. In the same limit, the
propagator  Eq.(\ref{totstressprop})
becomes in terms of the ${\cal A}$ fields,

\begin{eqnarray}
 G _{q << q_0} & = & - const. +  { { ( 1 - \nu )^2 \omega^2}
\over { 2 \mu^2 (1 + \nu)^2 ( 1 + \hat{\omega}^2) } }
\langle \langle  {\cal A}^L_{-1} | {\cal A}^L_{-1} \rangle \rangle  
\nonumber \\
& &  +  \; 
{ {2\omega^2} \over { 4 \mu^2 ( 1 + \hat{\omega^2})}}  
\langle \langle {\cal A}^T_{+1} | {\cal A}^T_{+1} \rangle \rangle
 \; +  \; { { 2 \omega^2} \over { 4 \mu^2}  }
\langle \langle  B^T_{-1} | B^T_{-1} \rangle \rangle 
\nonumber \\
& = & - const. + { 2 \over {\kappa}}
{ {q^2} \over { \omega^2 +  { {\kappa} \over {\rho}} q^2 } } 
\; +  \; { 1 \over {\mu} } \; 
{ { q^2 + 2 q^2_0 } \over { 2 \omega^2 + q^2 + 2q^2_0} }
\; +  \; { 1 \over {\mu} } 
{ {3 q^2 / 2  + 4 q^2_0 } \over { \omega^2 + 3 q^2 / 2 + 4q^2_0} }
\label{supercompr1}
\end{eqnarray}

and the spectral function can be read off,

\begin{eqnarray}
Im [  G _{q << q_0} ] & = & { {2 q^2} \over {\kappa}} \;
\delta \left( \omega^2 +  { {\kappa} \over {\rho}} q^2 \right)
\; +  \; { { q^2 + 2 q^2_0 } \over {\mu} } \; 
\delta \left(  2 \omega^2 + q^2 + 2q^2_0 \right)
\nonumber \\
& & \; +  \; { { 3 q^2 + 4 q^2_0 } \over {\mu} } \; 
\delta \; \left( \omega^2 + 3 q^2 / 2 + 4q^2_0 \right)
\label{supercompr}
\end{eqnarray}

\subsection{Discussion I: The scaling limit and the continuity principle}

Eq. (\ref{supercompr})  is one of the central results of this paper. 
It shows several highly interesting features. Most importantly, the
dislocation condensate or `dual of the crystal' is characterized
by a single massless collective mode. This mode is nothing else
than a true compression mode as is obvious from both the
velocity and the coupling constant, set both by the compression
modulus $\kappa$. The propagator is at long wavelength identical to
Eq. (\ref{Scom1}) associated with a `compression-only' solid 
characterized by $\nu=1$. Having understood the mechanics of the
dual Higgs mechanism this is not a surprise. At long wavelength
compression decouples from the dislocation sources, and the
compression mode is therefore not affected by the Higgs mass
coming from the dislocation Bose condensate.

The fact that this state is characterized by an isolated, massless
compression mode is sufficient condition to arrive at the following
far-reaching conclusion:\\
 
{\em The dislocation condensate is at the same time a conventional
superfluid.}\\

This statement rests on a general hydrodynamic argument predating
the discovery of off-diagonal long range order: it is originally
due to Landau\cite{Landau} and further elaborated by 
Feynman\cite{Feynman}. Landau argued that the superfluid fraction
could just be characterized by a `phonon', or more precisely, by
a propagating pure compression mode. The difference with a
normal fluid (carrying also sound) is in the complete vanishing
of the {\em shear viscosity} -- although a normal fluid does not 
carry a reactive response towards shear forces it will show a
dissipative response. From general hydrodynamic consideration
it follows that the flow of a fluid with vanishing shear viscosity
is {\em under all circumstances} 
a potential flow, and such a flow is (i) dissipationless and 
(ii) irrotational. In any hydrodynamics textbook one finds that
vorticity, dissipation and shear go hand in  hand. 

This hydrodynamical argument is not in conflict with ODLRO. The
phase mode of the conventional interacting Bose gas is just an
isolated compression mode. The novelty of our result is that
it demonstrates that Landau's hydrodynamical characterization
is the more fundamental one; conventional ODLRO is just one 
microscopic mechanism to realize this hydrodynamics. We 
demonstrate an alternative mechanism.  The `crystal-dual' 
presented here is characterized  by off-diagonal long range order
in terms of the dislocations. Every dislocation can be regarded as a 
composite
of an infinite number of constituent particles, and we took this
infinity particularly serious by insisting that the constituent
particles have disappeared altogether. In this sense, the overlap
with the conventional Bose gas is vanishing. The effect of the dual
Higgs condensate on the flow properties of the medium are
indirect. The `true' phonons of the original crystal are propagating
modes because of Goldstone protection. This Goldstone protection 
is not affected by the coherent dislocation Bose condensate: it
inherites its dissipationless nature from the elastic medium. However,
the dislocation condensate causes a Higgs mass in the shear `sector'
and an elastic medium characterized by massive shear and massless
compression sustains material flows which cannot be pinned while 
the flows have to be irrotational. In essence, it is like the
charge density wave sliding of Fr\"ohlich. The problem
with Fr\"ohlich's superflow is that any violation of Euclidean invariance
will restore dissipation: the impurity pinning. Pinning requires
shear rigidity, and the sliding mode/longitudinal phonon carries shear. 
However,  shear rigidity is vanishing in the dislocation
condensate. The impurity potential is screened by the dislocations,
and on scales exceeding the shear penetration depth $1 / q_0$ the
impurity potentials have disappeared. Hence, the dislocation condensate
turns the sliding mode in a superfluid mode  which cannot be pinned. At
the same time this fluid is irrotational. Only at energies exceeding
the shear mass gap, vortices can exist. Vorticity implies shear stresses.

Summarizing, our duality shows that the usual definition,

{\em a superfluid is a state of matter characterized by
off-diagonal long range order in term of the field operators
of the constituent bosons}

should be supplemented by,

{\em a superfluid is an elastic medium having lost its rigidity
against shear stresses due to the presence of a dual dislocation 
condensate,} 

to arrive at a complete characterization. We emphasize that both
viewpoints are equally correct. All one can discern in the scaling 
limit is the universal definition, as given by Landau:

{\em  a superfluid is a state of matter characterized by a low energy
spectrum which is exhausted by a propagating, massless compression mode.}

Since both `states' are hydrodynamically indistinguishable they
are actually the same state: by general principle it has to be 
possible to adiabatically continue the Bose-gas superfluid into 
our `order' superfluid. These are just limiting `microscopic'
descriptions of the same entity. Among others, this  also 
implies a `don't worry' theorem regarding the problem that interstitials
cannot be incorporated in the field theory. Because of continuity,
it has to be that these can be `smoothly' inserted in the theory,
driving the superfluid away from the order-asymptote which is not
a singular limit. 

To complete the argument we still have to demonstrate  that in
this irrotational fluid a  genuine Meissner effect should occur
when electromagnetic fields are coupled in. This we reserve for
section IX.

\subsection{Discussion II: the massive shear modes.}

At distances shorter than the shear penetration depth the
differences in the description become manifest, as follows 
directly from Eq. (\ref{supercompr}). From the study of $^4$Helium
a clear view has developed on the nature of the excitation
spectrum in a  `particle superfluid'. At higher energies, away
from the scaling limit, one finds besides an incoherent continuum
associated with particle-like excitations the roton-minimum. 
Following Feynman\cite{Feynman}, the roton is associated with crystal-like
correlations limited to length scales of order of the lattice
constant (the wiggles in the structure factor). The case we consider
is about a substance where the crystal-correlations extend to
length scales which are still very large compared to the
lattice constant. In the field theory, the lattice constant is sent
to zero
while the shear penetration depth is kept finite. We predict that
in this regime the system is characterized by two propagating, 
massive modes besides the massless compression mode,
see Eq. (\ref{supercompr}). The continuity argument given in the
above implies that these massive modes should be meaningful, also
away from the field theoretic $ a \rightarrow 0$ limit. Their existence
just requires that the crystalline correlations extend to a length
scale (shear penetration depth) which is substantially larger than
the lattice constant. There is every reason to expect that these massive 
shear modes are generic under these circumstances and we expect
that in a bosonic substance having a substantially weaker first order
crystal-superfluid transition than Helium these modes will show up.
One can even wonder if the roton seen in Helium can be interpreted as
a remnant of such a mode, in the case that $ 1 /q_0 $ starts to approach
the lattice constant.

What is the physical nature of the massive modes? It is easily checked
that both are exclusively carrying shear. The dual Higgs mechanism
acts at long wavelength to completely remove the compression content
from these modes, so that compression is entirely concentrated in the
massless mode. It is seen from  Eq. (\ref{supercompr}) that
both modes have a finite pole-strength at $q \rightarrow 0$ and these
would therefore show up in for instance inelastic scattering. Amusingly,
deep in the disordered state the mode content in the stress-photon 
representation becomes visible on `our side' of the duality. In the
solid, and in the regime $q > q_0$ (next section) only two phonons
are visible but when the disorder fields take over in the regime 
$q < q_0$, the photons take over.

It is also seen that the two shear modes are inequivalent. The $B^T_{-1}$
stress photon carries the natural velocity $\sqrt{ \mu / \rho}$ and Higgs
gap $q_0$. The  ${\cal A}^T_{+1}$ mode is, on the other hand, characterized
by a a velocity which is larger by a factor $\sqrt{3}$ and a gap which is
larger by a factor of two. How can this be, given that they both reflect
the same fundamental scales ($\mu$, $\rho$, $\hbar$)? Our analysis shows
that it originates in the way Lorentz invariance is broken, and since this
happens in a generic way in non-relativistic nature, the ratio of the
gaps and the velocities can be regarded as {\em universal}. The analogy
with electromagnetism which we introduced earlier is helpful in this
regard. The  $B^T_{-1}$ photon is magnetic-like while ${\cal A}^T_{+1}$
is electrical-like. Since Lorentz-invariance is broken these are no
longer equivalent even when $c=1$ and this anisotropy causes the modes to
be inequivalent in a universal fashion.

Summarizing, the analysis of the excitation spectrum of our 
topological-nematic superfluid in 2+1D has revealed features 
which should be far more general than the present case, since
they are all tied to general symmetry principles. The theory
is surprisingly predictive. All what is required is a bosonic
substance undergoing a second order (or sufficiently weak
first order) crystal-superfluid transition. By measuring the
phonons on the crystal side of the transition one can establish
the shear modulus $\mu$, the mass density $\rho$ and the Poisson
ratio $\nu$. The only free parameter left on the superfluid side
is the shear penetration depth $1 / q_0$ and this suffices to
extract the three distinct velocities and the two mass scales
associated with the three modes visible on the superfluid side.    

\subsection{Discussion III: intermediate scales.}

It is quite tedious to derive the  expressions for the 
full propagator in the regime of intermediate momenta and frequency,
and barely worth the effort. One can immediately infer what is
happening. Define a dimensionless gap scale,

\begin{equation}
\hat{q}_0 = { {q_0} \over {|p|} } = 
{ {q_0} \over { \sqrt{ q^2 + \omega^2} } }. 
\label{dimlessgap}
\end{equation}

It is immediately clear that in the regime $\hat{q}_0 << 1$, the
spectrum of the elastic state, characterized by a longitudinal-
and transverse phonon, should be recovered. Hence, when
$\hat{q_0} \simeq 1$ a crossover occurs from the compression- plus
two massive shear modes of the superfluid to the phonon spectrum.
At this crossover, the `electric' shear mode acquires compression
character, turning into the longitudinal mode. At the same time,
the `third stress photon' is rediscovered, and one expects this
third mode to gradually loose its spectral weight, to vanish 
completely at large momenta. 

It is easy to obtain a more detailed view on these matters by
considering the regime where the dimensionless gap $\hat{q_0}$
is small but finite. Consider the diagonalized action
Eq. (\ref{dualeffaction1}). Ignoring the magnetic shear mode,
this can be written as

\begin{eqnarray}
{\cal L}^{eff}_{\pm} ( {\cal B}) & = &  { 1 \over {4\mu ( 1 + \nu)} }
\sum_{\pm} \left( \omega^2 + ( 1 + { {\nu} \over 2} ) q^2 +
( 1 + \hat{\omega}^2) (1 + \nu ) q^2_0  \right. \nonumber \\
&  & \left.  \pm | p |^2
\sqrt{ ( 1 + \hat{\omega}^2 )^2
( {{\nu} \over 2} ) +  (1 + \nu ) \hat{q}^2_0 )^2 
- 2 \nu ( 1 + \nu ) ( 1 - \hat{\omega}^2)^2 \hat{q}^2_0 }
\right)  | {\cal B}_{\pm}|^2.
\label{effactplmi}
\end{eqnarray}

Expanding the square root up to order $\hat{q_0}^2$,

\begin{eqnarray}
{\cal L}^{eff}_{\pm} ( {\cal B}) & \simeq &  { 1 \over {4\mu} }
\left[ \left( \omega^2 \;  + \; q^2 + \; 8q^2_0 \; 
{ {\omega^2} \over { 2 \omega^2 +q^2} } \right) \; | {\cal B}_{+} |^2 
\right. \nonumber \\
& & \left. \; + \; \left({  { \omega^2 ( 1 - \nu)  +  q^2} 
\over { 1 + \nu} } 
+ \; 2q^2_0 \; 
{ {q^4} \over { (2 \omega^2 +q^2) ( \omega^2 + q^2)} } 
\right) \; | {\cal B}_{-} |^2 \right].
\label{effactplmi1}
\end{eqnarray}

One recognizes the `third photon' (${\cal B}_{+}$) and the longitudinal
phonon (${\cal B}_{-}$). The gaps `seen' by these stress photons are
multiplied by dimensionless ratios which depend on particular
combinations of spatial momenta and frequencies. However, it is immediately
clear from Eq. (\ref{effactplmi1}) that the poles will closely approach
the bare elastic stress-photon poles. To leading order in small $\hat{q}_0$
we can therefore evaluate the values of the frequency-momenta ratios 
multiplying $q^2_0$ by inserting for these ratio's their values at the
positions of the bare poles. We find 

\begin{eqnarray}
{\cal L}^{eff}_{\pm} ( {\cal B}) & \simeq &  { 1 \over {4\mu} }
\left( \omega^2 \;  + \; q^2 + \; 8q^2_0  \right) \; | {\cal B}_{+} |^2 
 \nonumber \\
& & \; + \;  { {1 - \nu} \over {4\mu (1 + \nu) } }
\left( \omega^2  \;  + \; { { q^2} \over { 1 - \nu}} 
+ \; { { 2 ( 1 - \nu) } \over
{ \nu } } \; q^2_0 \right)  \; | {\cal B}_{-} |^2.
\label{effactplmi2}
\end{eqnarray} 

It follows that the the third photon looses its spectral weight
in a `universal' way (i.e. independent of the Poisson ration $\nu$)
in much the same way as for instance a Bogoliubov excitation in
a BCS superconductor looses its weight upon exceeding the gap scale.
On the other hand, the  recovery of the longitudinal photon does 
depend on the Poisson ratio. Interestingly, one finds that for
vanishing Poisson ratio the above simple expansion becomes
singular. Vanishing Poisson ratio means that the compression and
shear moduli become equal, $\kappa = \mu$. What is going on at 
this special point?

As a fortunate circumstance, the $\nu = 0$ case can be easily evaluated
analytically. The reason is that in this case 
Eq. (\ref{elasacA}) is diagonal. The dual action simplifies to

\begin{eqnarray}
S (\nu =0) & = & { {\hbar} \over {4\mu}} \int dq^2 d\omega \left[
p^2 |B^T_1|^2 + ( p^2 + \omega^2) |B^T_{-1}|^2 + p^2 |B^L_{-1}|^2 \right.
\nonumber \\
& & \left. 
+ 2 q^2_0 (\; |B^T_{-1}|^2 + | B^T_1 -i \hat{\omega} B^L_{-1} |^2 \; )
\right]
\nonumber \\
 & = & { {\hbar} \over {4\mu} } \int dq^2 d\omega \left(
( 2 \omega^2 + q^2 + 2 q_0^2 ) |B^T_{-1}|^2  \right.
\nonumber \\
& & \left. +
( \omega^2 + q^2 + 2 q_0^2 ( 1 + \hat{\omega}^2) ) | {\cal A}^T_{+1}|^2
+ ( \omega^2 + q^2 )| {\cal A}^L_{-1} |^2 \right).
\label{supnuzerodiag}
\end{eqnarray}

With the exception of the $B^T_{-1}$ mode, one cannot yet read 
the mode spectrum from this expression because it involves a frequency
dependent gap $2 q^2_0 ( 1 + \hat{\omega}^2) = 2q^2_0 ( 2 \omega^2 + q^2 )
/ ( \omega^2 + q^2 )$. Let us consider therefore the 
propagator, simplifying considerably in the $\nu = 0$ case, 

\begin{equation}
G  =  { 1 \over {\mu} } - { {p^2} \over { 4 \mu^2} }
\left( 2 \hat{\omega}^2 \langle \langle B^T_{-1} |  B^T_{-1} \rangle \rangle \;
+  \; \langle \langle B^T_{+1} |  B^T_{+1} \rangle \rangle \; + \;
\hat{\omega}^2 \langle \langle  B^L_{-1} |  B^L_{-1} \rangle \rangle \right).
\label{nuzeroprop}
\end{equation}

Transforming to the ${\cal A}$'s,

\begin{eqnarray}
G  & =  & { 1 \over {\mu} } \; \left[ 1 \; - \; 
{ {2 \omega^2} \over  { 2 \omega^2 + q^2 + 2q^2_0} } \right. \nonumber \\
 &  & \left. \; + \; 
{ { ( 2 \omega^2 + q^2) ( \omega^2 + q^2 ) + 4 \omega^2 q_0^2 }
\over { ( \omega^2 + q^2 + 2q_0^2 ( 1 + \sqrt { 1 + { {q^2} \over {2 q_0^2}}}))
( \omega^2 + q^2 + 2q_0^2 ( 1 - \sqrt { 1 + { {q^2} \over {2 q_0^2}}}))} }
\right]
\nonumber \\
 & = & { 1 \over {\mu} } \; \left[  
{ {( q^2_0 + { {q^2} \over 2} )}
\over { \omega^2 + q^2 /2  + q^2_0} } \; + \;
{  { ( 2 q^2_0 +  { {q^2} \over 2} )}  \over {  
\sqrt { 1 + { {q^2} \over {2 q_0^2}}}} } \times
\right. 
\nonumber \\ 
& & \left. 
\left( { {  \sqrt { 1 + { {q^2} \over {2 q_0^2}}} + 1}
\over {\omega^2 + q^2 + 2q_0^2 ( 1 + \sqrt { 1 + { {q^2}} 
\over {2 q_0^2}}})}   \; + \; 
{ {  \sqrt { 1 + { {q^2} \over {2 q_0^2}}} - 1} \over
{( \omega^2 + q^2 + 2q_0^2 ( 1 - \sqrt { 1 + { {q^2} \over {2 q_0^2}}}))}} 
\right) \right].
\label{nu0supprop}
\end{eqnarray}

It immediately follows that
\begin{eqnarray}
Im [ G ]  & = &  { 1 \over {\mu} } \; \left[
( q^2_0 + { {q^2} \over 2} ) \delta ( \omega^2 + { {q^2} \over 2} + q_0^2 )
+ { { 2 q^2_0 +  { {q^2} \over 2} }  \over {  
\sqrt { 1 + { {q^2} \over {2 q_0^2}}}} } \right. \nonumber \\
&  & \left.  \times
\left(  (  \sqrt { 1 + { {q^2} \over {2 q_0^2}}} + 1 )
\delta ( \omega^2 + q^2 + 2q_0^2 ( 1 + \sqrt { 1 + { {q^2} \over {2 q_0^2}} }) 
\right. \right. \nonumber \\
& & \left. \left.  + (  \sqrt { 1 + { {q^2} \over {2 q_0^2}}} - 1 )
\delta ( \omega^2 + q^2 + 2q_0^2 ( 1 - \sqrt { 1 + { {q^2} \over {2 q_0^2}}}) 
\right) \right],    
\label{nu0supspec}
\end{eqnarray}
providing a closed solution for the spectrum.

\begin{figure}[tb]
\begin{center}
\vspace{10mm}
\leavevmode
\epsfxsize=0.6\hsize
\epsfysize=10cm
\epsfbox{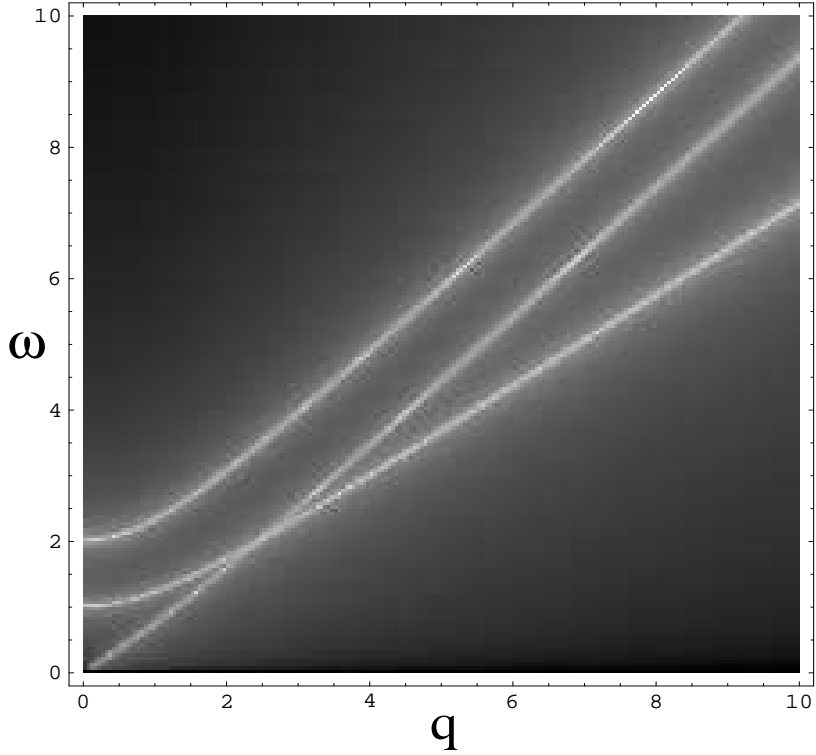}
\end{center}
\caption{The spectral function derived from the elastic propagator of
the Coulomb nematic superfluid
for the special case that the Poisson ratio is vanishing ($\mu = \kappa$).
A small artificial broadening is added and the grey scale indicates the 
magnitude of the spectral function. One recognizes at small momentum $q$ 
the massless compression (or superfluid phase-) mode, and the two massive 
shear modes. For accidental reasons, the superfluid modes do not merge
with the longitudinal and transverse phonons of the solid at large momenta,
and the three modes remain visible. This effect can be traced 
to a accidental
degeneracy, see text.}
\label{spfiesheacom}
\end{figure}

In Fig. (\ref{spfiesheacom}) the spectral function of the $\nu = 0$
case is shown in full, in units of $q_0 = 1$. In the long wavelength
regime ($q << 1$) this is according to the expectations: one finds
the massless  compression mode and the magnetic- and electric massive
shear photons with a gap ratio of two. At high frequency, $q >> 1$,
a surprise occurs: all three stress photons remain visible, while
the `longitudinal phonon' and third mode are separated by a fixed frequency
$\sim q_0$, sharing the total spectral weight in the `electrical' sector. 
This phenomenon has a simple explanation. In the general case the
poles as function of real frequency will be determined by
$\omega \sim \sqrt{ q^2 + \alpha q^2_0}$ with $\alpha$ a number of order
unity at large frequencies. For $q^2 >> q^2_0$ this can be written
as $\omega \sim q^2 + (\alpha /2 ) q^2_0 / q$ and the mass gap will therefore
not exert influences in the large frequency regime. The specialty of
the $\nu = 0$ case is, however, that the velocity of the longitudinal-
and third stress photons is the same, see Eq. (\ref{supnuzerodiag}). 
In other words the mode-coupling caused by the gap is at resonance, 
regardless the measuring frequency. Accordingly, it follows from
Eq. (\ref{nu0supprop}) that at high frequencies the `electrical' poles 
are at $\omega_{\pm} \sim q \sqrt{1 \pm \sqrt{2} (q_0 / q) + 2 (q_0 /q)^2}
\simeq q \pm q_0 / \sqrt{2}$, split by a fixed amount regardless the
frequency.

\section{Breaking rotational symmetry: is it a nematic or a smectic?}

As we already announced, the ordered quantum-nematic state is 
a straightforward extension of the Coulomb nematic discussed
in detail in the previous section. As we explained in section VI B,
the shear Higgs mass is multiplied by a factor  which depends on
the orientation in (spatial) momentum space $\vec{\hat{q}}$,

\begin{equation}
q^2_0 (\vec{\hat{q}}) = \tilde{q}^2_0 \left( 1 
+ 2 \hat{q}^2_y Q_{xx} + 2 \hat{q}^2_x Q_{yy} - 4  \hat{q}_x 
\hat{q}_y Q_{xy} \right).
\label{anisomass}
\end{equation}

To gain some intuition, let us consider 
the case that the director is 
`completely polarized'. As we explained in section III, $Q_{ab} = n_a n_b
- 1/2 \delta_{a b}$ where the vectors $\vec{n}$ are $O(2)$ unit rotors 
parameterizing the Burgers vectors. `Complete polarization' now means that
the Burgers vectors are exclusively oriented in one direction. Under this
condition the $Q$'s can be parameterized as

\begin{eqnarray}
\hat{n}_x & = & \cos{\eta} \nonumber \\
\hat{n}_y & = & \sin{\eta} \nonumber \\
Q_{xx} & = & {1 \over 2}  \cos{ (2\eta) }  \nonumber \\
Q_{yy} & = &  - {1 \over 2} \cos{ (2\eta)}  \nonumber \\
Q_{xy} & = & Q_{yx} = - {1 \over 2} \sin{ (2\eta)}. 
\label{Qparamet}
\end{eqnarray}

Let us parameterize $(\hat{q}_x, \hat{q}_y) = \left( \cos (\phi_{\bf{q}}),
 \cos (\phi_{\bf{q}}) \right)$, as in section IV. Eq. (\ref{anisomass}) 
becomes

\begin{equation}
q^2_0 (\vec{\hat{q}}) = \tilde{q}^2_0 \left( 1 - \cos \left[ 2 ( \phi - \eta ) 
\right] \right).
\label{fullpollmass}
\end{equation}

One observes that the Higgs mass acquires the correct $\pi$ periodicity
in momentum space, as required by the $\pi$ periodicity of the director.
Secondly, when the stress-photon has a momentum parallel to the director
such that $\phi - \eta = 0 \; mod (\pi)$ the {\em Higgs mass is vanishing}
while the Higgs mass is maximal when the stress-photon momentum is
at a right angle relative to the director,  $\phi - \eta = \pi/2 \; mod(\pi)$.
This makes sense. Because of the glide principle, the dislocation `screening'
currents are running parallel to the director. The shear `electrical-' and
`magnetic field strengths' associated with the stress photons are transversal
to their propagation direction. Therefore, they acquire a maximal Higgs mass
when their propagation direction is transversal to the direction of the
screening currents/director. This is a most interesting result. It shows
that the the spectrum of stress
photons is for every orientation in momentum of the form as explained
in the previous section, except that the Higgs mass has now become a
function of orientation. Stronger, the Higgs mass is vanishing
at the two  points `parallel' (with regard to the photon propagation direction)
to the director. At these points, the shear penetration depth is diverging 
and the phonons of the elastic medium re-emerge.

Let us re-emphasize that this is  just a consequence of the glide principle.
$\langle Q_{a b} \rangle \neq 0$ means that the Burgers vectors acquire
an orientational order.  As we discussed, to keep compression
rigidity decoupled from the dislocations one has to impose an absolute
glide condition on the dynamics of the dislocations. In other words,
dislocations can only delocalize in the direction 
of their oriented Burgers vectors. Their kinetic energy translates into
the shear Higgs mass and the result is that in the directions precisely
perpendicular to the director shear is decoupling from the dislocation
condensate, and the phonons re-emerge.

Let us re-emphasize that is  not correct to view this condensate
as a quasi one dimensional form of matter, despite the fact that
the dislocations currents are spontaneously oriented. The reasons
are given in section III and Appendix B. The dislocations can be
created and annihilated everywhere in the 2D space and this determines
the dimensionality of the effective field theory. Accordingly,
 the disorder field $\Psi$ parameterizing
the dislocation superfluid is just a conventional 2+1D GLW order parameter
field. The `dimensional'
reduction resides in the way its rigidity is transferred to the stress
photons. Their screening requires a real motion of the dislocations
(shear screening currents) and these are guided by the director as
implied by the glide constraint, causing an anisotropic shear Higgs mass.
 
More than anything else discussed in this paper, this `Higgs nematic'
should be considered as an entity which only carries a meaning in the
field theoretic description. As we repeatedly emphasized, in real 
condensed matter systems there will always be a finite density of
interstitial defects in the fluid states. These interstitials will
directly corrupt the glide constraint; although is not clear to us
how to incorporate such an interstitial gas in the present formalism
it is clear that the presence of interstitials will immediately liberate
climb. When climb is possible, dislocations will carry kinetic energy
perpendicular to their Burgers vectors and the result has to be
that the shear Higgs mass is finite in all directions. We can safely
conclude that the quantum nematic derived from normal matter will
be a superfluid (i.e. finite shear penetration depth in all directions)
although one which might be highly anisotropic. Upon closely approaching
the `order limit' it will be a much better superfluid in the direction
parallel to the director than in the perpendicular direction. We will
discuss this further in the context of the superconductor, section IX.

Another caveat is that a-priori the director is a fluctuating quantity,
as we discussed at length in section III. Considering Eq.'s (
\ref{anisomass} - \ref{fullpollmass}) in a naive fashion, fluctuations
of the director seem to  imply that a mass gap would open up everywhere
at the moment that these fluctuations would become noticeable. Local
fluctuations would render $\langle | Q_{a b} | \rangle < 1$ and this
implies $q^2_0 (\vec{\hat{q}}) > 0$ for all $\hat{q}$'s. This cannot be correct,
as can be inferred from simple physics considerations. As long as the director
is ordered a  correlation length is present. At distances shorter
than this length the director is fluctuating and thereby the Burgers
vector undergo directional fluctuations. Hence, at these small scales 
shear screening currents flow also in `wrong' directions. However, 
at length scales exceeding the correlation length director fluctuations
become invisible and at these large scales only dislocation currents 
can flow which are precisely conforming to the director. Clearly, 
the fate of the phonons at the massless points is about macroscopic
distances and here the director should rule in an absolute fashion.
This is of course no more than a qualitative argument and this interesting
problem deserves a more careful consideration, which we reserve for future 
study.

Despite these reservations, one can take the field theory at face value,
to wonder what our finding might mean in general. This state is quite
peculiar: a superfluid in all directions, except for precisely two
orientations where it behaves like a solid. One could argue that it is
semantically more correct to call it a `smectic'. It should be emphasized,
however, that this `glide smectic' has nothing to do with conventional
(classical) smectics and the quantum generalization thereof recently
introduced by Kivelson and Lubensky {\em et al.} (also called the `sliding
phases')\cite{kivemfrad,slidingph} Invariantly, one starts out with 
special microscopic conditions
such that the microscopic constituents want to form fluid- (smectic A)
or algebraically ordered (smectic B, Luttinger liquids) layers. These
layers then form a periodic stack, breaking the translational symmetry
in the perpendicular direction. The main challenge is then to show that
the layers stay fluid despite the symmetry breaking in the third direction.
The rigidity of these smectics is typically governed by elastic actions
of the form (3+1D, $z$ is the stacking direction),

\begin{equation}
S \sim \int d\Omega \left[  B  (\partial_z u )^2  + 
\rho (\partial_{\tau} u )^2 + K \left( \partial^2_x + \partial^2_y \right)^2
+ \cdots \right]
\label{smecticact}
\end{equation}

carrying only a propagating, longitudinal phonon mode in the stacking 
direction. One notices that this is very different from our `glide 
smectic' where the full phonon response (longitudinal- and transversal
modes) of the solid is recovered in the `ordered' direction.

In conclusion, it appears that we have identified a truly novel
state of quantum matter. It can only be literally realized in
an unphysical limit (absence of interstitials). However, this
limit can at least in principle be approached arbitrarily closely
and the `glide smectic' is potentially of relevance for the
physics of the cross-over regime. Starting from the ultraviolet,
one expects first to see the physics of the crystal (the phonons),
to subsequently find that the phonons turn into the stress
photons of the superfluid in one direction while they
survive in the perpendicular direction. Only at the largest scales
the latter also turn into the modes of the superfluid. From the
theoretical side, the construction gives away a curious motive which
makes one wonder if it might be of relevance elsewhere. In the
conspiracy of spontaneous rotational symmetry breaking and the
glide principle, both associated with the topological charge of
the dislocation, a {\em dynamical compactification mechanism}
has emerged. One can equally well view the above as an example
of curling up an extra dimension. 
An observer which can only exchange information with other observers
by employing shear stresses is locked up into the dimension perpendicular
to the director in the static limit, to only infer about the existence of
the extra dimension by investing a large energy.

\section{The superconductor as the dual of the
bosonic Wigner crystal.}

Up to this point we have focussed on elastic matter which is
decoupled from other gauge fields. The outcome is that this
neutral `crystal' dualizes into the neutral superfluid. A
next issue is what happens when the crystal is coupled to
other gauge fields, and 
the gauge fields of interest is of course electromagnetism.
The starting point is the electromagnetically charged crystal
which one could call a charge density wave or a Wigner crystal.
Traditionally, the name Wigner crystal seems reserved to the
crystalline state of electrons realized at low densities in
the continuum. Such a crystal carries a fermionic statistic and
is thereby beyond the scope of this paper. However, adding the
prefix `bosonic' one might want to interpret it as a crystal
composed of electromagnetically charged bosons, for instance 
the charged ordered state of Cooper pairs breaking spontaneously
translational symmetry.

We asserted in the previous chapter that the dual of the neutral
crystal is a superfluid. If correct, this should imply that {\em the
dual of the bosonic Wigner crystal has to be a superconductor}.
Symmetry does not allow for any other outcome. Upon coupling 
the superfluid matter fields to the electromagnetic gauge fields,
the only possible outcome is the Higgs-Meissner phase.

The case can be made even more rigorous. We showed in section
VII that the neutral case is characterized by an isolated
propagating compression mode. Wen and Zee\cite{Wen} presented
some time ago a theorem demonstrating that such a state has to
exhibit an electromagnetic Meissner effect when such a medium
acquires electrical charge. Their argument is quite
elementary. A compression mode can be dualized in a pair of
compression `photons' and these are coupled to the EM photons.
Integrating the former results in a mass gap for the latter.

A-priori, it is far from obvious that the Meissner phase will emerge
in the dislocation condensate. 
Consistency requires that the field theory should only know about 
the lowest order derivatives (i.e., linear elasticity) and in this
order the displacement fields only couple to electrical fields and
not to magnetic fields. How can this generate Meissner? The next
puzzle is that the dual of the crystal is a Bose condensate of
dislocations, but it is easily demonstrated that the dislocations
are not directly coupled to electromagnetism. Somehow, in order for
the dual to be a superconductor, the electromagnetic
screening currents have to be carried by the remnants of the
elastic medium, since the dislocations themselves are incapable of carrying
electromagnetic currents.

As we will show, this riddle has an amazing resolution. We will present
a Higgs mechanism which is completely different
from what is found in the textbooks. At the same time, the continuity
arguments we used in the previous section apply equally well in this case,
meaning that this interpretation is nor better and neither worse than the
textbook one: it just offers an equivalent viewpoint on the nature of the
superconducting state as far as the scaling limit is concerned,
acquiring only a different meaning when one goes away from the
scaling limit. This new mechanism can be summarized as follows:

{\em The Meissner effect lies in hide in the bosonic Wigner crystal,
to get liberated when shear rigidity becomes short ranged.}

The full theory of the electrodynamics of the Wigner crystal and its
dual `order superconductor' is rather tedious. However, using our helical
projections the problem factorizes in a magnetic sector where the
Meissner effect resides and an electrical sector responsible for the
physics of the plasmons. The magnetic sector is quite simple technically
and will be discussed in full here. The electrical sector involves some
tedious algebra and will be discussed in a future publication.     
 
\subsection{Basics: electrodynamics and linear elasticity.}

Let us first  review the standard derivation of the coupling between
the E.M. field and the charged crystal. We start with the atomistic view 
on the elastic medium as a composite of charged particles. Every
individual (non-relativistic) particle positioned at $\vec{R} = (R_x, R_y)$
carries a charge $e$ and it interacts with the field according to,

\begin{equation}
S_{EM} = e \int d\tau \left( A_{\tau} ( \vec{R} ) - { 1 \over c}  
A_a (\vec{R}) \partial_{\tau} R_a \right)
\label{startEM}
\end{equation}

For convenience, we use units of time measured in units of length with
a conversion factor given by the phonon velocity $c^2_{ph} = 2 \mu / \rho$.
The velocity of light $c$ in Eq. (\ref{startEM}) thereby corresponds with
the ratio of the light velocity to the phonon velocity, which will turn
into the true light velocity when the phonon velocity is reinserted in
the final outcomes.

Consistency with linearized quantum-elasticity requires that only
the leading order in the gradient expansion for the phonon-photon coupling   
should be kept. Writing $\vec{R} = \vec{R}_0 + \vec{u}$ and expanding
to lowest order,

\begin{equation}
S_{EM} =  \int d^2x d\tau \left[ (n_e e) u^a \partial_a A_{\tau} - ( { {n_e e} 
\over c} )
A_a \partial_{\tau} u^a + { 1 \over {16 \pi c} } F_{\mu \nu} F^{\mu \nu} 
\right]
\label{landaulif}
\end{equation}

where we explicitly included the Maxwell term of the electromagnetic 
fields. The above amounts to the simple statement that for infinitesimal
displacements the electrical fields exert a force on the displacements.
The coupling to magnetic fields emerge only in the next order of the
gradient expansion and these torque stresses belong to the realms of second 
gradient elasticity, to be omitted in the limit that the lattice constant
is sent to zero. 

Let us now reconsider the transformation to stress representation, keeping
track of the coupling to electromagnetic fields. Including Eq. (\ref{landaulif})
in the duality transformation presented in section IIIB, we find after the
H-S transformation, and reshuffling of derivatives,

\begin{eqnarray}
S  & = & \int d\Omega \left[ -\frac{1}{2} \sigma^{a}_{\mu}
 C_{\mu \nu ab}^{-1} \sigma^{b}_{\nu} + 
\sigma^{a}_{\mu} 
\partial_{\mu} u_{P}^{a} + [ u^{a}_{P} \partial_{a} A_{\tau} - A_{a} 
\partial_{\tau} u^{a}_{P} ] \right. \nonumber \\
& & \left. + u^a \partial_{\mu} ( - \sigma_{\mu}^a + A_{\tau} \delta_{\mu, a} +
A_a \delta_{\mu, \tau} )\; + \;  { 1 \over {16 \pi c} } F_{\mu \nu} F^{\mu \nu} 
\right]
\label{HSEM}
\end{eqnarray}

By integrating over the smooth configurations $u^a$ 
it follows that all components
of the quantity,

\begin{equation}
\tilde{\sigma}_{\mu}^{a} \equiv \sigma_{\mu}^{a} - A_{\tau} \delta_{\mu,a}
 - A_{a} \delta_{\mu,\tau}.
\end{equation}

are separately conserved,

\begin{equation}
\partial_{\mu}\tilde{\sigma}^{a}_{\mu}=0.
\label{emeqmot} 
\end{equation}

The fields $\sigma^{a}_{\mu}$ are again
the components of the stress tensor, now
satisfying the equation of motion
$-\partial_{\mu} \sigma^{a}_{\mu} = f_{a}$ 
where $f_{a}$ is the $a-th$ component
of the net applied external force density, here the
electrical field density.
In the present dynamical context this just corresponds
with the equations of motions in the presence of an electrical field.
The electrical fields $\vec{E}$ are

\begin{equation}
E_a = - { 1 \over c} \partial_{\tau} A_a - \partial_a A_{\tau}
\label{efields}
\end{equation}

and using the stress-strain relations Eq. (\ref{stressstrainiso})
for the isotropic medium one finds explicitly the familiar
real time equations of motions for the displacement fields,

\begin{eqnarray}
\rho \partial_t^2 u_x & = & \left( ( \kappa + \mu ) \partial^2_x +
\mu \partial^2_y \right) u_x + \kappa \partial_x \partial_y u_y
+ n_e e E_x \nonumber \\
\rho \partial_t^2 u_y & = & \kappa \partial_x \partial_y u_x +
\left( \mu \partial^2_x + ( \kappa + \mu ) \partial^2_y \right) u_y
+ n_e e E_y
\label{eqnmotexpl}
\end{eqnarray}

\subsection{Photons and stress photons.}

Up to this point it is of course standard. In the mean time, the
reader should have recognized that the stress gauge fields,
Kleinert's invention, have a remarkable capacity to generate
deep insights. As we perceive it, the remarkable powers of
the concept become fully visible when one in interested in
the faith of the electromagnetic photon in the dual condensate.

It starts out with the simple observation that since the full stress
tensor  $\tilde{\sigma}$  (including the E.M. forces) is divergenceless,

\begin{equation}
\tilde{\sigma}_{\mu}^{a} = \epsilon_{\mu \nu \lambda} 
\partial_{\nu} B^{a}_{\lambda}
\end{equation}

The stress gauge fields therefore `include' the electromagnetic forces.
The elastic components of the stress fields can therefore be written as

\begin{equation}
\sigma^a_{\mu} = \epsilon_{\mu \nu \lambda}  \partial_{\nu} B^{a}_{\lambda}
+ A_{\tau} \delta_{\mu, a} + A_a \delta_{\mu, \tau}.
\label{sigmabare}
\end{equation}

The local conservation law on the total stress can be imposed on the
action Eq. (\ref{HSEM}) by inserting this identity. After some algebra,

\begin{eqnarray}
S  & =  & \int d\Omega \left[ - { 1 \over 2} 
\left(   \epsilon_{\mu \nu \lambda}  \partial_{\nu} B^{a}_{\lambda}
+ A_{\tau} \delta_{\mu, a} + A_a \delta_{\mu, \tau} \right) \;
C^{-1}_{\mu \mu' a b} \;  \left(   \epsilon_{\mu' \nu' \lambda'}  
\partial^{\nu'} B^{b}_{\lambda'}
+ A_{\tau} \delta_{\mu', b} + A_b \delta_{\mu', \tau} \right) \right.
 \nonumber \\
& & \left. \; + \; i B^a_{\mu} J^a_{\mu} 
\; + \; { 1 \over {16 \pi c} } F_{\mu \nu} F^{\mu \nu} \right].
\label{dualactEM}
\end{eqnarray}

One finds that the dislocation currents $J^a_{\mu}$ are sources for
just the `total stress' gauge fields $B$. Introducing the quantity
(`elasticity dressed' electromagnetic fields),

\begin{equation}
\Phi^a_{\lambda} = (n_e e) \; C^{-1}_{ \mu \mu' a b} \epsilon_{\mu \nu \lambda}
\partial_{\nu} \left( A_{\tau} \delta_{\mu', b} + { 1 \over c} 
A_b \delta_{\mu', \tau} \right)
\label{dressedEMfield}
\end{equation}

we find that the action can be rewritten in the simple form,

\begin{eqnarray}
S & = & \int d^2x d\tau \left[ - {1 \over 2} \left( 
\epsilon_{\mu \nu \lambda}  \partial_{\nu} B^{a}_{\lambda} \right)
C^{-1}_{ \mu \mu' a b}  \left( 
\epsilon_{\mu' \nu' \lambda'}  \partial_{\nu'} B^{b}_{\lambda'}) \right)
\; - \; B^a_{\mu} \left( i J^a_{\mu} + \Phi^a_{\mu} \right) \right.
\nonumber \\
& &  \left. \; + \;
{ {n_e e^2} \over 2} \left( A_{\tau} \delta_{\mu,a} + { 1 \over c}  
A_a \delta_{\mu, \tau}) \right) 
C^{-1}_{ \mu \mu' a b} \left( A_{\tau} \delta_{\mu',b} + { 1 \over c}  
A_b \delta_{\mu', \tau} \right) \; + \;  
{ 1 \over {16 \pi c} } F_{\mu \nu} F^{\mu \nu} \right]
\label{fullEMdualaction}
\end{eqnarray}

The first- and last term are the now familiar stress- and electromagnetism
Maxwell terms. The second term shows that the dressed EM fields $\Phi$ 
couple minimally to the stress-gauge fields. It is also seen that the
EM fields do not couple {\em  directly} to the dislocations: terms
of the form $A J$ have cancelled out and in this sense are dislocations
electrically neutral. However, both the dislocations and the EM potentials
couple to the stress gauge fields. Although the direct coupling is absent,
dislocations do interact with EM fields, albeit in an indirect way by 
the influences they both exert on the intermediary elastic medium.

The real surprise is in the third term. By just rewriting the action in terms
of the stress gauge fields, a {\em Meissner term has emerged!} Consider
the space like E.M. potentials occurring in the third term. 
Because of the delta functions,
only the time like components of the inverse elastic tensor matter and
 $C^{-1}_{\tau \tau a b} = \delta_{a b} / \rho$. In other words, this action
contains the term ${ {n^2_e e^2} \over  {\rho c^2} } ( A^2_x + A^2_y )$.
The prefactor carries the dimension of inverse area,

\begin{equation}
 { 1 \over {\lambda_L^2} } = {4\pi {n^2_e e^2} \over  {\rho c^2} }
\label{londonlength}
\end{equation}

and $\lambda_L$ is recognized as the London penetration depth, 
expressed in the relevant dimensionful quantities characterizing this
problem. This is remarkable. Just by representing the problem of
the charged elastic medium in terms of stress gauge fields, 
the electromagnetic sector acquires automatically a Meissner term.
In the particle language one has to work hard to discover the Meissner
as a ramification of the off-diagonal long range order. In this
order language, when one knows to use the natural representation
(stress gauge fields) Meissner is around all along, and the problem 
is actually to get rid of it! In the absence of dislocations 
$J^a_{\mu} = 0$ it seems to be still present. However, this corresponds with
the elastic/crystalline state and the crystal is surely not a 
superconductor; the Meissner term has to vanish from the effective
action describing the electrodynamics of the crystal. As we will 
show, this is indeed the case. The E.M. photons couple linearly to the
stress photons via the term $B \Phi$. The stress photons have to be integrated
to arrive at the effective action for the electrodynamics. As
we will show, these integrations produces a counter term which is
precisely compensating the Meissner term in the bare action. 
However, when the shear photons acquire their Higgs mass due to the
presence of the dislocation condensate, this compensation is no longer
complete and a true Meissner term remains demonstrating that the
dual state is a superconductor. The reader can now appreciate the
meaning of the metaphor we presented earlier: in the crystal, the
Meissner effect is around (Eq. \ref{fullEMdualaction}), but it is lying
in hide because it is `eaten by the phonons', to get liberated
in the dual fluid, because of the mass gap in the stress-photon sector
associated with the finite range of shear.

\subsection{The effective electromagnetic action.}

The expressions Eq. (\ref{dressedEMfield}, \ref{fullEMdualaction})
appear at first sight as quite complicated, although they represent
no more than the coupling of phonons and photons in the Wigner
crystal (ignoring the dislocations). Matters simplify considerably
by using the helical projections introduced in section IV. To
establish the connections with the familiar context of electromagnetism,
it is easily shown that the magnetic- ($B^2$) and electrical ($E^2$)
energy densities in momentum space  are simply given by,

\begin{eqnarray}
B^2 (\vec{q}, \omega) & \sim & q^2 | A^{-1} |^2 \nonumber \\
E^2 (\vec{q}, \omega) & \sim & (\omega/ c)^2 | A^{-1}|^2 +
( (\omega/c)^2 + q^2 ) |A^{+1}|^2 
\label{eandbfields}
\end{eqnarray}

and the electromagnetic Maxwell action becomes

\begin{eqnarray} 
S_{Maxwell}  & = & { 1 \over {16 \pi c}} \int d\Omega F_{\mu \nu} F^{\mu \nu}
\nonumber \\
 & = & { 1 \over {16 \pi c}} \int { { d^2 q d\omega} \over { (2\pi)^3 } }
( \omega^2 + c^2 q^2 ) \; \left( |A^{+1}|^2 \; + \; |A^{-1}|^2 \right). 
\label{EMmaxwell}
\end{eqnarray}

The total action for the Wigner crystal becomes

\begin{equation}
S_{tot} = S_{Maxwell} + S_{stress} + S_{Meissner} + S_{int},
\label{totalactionsch}
\end{equation}

where $S_{stress}$ is the `Maxwell' action for the stress gauge fields,
Eq. (\ref{elasticaction}), to be augmented with the dislocation Higgs
mass, Eq. (\ref{dualeffaction}) when we want to address the superconductor.
We still have 
to derive the `pseudo-Meissner' action $S_{Meissner}$  and the coupling 
between the stress- and the EM gauge fields $S_{int}$ in terms of the
projected fields.

The pseudo-Meissner term reads

\begin{eqnarray}
{\cal L}_{Meissner} & = & n^2_e e^2 ( C^{-1} )_{\mu\nu ab}  ( A_{\tau} 
\delta_{a, \mu}
- ( 1/c) A_a  \delta_{\tau, \mu} )
( A_{\tau} \delta_{b, \nu}
- ( 1/c) A_b  \delta_{\tau, \nu} ) \nonumber \\
      & = &  n^2_e e^2 \left[ {1 \over { \rho c^2} } | A^{-1}|^2 \; +
\left( { {( 1 - \hat{\omega}^2 )} \over {\kappa} } + 
{ {\hat{\omega}^2}  \over { \rho c^2} } \right) | A^1|^2 \right].
\label{Spseudomeissner}
\end{eqnarray}

In the non-relativistic limit $q >> \omega / c$, we recognize
that the mass in the electrical sector $\sim |A^1|^2$ with the
electrical screening length $q^2_{p} = n^2_e e^2 / \kappa$, 
which is set by the compression modulus and will translate
in a plasmon-frequency via $\omega^2_p = n_e e^2 / m =
c^2_{\kappa} q^2_p$ with the compression velocity $c_{\kappa}
= \sqrt{ \kappa / \rho }$. As we already discussed, the 
surprising feature is the term $ |A^{-1}|^2 / \lambda^2_L$,
showing that the magnetic vector potential acquires
a genuine Higgs mass.

The stress- and electromagnetic gauge potentials are linearly
coupled and in order to arrive at meaningful statements regarding
the electrodynamics one has to explicitly integrate out the stress 
fields. These couplings read in terms of the projected fields,  

\begin{equation}
S_{int} = n_e e q \left( { {-1} \over { \rho c} } B^T_{-1}{ A_{-1}}^*
 - \hat{\omega} (\; { 1 \over { \rho c}} \; - \; { 1 \over {2 \kappa} }
\; )  B^L_{-1} {A_{1}}^* -  { i \over { 2\kappa} } B^T_{1} {A_{1}}^* \right)
\label{SBA}
\end{equation}

and together with $S_{stress}$ the $B$ fields can be integrated out.
Although these integrations are in principle straightforward, the
problem is in practice quite complicated, especially so in the presence
of the dislocation Higgs mass, and we leave the full analysis to a
future publication. However, these problems all reside in the
electrical sector (i.e. the plasmons), and as a pleasant 
circumstance the magnetic sector is simple. We observe 
that in all pieces of the total action Eq. (\ref{totalactionsch})
the problem factorizes in a `magnetic' part in terms of 
$B^{-1}_T$ and $A^{-1}$ and an electrical part in terms of the
other helical components. Collecting all the magnetic pieces,

\begin{eqnarray}
S_{tot} & = & S_{elec} + S_{mag} \nonumber \\
S_{magn} & = & \int { { d^2 q d\omega} \over { (2\pi)^3 } } \left[
{ 1 \over {16 \pi c^2}} ( \omega^2 + c^2 q^2 + c^2 / \lambda^2_L ) \;  
|A^{-1}|^2
\; - \; { {n_e e q}  \over { \rho c} } B^T_{-1} {A_{-1}}^* \right. \nonumber \\
&& \left.
+ \; { 1 \over { 4 \mu} } ( 2 \omega^2 + q^2 + 2q^2_0 ) |B^T_{-1}|^2
\right]
\label{magneticaction}
\end{eqnarray}

The $B$ fields have to be integrated to arrive at the
electrodynamics of the crystal or the dual fluid.
The integration rule is as usual for complex variables (Fourier components),

\begin{equation}
\int d Re\{ l\}dIm\{l\} 
e^{- \beta {\bf l}^2 + i ({\bf l} {\bf a^*}+c.c.)} = e^{ - { { {\bf| a|}^2} \over
{ \beta} } },
\label{gaussianint}
\end{equation}

where the standard unimportant 
multiplicative has been omitted.
Setting $\beta$ to be the propagator of the $B$ field, 
$a = i ( n_e e q ) A^{-1} / \rho c$, and recalling that we use
the velocity convention $c^2_{ph} = 2 \mu / \rho \equiv 1$,
we find for the magnetic piece of the effective electromagnetic action,

\begin{equation}
S_{eff, magn.} =  \left[ { {n^2_e e^2} \over {\rho c^2} } \;
\left( 1 - { { q^2} \over { 2 \omega^2 + q^2 + 2q_0^2 } } \right)    
 + { 1 \over  { 16 \pi } } \left( 
{ { \omega^2} \over {c^2}} + q^2 \right) \right] \;  |A^{-1}|^2.
\label{effmagnetism}
\end{equation}

\subsection{The Meissner effect and the screening current oscillations.}

Eq. (\ref{effmagnetism}) is our second central result because
it demonstrates that the dislocation Bose condensate is at the
same time a conventional electromagnetic Meissner state: it offers
a complementary way to understand the phenomenon of
superconductivity. 

Magnetic flux expulsion is associated with the vacuum and one should
therefore consider the meaning of the action Eq. (\ref{effmagnetism})
in the limit $\omega \rightarrow 0$. Define the shear penetration
depth $\lambda_S$ as, 

\begin{equation}
\lambda^2_S \equiv  { 1 \over { 2 q_0^2} }
\label{sheardepth}
\end{equation}

Recalling the definition Eq.(\ref{londonlength}) for the London 
penetration depth $\lambda_L$

\begin{equation}
S_{Magn, \omega=0}  = \left[ { 1 \over {\lambda^2_L}}
{  { 1 \over { ( q \lambda_s )^2} } \over 
{ 1 +  { 1 \over { ( q \lambda_s )^2} } }} \; + \;  q^2 \right]
\frac{ |A^{-1}|^2}{16\pi}
\label{statBaction}
\end{equation}

and the propagator for static  magnetic fields follows,

\begin{equation}
G_{magn} ( \omega =0 )  = { 1 \over { q^2 \; + \;  
{ 1 \over {\lambda^2_L}} {  { 1 \over { ( q \lambda_s )^2} } \over 
{ 1 +  { 1 \over { ( q \lambda_s )^2} } }} } }
\label{statBprop}
\end{equation}

First consider the Wigner crystal,
characterized by the shear penetration depth being infinite: $\lambda_S 
\rightarrow \infty$.
It follows directly from Eq. (\ref{statBprop}) that the Meissner 
term cancels. We recover the expected 
property of crystalline matter that it does not couple to static
magnetic fields. 

However, the above reveals a deep insight which we believe is far 
more general than the specific case we have analyzed. We perceive
it as a sharpening of Feynman's ideas we referred to earlier. Let
us formulate it as an explicit conjecture:

{\em It is sufficient condition for the Meissner phase to occur 
that a bosonic and electromagnetically charged
elastic medium acquires a finite penetration length for the
mediation of shear stresses.}

We cannot prove this conjecture for the general case. However,
it is obvious for the present case. At length scales large compared 
to the shear penetration depth, $q \lambda_S << 0$ the propagator
becomes $ 1 / ( q^2 + 1 / \lambda^2_L$), demonstrating that in real
space magnetic fields decay like $\exp{( - r / \lambda_L)}$, corresponding
with a conventional Meissner effect. However, at distances
small compared to the shear penetration depth,  $q \lambda_S >> 0$,
the medium rediscovers its heritage as a solid and the magnetic
field can freely propagate, $G = 1 / q^2$.

There is no doubt that the dislocation condensate is a conventional
superconductor and let us once again emphasize the great differences
between this Higgs mechanism and the conventional understanding based
on the gas limit. Starting out from the Bose gas, the central wheel
is the macroscopic quantum entanglement of the boson world-lines,
translating into off-diagonal order in terms of the boson field
operators, which in turn by minimal coupling causes the vector 
potentials to become pure gauge. Starting out from the crystal,
just by re-parameterizing the dynamics in terms of stress photons
a Meissner term is generated. When shear is massless as in the
crystal, the shear photons eat the Meissner. However, when 
the shear photons acquire a mass, necessarily involving a
condensate of dislocations being the unique agents destroying 
translational invariance and thereby shear rigidity, shear
photons are eaten themselves at long distances with the effect
that the hidden Meissner turns into a physical Meissner, giving
the electromagnetic its mass. It seems, all what is required
for this mechanism is the existence of a stress photon representation,
which is  general, and the fact that their shear content acquires
a mass in the fluid, which is also general. This gives reason
to believe that the above conjecture might well be a theorem.

The differences between the conventional and our `order' superconductor
should become more manifest when one measures shorter
distances and -times. The key difference is in the shear-length,
an additional scale which is central  in the `order' 
description. The gas limit can be viewed as the special case 
of the order superconductor where the shear length has shrunk
to the lattice constant. One notices that shear is never
mentioned in the textbook treatments of the quantum fluids. 
The present theory can be viewed in this regard as an extension
because we can address the question if the presence
of this additional length scale causes new phenomena.

One expects something to happen when the shear length becomes
 larger than the London length. In
practice, the London length is quite large (of order of microns)
while it is natural to expect that deep inside the superfluid/conductor
the shear length will be of order of the lattice constant. In
order to enter the regime where the shear length becomes the
largest length scale it appears as a requirement that the
superfluid/conductor undergoes a second order transition into
the crystalline state. This condition can in principle be
satisfied when the superfluid/conductor carries a nematic order,
as we discussed in previous sections. Since the shear length is diverging
at this transition, by approaching it sufficiently closely one
will always enter a regime where the shear length is the largest
length scale.

In this regime, a new effect is found which does not seem have
an analogy elsewhere in physics. Let
us consider the propagator Eq. (\ref{statBprop}) 
in real space. The inverse of the propagator
Eq.(\ref{statBprop}) is,

\begin{equation}
G^{-1} = q^{2}+ \frac{1}{\lambda_{L}^{2}(1+ \lambda_{S}^{2}q^{2})}
\label{invstatprop}
\end{equation}

in case of the `normal' superconductor, characterized by
$\lambda_{L} > 2 \lambda_{S}$, the poles lie on the imaginary axis at,

\begin{eqnarray}
q_{\pm} &  = & \pm i m_{\pm}  \nonumber \\ 
m_{\pm} & =  & \frac{\lambda_{L}^{2} \pm \sqrt{\lambda_L^{4}-4 
\lambda_{L}^{2} \lambda_{s}^{2}}}{2(\lambda_{L} \lambda_{s})^{2}}.
\label{magnpoles}
\end{eqnarray}

Closing the contour in the upper half plane in the Fourier transform
to real space, these lead to  
the standard exponential decay of the real space correlator 
$G (r) \sim e^{- r/\lambda_L}$.

For the interesting regime where the shear length exceeds 
the London penetration depth ($\lambda_{L} < 2 \lambda_{S}$),
the situation changes drastically. 
At $\lambda_{L}= 2 \lambda_{S}$ the four poles merge
in pairs and then bifurcate once
again- this time along a circle.
The point $\lambda_{L}= 2 \lambda_{S}$ 
signifies the disorder line. For $\lambda_{L} < 2 \lambda_{S}$,
the four poles lie on a circle $q$ in the complex
plane with radius $R$ and phase $\theta$,

\begin{eqnarray}
q & = & \pm R e^{\pm i \theta} \nonumber \\
R & = & \frac{1}{\sqrt{\lambda_{L} \lambda_{S}}} \nonumber \\
\theta & = & \frac{1}{2} \cos^{-1}(-\frac{\lambda_{L}}{2 \lambda_{S}}) 
\label{poles1} 
\end{eqnarray}

Hence, the poles are no longer on the imaginary axis, and one
anticipates an oscillatory behavior of the real space propagators.  
To gain insight, let us first perform the Fourier transformation in 
one space dimension. The real space propagator is given by the
exact expression, 

\begin{eqnarray}
G(r) & = & - \pi \sqrt{\frac{\lambda_{L}}{\lambda_{S}}} 
\exp[- r \sqrt{ {  { 1+ \frac{\lambda_{L}}{2 \lambda_{S}}
} \over { 2 \lambda_S \lambda_L} } }] \nonumber
\\
& & \times \frac{1}{\sqrt{1- \frac{\lambda_{L}^{2}}
{4 \lambda_{S}^{2}}}}  \Big[(\lambda_{S}+\lambda_{L}) 
\sqrt{\frac{1-\frac{\lambda_{L}}{{2 \lambda_{S}}}} {2 \lambda_S \lambda_L}} 
\sin(r \sqrt{\frac{1- (\lambda_{L}/(2 \lambda_{S}))}{2 \lambda_{L} 
\lambda_{S}}}) \nonumber
\\ 
& & + (\lambda_{S}- \lambda_{L})  
\sqrt{\frac{1+ \frac{\lambda_{L}}
{{2 \lambda_{S}}}} {2 \lambda_S \lambda_L} } 
\cos(r \sqrt{\frac{1- (\lambda_{L}/(2 \lambda_{S}))}
{2 \lambda_{L} \lambda_{S}}}) \Big].
\label{oneDoscilprop}
\end{eqnarray}

This simplifies in the case that the shear length is much larger than
the penetration depth ($\lambda_S >> \lambda_L$) to,

\begin{equation}
G ( r ) = - \pi e^{ - \frac{ r } {\sqrt{2\lambda_S \lambda_L} } }
\sin( \frac{ r } {\sqrt{2\lambda_S \lambda_L}} + \frac{\pi} {4} )
\label{oneDoscilpropas}
\end{equation}

This is a remarkable result. It shows that that the true
magnetic penetration depth corresponds with the geometrical mean
of the shear- and the London length, $\lambda_M = \sqrt{ 2 \lambda_S
\lambda_L}$.  Even more surprising, (on the scale of this
penetration depth, the electromagnetic screening currents have an
{\em oscillatory} nature. Upon entering the superconductors,
one encounters first a layer of screening currents, than weaker
anti-screening currents, etcetera.  

This is of course not a pathology of the one dimensional case --
it is just a generic consequence of the location of the poles
of the propagator in the complex plane. In two space dimensions
the real space propagator reads,

\begin{equation}
G(r) = \frac{1}{2 \pi (\lambda_{L} \lambda_{S})^{2}} \Big[ a_{+} K_{0}(m_{+}r) + a_{-} K_{0}(m_{-}r) \Big]
\label{2Dmagnprop}
\end{equation}

with
 
\begin{eqnarray}
m_{\pm}^{2} & \equiv &
\frac{\lambda_{L}^{2} \pm i \sqrt{4 \lambda_{L}^{2} \lambda_{S}^{2
} - \lambda_{L}^{4}}}{2(\lambda_{L} \lambda_{s})^{2}} \nonumber
\\
a_{-} & \equiv & \frac{\lambda_{L}^{2} - m_{-}^{2} (\lambda_{L} 
\lambda_{S})^{2}}{m_{+}^{2}-m_{-}^{2}} 
\nonumber \\ 
a_{+} & \equiv & \frac{\lambda_{L}^{2}-m_{+}^{2}(\lambda_{L} 
\lambda_{S})^{2}}{m_{-}^{2}-m_{+}^{2}}
\end{eqnarray}

with the Bessel function
 
\begin{eqnarray}
K_{0}(x) = \int_{0}^{\infty} dt \frac{\cos xt}{\sqrt{1+t^{2}}}
\end{eqnarray}

arising from the integral

\begin{eqnarray}
\int\frac{d^{2}k}{(2 \pi)^{2}} \frac{\exp[i \vec{k} \cdot \vec{r}]}{k^{2}+m^{2}} = \frac{1}{2 \pi} K_{0}(mr).
\end{eqnarray} 

\begin{figure}[tb]
\begin{center}
\vspace{10mm}
\leavevmode
\epsfxsize=0.9\hsize
\epsfysize=10cm
\epsfbox{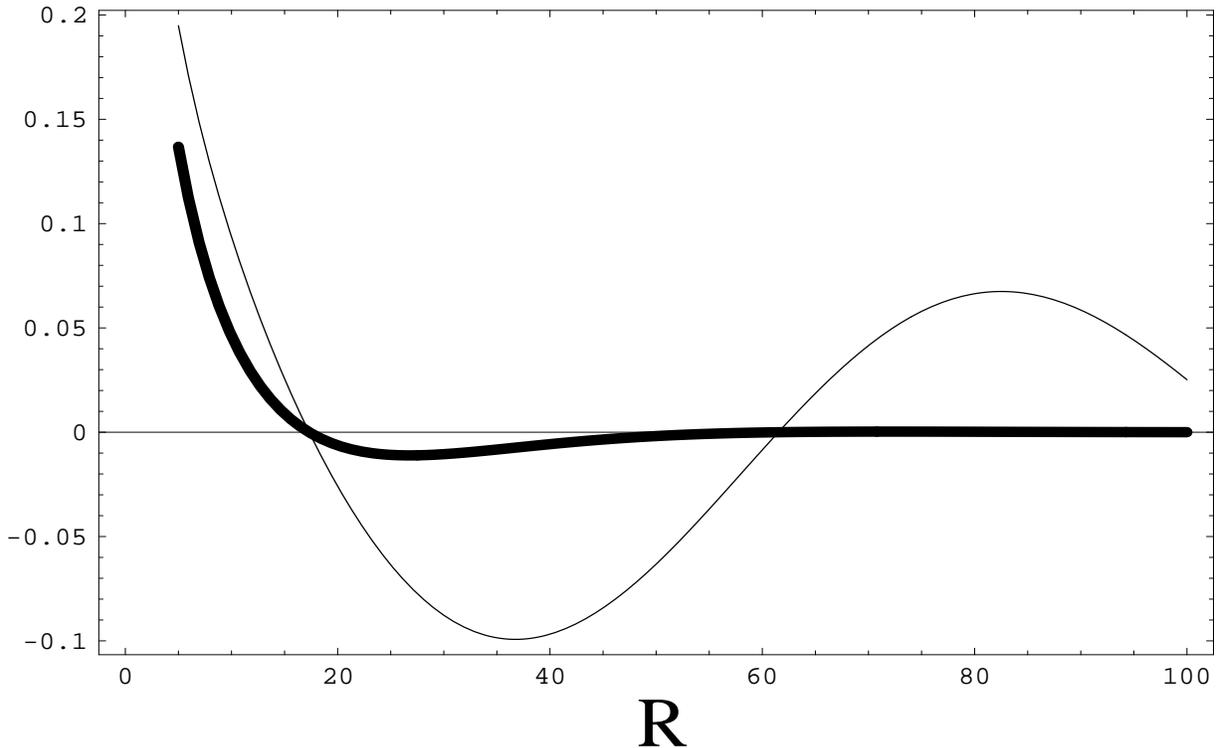}
\end{center}
\caption{The behavior of the static magnetic propagator in 2 space
dimensions, indicating how magnetic fields vanish upon entering
the `order superconductor'. The unit of length is set by 
the London penetration depth $\lambda_L = 1$ and we consider
here the case than the shear penetration depth is large, $\lambda_S = 100$,
corresponding with the situation  that the magnetic fields are screened
by matter which is rediscovering its solid nature. The thick line corresponds
with the propagator and to amplify the screening oscillations we have divided
out the exponential factor $\exp{(\sqrt{2\lambda_L \lambda_S})}$ (thin line).
The real magnetic penetration depth is $\sqrt{2\lambda_L \lambda_S}$ and one
notices that on the same scale the response oscillates between diamagnetic-
and paramagnetic behaviors.} 
\label{lineoscillations}
\end{figure}

\begin{figure}[tb]
\begin{center}
\vspace{10mm}
\leavevmode
\epsfxsize=0.9\hsize
\epsfysize=10cm
\epsfbox{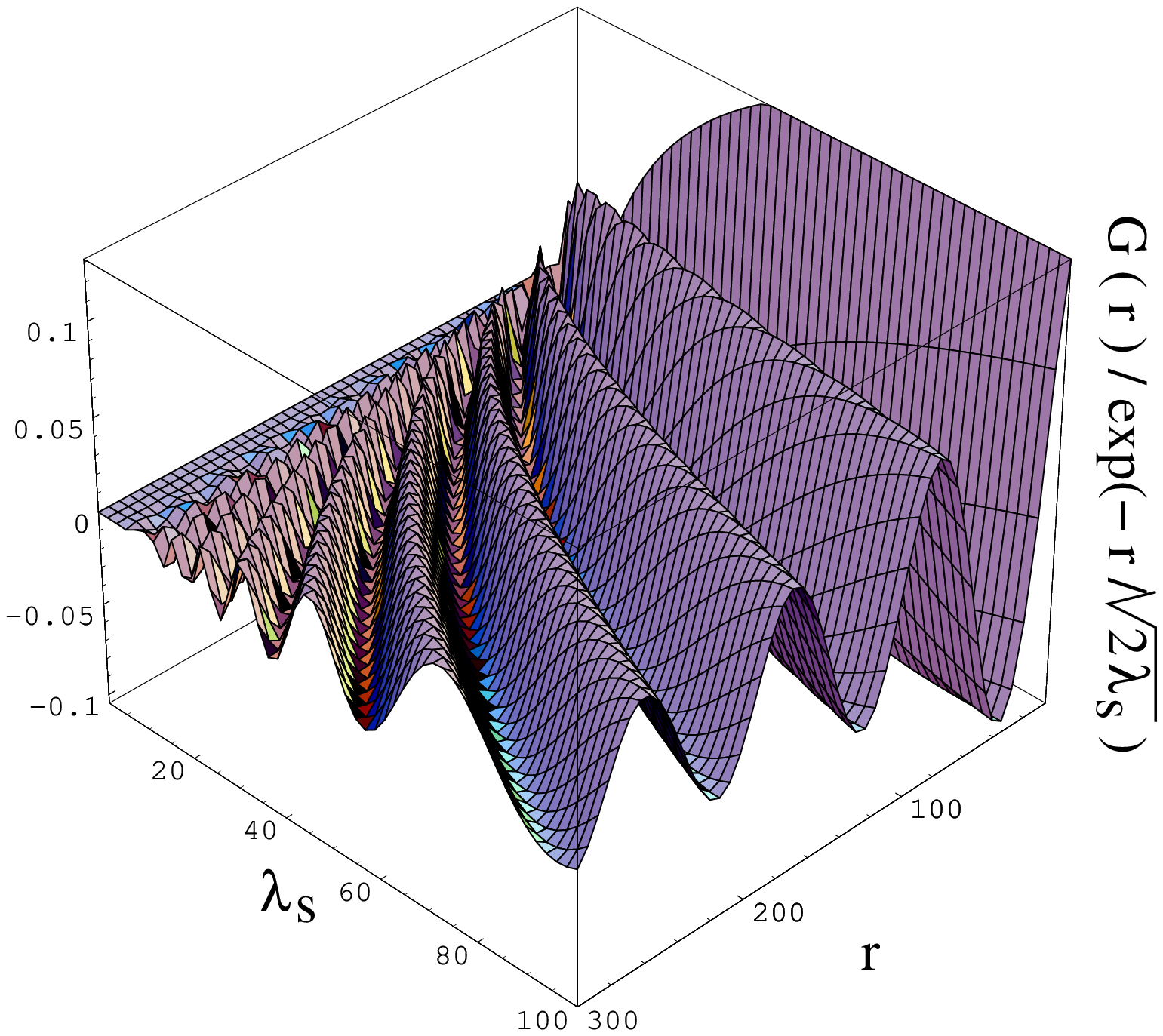}
\end{center}
\caption{Same as the thin line in Fig. (\ref{lineoscillations}) except
that we now show how the oscillations develop as function of the
shear length $\lambda_S$, again using $\lambda_L = 1$ as unit of length.}
\label{oscillations3D}
\end{figure}

In Fig.'s  (\ref{lineoscillations},\ref{oscillations3D})  
we have plotted some representative cases and it is seen
that the 2D propagator behaves in the same way as in one dimensions.
In Fig. (\ref{lineoscillations}) we consider the case
that $\lambda_S = 100 \lambda_L$, deep in the interesting regime
where the shear penetration depth is much larger than the
London length. The thick line shows the behavior of the
magnetic propagator Eq. (\ref{2Dmagnprop}) directly. Since
the scale setting the exponential envelope is the same 
as the one setting the oscillation period $\sim \sqrt{2\lambda_L 
\lambda_S}$ the actual magnitude of the anti-screening currents
is quite small. With some effort one can just recognize 1.5 full
oscillations. We therefore also show the outcome after dividing
out the exponential envelope $\sim \exp{( - r / \sqrt{ 2 \lambda_L
\lambda_S})}$ (thin line) and this shows very clearly the oscillations.
In figure (\ref{oscillations3D}), we follow the same strategy,
now showing how the oscillations develop as function of increasing
$\lambda_S / \lambda_L$. In accordance with the 1+1D result, we find
that the oscillations show up only when $\lambda_S > \lambda_L$,
while their magnitude is increasing when $\lambda_S$ grows. In addition,
one can clearly see that the period of the oscillations $\sim 
\sqrt{2\lambda_L \lambda_S}$.

What is the physical implication of these findings? In the regime
where the oscillations occur, the magnetic screening 
has to take place at lengths smaller than the shear penetration depth.
In the beginning of this section we stated that at these scales 
the medium is rediscovering that it is actually more like a solid and
that magnetic fields should freely penetrate. On closer inspection, 
however, the answer turns out to be more subtle. In the presence
of the dislocation condensate, the solid is actually not quite 
recovered. At distances where the medium can carry shear
stresses, it has still sufficient fluid character in order to
build up the currents needed to screen the magnetic field. It
is doing so less efficiently than in the true fluid regime
characterized by the London penetration depth. The magnetic
penetration length is now given by the geometrical mean 
$\sqrt{ \lambda_S \lambda_L}$ which is large compared to
the London length, although still small compared to $\lambda_S$
itself. The magnetic fields can be completely screened by the
`shearing' fluid.    

The most surprising feature are the current oscillations. Is 
there any further physical interpretation available? 
At first sight one might be tempted to draw 
analogies with Friedel oscillations, to find out on closer
inspection that there is little relationship. We have failed
to find any other analogy for this phenomenon in terms
of the physics which is known to us. We therefore consider it
as an irreducible mathematical fact and thereby as
the central prediction of this paper, given its counterintuitive 
nature.

\subsection{Magnetic screening in the ordered nematic.}

Up to this point we have focussed on the simpler  case of
the Coulomb nematic. However, it is easily deduced what
happens with the magnetic screening in the ordered nematic
state. Although unphysical, it is entertaining to consider
what happens in the `glide smectic' of section XIII. We 
learned that the shear penetration depth is actually diverging
in the direction perpendicular to the director. This in turn
implies  that in the direction parallel to the director the
magnetic penetration depth is just the London length while
the magnetic penetration depth becomes infinite in the perpendicular
direction.

Considering the type II state, this has the amusing consequence
that vortices actually turn into lines or `stripes'. This sounds 
far-fetched but a very similar phenomenology is expected in the
Luttinger-liquid smectics discovered by Kivelson et al.\cite{slidingph}.
It is just expressing the fact that the effectively one dimensional screening
currents are not able to screen the magnetic fields. 

More realistically, one would expect that because of the liberation
of climb the shear penetration depth might be very anisotropic
but still finite in the climb direction. This in turn implies that
the magnetic penetration depth is strongly anisotropic, and the 
vortices in the type II state should have a strongly elongated form.
At least in the 123 high Tc superconductors a sizable anisotropy
of this kind has been observed. Still, these anisotropies are
quite moderate implying that if nematic order of this kind is 
present the system has to be relatively far removed from the order
limit.

\section{Conclusions}

Nearing the end of this exposition, we are left with the
feeling that this paper should have been written some
fifty years ago in the era that the fundamental theory
of superconductivity was developed. Although our initial aim was to
shed light on the nature of nematic quantum fluids, we
were forced in the process to reconsider the fundamentals of the
theory of bosonic quantum fluids. We discovered a way to
understand  these fluids which is 
strikingly different, but complementary to
the theory found in the textbooks. The superconductor
and the superfluid can be understood as duals of
the crystalline state, the state realized when a
dislocation condensate has destroyed the translational
symmetry breaking governing the elastic state.

In order to keep the mathematics under control,
we could only develop the theory in full in a 
limiting case. This limit can be viewed as the
limiting case of `fluctuating order' (or strong
correlations) in a bosonic fluid, and is defined 
through the requirement that  interstitial degrees
of freedom do not exist. This condition can never
be satisfied in fluids composed of atomistic matter,
and the theory is thereby not literally applicable
to condensed matter systems. However, knowledge
of limits is always useful. Stronger, as we argued
repeatedly, since our `order' superconductor can
be adiabatically continued to the Bose-gas limit
(of course, modulo nematic `obstructions') the
presence of interstitial (Bose-gas like) excitations
can only change matters quantitatively and not qualitatively.
In sufficiently strongly correlated Bosonic fluids, 
the physics discussed in this paper can be closely
approached in a finite energy crossover regime, to
turn into a truly conventional superconductor only at
the lowest energies.

A second main theme has been the development of
a topological quantum theory addressing zero
temperature nematic order. With regard to the
`order superconductivity' theme nematicity has
taken the role of auxiliary device, simplifying
the theory. We believe it is in this regard not
essential. We did not address explicitly  the 
transition from the nematic- into the isotropic 
quantum fluid driven by the condensation of
disclinations. It remains to be proven that this
isotropic state is in the scaling limit indistinguishable
form a conventional superconductor/superconductor. 
However, there is every reason to believe that this
is the case. Relative to the stress photons, the
disclinations live `two gradients higher' (Kleinert's
double curl gauge field construction). These cannot
couple to the massless compression mode of the superfluid,
and they will not alter the electromagnetic Higgs mechanism
presented in section IX. To the best of our present 
understanding, the disclination condensate will just
destroy nematicity, be it of a topological- or fully
ordered kind, in the way it is envisaged in the 
Lammert-Toner-Rokshar gauge theory\cite{Lammert}.

Nevertheless, we perceive our findings regarding the nature of
nematic orders  as quite interesting. A main result is the
de-mystification of the topologically ordered nematic of
Lammert et al.\cite{Lammert}. Although quite mysterious
starting from the conventional `lengthy molecule gas' perspective,
this Coulomb nematic appears as a natural possibility in
the field-theoretic framework. We emphasize that our arguments
are in this regard of equal relevance in the 3D classical melting
context. Topological ordered states of this
kind are quite elusive. Experimentalists can only infer 
their existence through the presence of second order phase
transitions taking place in the absence of manifest order
parameters. In addition, one could hope to observe 
highly energetic quantized disclinations with strange
multiplicities -- in the absence of a manifest nematic
order it is very hard to force them in by external means.
It makes us wonder, could it be that states of this kind
are around both in the classical- and the quantum context,
being overlooked by experimentalists because of the lack
of awareness?     

The other novelty is the ordered nematic state which we found
to behave more like a smectic (the `glide smectic' of section
VIII). This is a state having no analogy elsewhere. It rests
on the unique dynamical principle of glide which on the quantum
level drives a literal dynamical  compactification, a reduction
of dimensionality driven by spontaneous symmetry breaking. At
the same time, since this state will be destroyed immediately
by the presence of interstitials, it is not of relevance to
condensed matter physics. As with more issues raised by this
work, we wonder if this mechanism  might be of use in the
context of fundamental physics (cosmology and high energy 
physics). We will return to this issue later.

Let us now turn to the hard question is. 
Is our `order superconductivity' of
relevance to nature? Fact is that no experimental evidence is
available supporting any of the hard predictions presented in this
paper (the quantum nematic states, the spectrum of the superfluid,
the electromagnet screening current oscillations). One statement
can be made easily: if it exists, it is not particularly abundant
and not easily accessible by condensed matter experimentation. As
we emphasized repeatedly, all what is needed is a Bosonic quantum
fluid characterized by a shear penetration depth which is large
compared to the lattice constant. For detailed microscopic
reasons such a condition is not easily satisfied starting out with
electrons and/or atoms.  There is no doubt
that conventional (BCS) superconductors, as well as the Bose-Einstein
atomic gases approach the ideal quantum gas limit very closely.
$^4$Helium is in principle a more fruitful territory. As
we already discussed, the roton might be viewed as a first signal
of the approach in the direction of the order superfluid, although
it is still far removed from the massive shear modes we discussed
in section VI. An open issue is how matters look like in Helium
layers. Reduction of dimensionality surely helps `our' physics
and after all there is an indication for a second phase 
transition suggesting the presence of hexatic order\cite{Stoof,superhexatic}.
One would like to measure the excitation spectrum using inelastic neutron
scattering to look for  the massive shear modes, but this is obviously
a very hard experiment.

Natural candidates for the order superconductivity are the under-doped
high Tc superconductors. Evidences have been accumulating that the
superconducting state is in a tight competition with a (anisotropic)
`Wigner crystal' (stripe phase) (ref.\cite{flucorder} and references
therein). A serious concern is that this stripe phase is itself
formed because of strong lattice commensuration effects. It is a 
state which is uniquely associated with doping the Mott-insulator.
The Mott-insulator itself is best understood as an electronic 
density wave which is commensurately pinned by the lattice. The
stripe phase can be seen as a higher order commensurate state with
the charges stripes to be interpreted as the
discommensurations forming a lattice\cite{ZaanenGunnarsson}. A case
can be made that due to fluctuations the effects of the lattice potential
can be much weakened so that at long wavelength the stripe phase might
closely approach an effective Gallilean
invariance\cite{philmag,Zaanenstripefluids}.
If this is the case, dislocations might become cheap as compared to
interstitials, and the conditions might be right for the emergence
of quantum nematics. Although controversial, recent STM work by
Kapitulnik and coworkers is suggestive of a strongly dislocated 
character of a partially pinned stripe phase\cite{Kapitulnik}. 
There is also counter-evidence. The recent observation of striped `halos'
surrounding the vortex cores in the type II state of the cuprate
superconductors has added credibility to the notion that the superconductor
is in a tight competition with the stripe phase\cite{SeamusDavis,LakeAeppli}.
However, these data are quite successfully explained in terms of
theories based on supersolids\cite{Sachdevsuso,Zhangsuso}. In fact,
from the present work another counter argument follows. As we discussed
in section IV, in the quantum nematic state one would invariably expect the
vortices to have strongly elongated shapes, and this is not 
observed.

Although the focus of this paper has been on condensed matter physics,
a next open question is to what extent our findings can be of use in
the context of fundamental theories, either in a metaphorical- or even
in a literal sense. We already referred to the intriguing connections
between generalized elasticity and gravity, especially promoted 
by Kleinert\cite{KleinertII,kleinertgrav2,kleinertgrav3}. The theory
can be reformulated in the language of differential geometry, starting
with the identification of the metric $g_{ij} = \delta_{ij} + w_{ij}$.
It can then be demonstrated that dislocations and disclinations take
the roles of torsion- and curvature sources, respectively. A first
requirement for the identification of space-time with an elasticity-type
field theory is that the latter should be manifestly Lorentz-invariant,
at least at long distances. This implies that one has to consider 
isotropic elasticity in D+1 dimensions, such that the physics along
the time axis is the same as that on the time slice. Amusingly,
the relativistic generalization of the Coulomb nematic state as
discussed in this  paper turns out to be indistinguishable at large
wavelength from the theory of general relativity\cite{KleinertZaanen}!
A crucial ingredient is that in the relativistic medium the glide
principle cannot exist. The reason is that, say, in three classical
dimensions dislocation lines can have arbitrary orientations relative
to their Burgers vectors (screw- versus edge dislocations). It is
a fundamental requirement for glide that a time axis can be singled out.
As we explained in section III, glide is fundamentally related to the
neutrality of dislocations relative to compression. When glide is absent,
dislocations will carry compression charge, and the result is that in
the relativistic Coulomb nematic both shear and compression acquire
a Higgs mass. The resulting medium is one which is only carrying 
curvature rigidity and such a `liquid crystal' has the same long-wavelength
properties as the space-time according to Einstein.

Our present work suggest some other alleys which might be worth 
investigating further with an eye on problems in fundamental physics.
We already mentioned the intriguing feature that in the `glide smectic'
a dimension is compactified in the stress-photon universe due to
the conspiracy of spontaneous breaking of the rotations in the embedding
space, and the glide principle. One of the great unsolved problems
in string theory is dynamical compactification. Our mechanism is 
stand alone in the sense that it is field-theoretical. It makes us
wonder if some appropriate generalization of the basic mechanism in
the much richer context of string theory can be formulated.

A final issue is the Higgs mechanism of high energy physics, believed
to be responsible for the origin of non-radiative rest mass. It is
incorporated in the Standard Model in the form of an effective field
of the Ginzburg-Landau-Wilson type. A wide open question is, is this
field fundamental or does it parameterize a collective property of,
say, Planck scale matter? The only analogy available used to be the
Bose gas, and the notion that the Higgs field is coming from a Bose
gas of particles living at the Planck scale is for good reasons not
very popular. Our `order superconductor' offers an alternative 
perspective. It shows that the Higgs mass of the gauge fields
is a secondary effect,
driven by a primary condensation associated with symmetry restoration.
Given that the latter has intriguing connections with the structure
of space-time\cite{KleinertZaanen}, it suggest an unexpected potential
connection between the Higgs phenomenon and gravity.

{\em Acknowledgments.} 
 This work profited directly from insights by F.A. Bais,
L. Balents, M.P.A. Fisher, E. Fradkin, S.A. Kivelson,
D.H. Lee, L. Pryadko, T. Senthil, Z. Tesanovic, X.-G. Wen,
A. Zee, S.C. Zhang, and especially H. Kleinert. The work by ZN and 
SIM at Leiden University 
was supported by the Netherlands Foundation for Fundamental Research 
on Matter (FOM) and the work of SIM in Moscow by the 
RFBR grant 02-02-165354.

\appendix 
\section{Abelian Higgs duality in 2+1D.}
\label{App1}

The duality discussed in this paper is closely related to
the Abelian Higgs duality. This is an important duality
which has been studied in great detail and has many applications
(see, e.g., 
ref.'s \cite{FisherLee,Fazio,banks,KleinertI,Sudbo,Senthil,Tesanovic}).
Here we will just review the essence of this duality. For an even
more minimal exposition, we refer to a paper by Zee\cite{Zee}.

Consider the Hamiltonian describing e.g. a 2D array of
Josephson junctions in the absence of dissipation and
neglecting the coupling to electromagnetic fields (in
fact, this describes a neutral superfluid),

\begin{equation}
H = Q \sum_i (n_i)^2 + J \sum_{<ij>} \cos (\phi_i - \phi_j)
\label{phasedyn}
\end{equation}

where the second term describes the Josephson coupling ($\phi_i$)
is the superfluid phase on island $i$) and the first term the
`charging' energy ($n_i$ is number density). This is quantized
because of number-phase conjugation $\left[ n_i, \phi_j \right] =
i \delta_{ij}$. The Lagrangian is, in relativistic short hand,
setting the phase velocity 1,

\begin{equation}
{\cal L} = { 1 \over g} \left( \partial_{\mu} (\phi + \phi_V) \right)^2
\label{phasedynL}
\end{equation}

where the coupling constant $g = Q/J$ while $\phi$ is a (compact)
$O(2)$ field: $\phi \rightarrow \phi + 2\pi n, n \epsilon Z$. This
implies that in principle the phase field can have multivalued 
configurations, and these are lumped in $\phi_V$. One can make this 
further explicit by the Villain construction $\phi_V = 2\pi n$,
treating $\phi$ as non compact and including the $n$'s in the path
integral measure. With some effort it can be shown that this 
regularization of the XY problem is qualitatively correct
and quantitatively quite accurate\cite{KleinertI}.

\begin{figure}[tb]
\begin{center}
\vspace{10mm}
\leavevmode
\epsfxsize=0.6\hsize
\epsfysize=10cm
\epsfbox{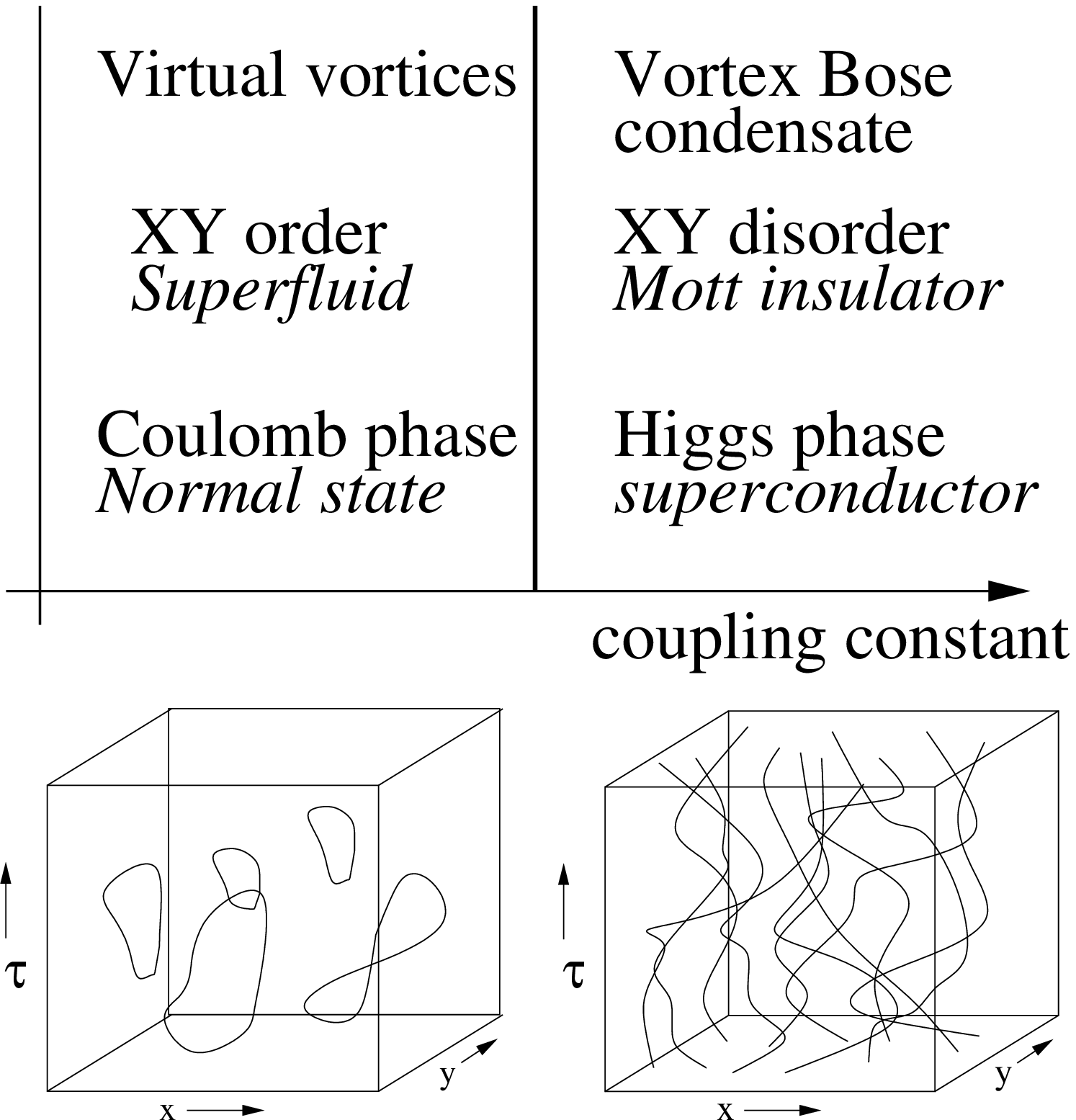}
\end{center}
\caption{Illustration of the physical meaning of the Abelian Higgs
duality and the various interpretations one can ascribe to the states
on both sides of the duality. On the `ordered' side vortices form
closed loops in space time, and these have blown out on the disordered
side forming a tangle corresponding with a Bose condensate. One can
either interpret the ordered side as a neutral superfluid (disordered
side a Mott-insulator) or the disordered side as a superconductor (ordered
side normal state with massless photons).}
\label{higgspicture}
\end{figure}

It is obvious
that a quantum-phase transition will occur.  For 
$g << 1$ $\phi$ will order, describing the superfluid while for
large $g$ number fluctuations are suppressed and a  state with
a fixed particle number per site is a Mott-insulator, see Fig.
(\ref{higgspicture}).

The topological excitation of superfluid order in 2+1 dimensions
is a particle,  the well known `vortex pancake'. It carries a
topological invariant on the time-slice defined by the circuit integral,

\begin{equation}
\oint d \phi = 2\pi N
\label{winding}
\end{equation}

 where $N$ is the winding 
number, or vorticity. Since the winding number is conserved,
the vortex particle corresponds with a world-line in space-time.
Using Stokes theorem, a conserved vortex current is derived
which expresses the non-integrability of the field $\phi$,
collected in the piece $\phi_V$,

\begin{eqnarray}
J^V_{\mu} & = & \epsilon_{\mu \nu \lambda} \partial_{\nu}
\partial_{\lambda} \phi_V \label{Vcurrent} \\
\partial_{\mu} J^V_{\mu}&  = &  0
\label{Vconser}
\end{eqnarray}

where the time component ($J_0$) corresponds with the vortex
density. Introducing an auxiliary field $\xi_{\mu}$, 
Eq. (\ref{phasedynL}) becomes after a Hubbard-Stratanowich
transformation,

\begin{equation}
{\cal L} = g \xi_{\mu} \xi_{\mu} + i \xi_{\mu} \partial_{\mu}\left( \phi 
+ \phi_V \right)
\label{hubstr}
\end{equation}

and it is observed that in this stage the coupling constant $g$ has
already been inverted, meaning that weak coupling in $\phi$ is strong
coupling in $\xi_{\mu}$ and vice versa.
Since $\phi$ is by definition integrable the derivative can be shifted,

\begin{eqnarray}
L & = & g \xi_{\mu} \xi_{\mu} + i \xi_{\mu} \partial_{\mu} \phi_V
+ i \xi_{\mu} \partial_{\mu} \phi \label{split1} \\
 & = & g \xi_{\mu} \xi_{\mu} + i \xi_{\mu} \partial_{\mu} \phi_V
- i \phi \partial_{\mu} \xi_{mu}
\label{split2}
\end{eqnarray}

$\phi$ appears as a Lagrange multiplier and upon integration
it follows that $\xi_{\mu}$ is conserved and it can therefore
be written as the curl of an `emerging'  $U(1)$ gauge field $\vec{A}$,

\begin{eqnarray}
\partial_{\mu} \xi_{\mu} & = & 0 \nonumber \\
\xi_{\mu} & = & \epsilon_{\mu \nu \lambda} \partial_{\nu} A_{\lambda} 
\label{auxgauge}
\end{eqnarray}

this describes the conservation of the superfluid current.
Inserting Eq. (\ref{auxgauge}) in the remaining part of
Eq. (\ref{split2}),

\begin{eqnarray}
L & = & g F_{\mu \nu} F^{\mu \nu} + i \epsilon_{\mu \nu \lambda} 
\partial_{\nu} A_{\lambda} \partial_{\mu} \phi_V \nonumber \\
 & = & g F_{\mu \nu} F^{\mu \nu} + i A_\mu J^V_{\mu}
\label{dualL}
\end{eqnarray}

where $F_{\mu \nu} = \partial_{\mu} A_{\nu} - \partial_{\nu}
A_{\mu}$, the field strength of the $\vec{A}$ field, while the
second line follows from the first modulo a boundary term.

Hence, the vortex particle as described by the current 
Eq. (\ref{Vcurrent}) acts like a source for the superfluid
gauge-field and the latter behaves like a electromagnetic field
on the vortex particle! Reversely, one might also want to claim 
that quantum-electromagnetism is nothing else than an ordered XY magnet
in disguise. The only reason for this to be special for 2+1D is that
in this dimension vortices are like particles. In 3+1D vortices become
strings and this is somewhat hard to reconcile with the point like
character of the particles.

The hard work has been done and the
remainder rests on standard notions of field-theory. Vortices 
are sources of kinetic energy (charging energy in the phase-dynamics
interpretation). Upon increasing the coupling constant, closed loops
of vortex lines will appear (virtual fluctuations) and these loops
will steadily increase their characteristic size. When the 
coupling constant exceeds a critical value, a `loop blow out'
transition will occur. The vortices turn into real particles
while their world-lines will entangle forming a Bose condensate.  
Eq. (\ref{dualL}) describes a single vortex-worldline and in
the disordered phase one is dealing with an interacting
system of bosonic vortex particles. This system is just
a system of bosons which is charged relative to an electromagnetic
field, and in the neighborhood of  the quantum phase transition
this is described in terms of a Ginzburg-Landau-Wilson action,

\begin{equation}
{\cal L}_{dual} = | ( \partial_{\mu} - i A_{\mu} ) \Phi^V|^2 + m^2 |\Phi^V|^2
+ w | \Phi^V |^4 + g F_{\mu \nu} F^{\mu \nu}
\label{GLdual}
\end{equation}

This  disorder field theory can  be derived explicitly 
(see appendix B).
The mass term ($m^2 | \phi^V |^2$) corresponds with the 
core-energy/meandering kinetic energy balance of the vortex loops.
The proliferation of vortices is
encoded by $m^2$ changing sign. The $\Phi^4$ term represents the short
range (hard-core) repulsions while the coupling to the auxiliary
gauge field describes the long-range Coulombic interactions
between the vortices as mediated by  the superfluid fields.
Eq. (\ref{GLdual}) describes a charged superfluid and for
$m^2 < 0$ (the phase disordered state) this enters the Meissner phase
where the superfluid `photon' is expelled. Hence, there are no
low lying excitations and this superconducting condensate of
vortex-bosons describes precisely the Mott-insulator! 

\section{Glide constraints, loop gasses and GLW field theory.}
\label{App1a}

It is a central result in quantum-field theory and statistical physics
that the Ginzburg-Landau-Wilson `$\phi^4$' theory describes
an ensemble of fluctuating, interacting lines. In the present context
we are dealing with the special circumstance that the dislocation 
worldlines can only meander in  planes spanned by the Burgers vectors 
of the dislocations  and the time direction,  due to the glide 
constraint. In section IIIC we  use Landau's original argument to
deduce the structure  of the field theory 
describing the dislocation condensate,
postulating that the gradient terms follow from a conjugate
momentum acting on a `wavefunction'. From this argument it follows that
the condensate field is just a conventional one, while the effects of
glide are entirely absorbed in the couplings to the stress gauge fields.
Starting from a loop gas perspective, this is less obvious. Why should
the condensate be described by a 2+1D GLW field, while the dislocations
might even decide to move in just one direction? Shouldn't this be an
effective 1+1D problem? Such an expectation is based on a flawed 
understanding of the loop gas - GLW mapping: the order parameter field
$\Psi$ just keeps track of the fact that dislocation loops can appear
everywhere in space time, and only the couplings to the gauge fields
keep track of the direction of the motions. To make this explicit,
we will first shortly review the standard loop gas-GLW mapping, to
subsequently analyze the (exaggurated) case that Burgers vectors
can only point along $x$ and $y$ directions.

The standard mapping is discussed in great detail in the first part of
the Kleinert volumes\cite{KleinertI}, and here we review just the
essence of the mapping using a statistical physics language. 
Defect-antidefect worldline loops behave
like random walkers and it is convenient to define these on a 
lattice. The partition function for one closed loop is given 
by 

\begin{equation}
Z_1  =  \sum_{{\bf x}, N} { { (2 D)^N } \over N}
P ( {\bf x}, {\bf x}, N ) e^{- \beta \epsilon N} 
\label{oneloopZdef}
\end{equation}

$\epsilon$ is the core energy (energy/unit length) of the defect,
$\beta$ inverse temperature (coupling constant in the quantum 
case). The
factor $(2D)^N$ ($D$ is space-time dimensionality)
accounts for the number of spatial configurations
of a chain of length $N$ and the factor $N^{-1}$ ensures that one
particular loop is only counted once. $P(  {\bf x}, {\bf x}, N )$
is the {\em return probability}, i.e. the probability that the
random walker returns to its point of departure so that it forms
a closed loop. This probability is governed by a discrete diffusion
equation ($a$ is the lattice constant, $\bar{\Delta}_{\mu},  \Delta_{\mu}$
 backward- and forward finite difference operators, respectively),

\begin{equation}
\bar{\Delta}_N P  ( {\bf x}, {\bf y}, N ) = { a \over {2D} }
\sum_{\mu} \bar{\Delta}_{\mu} \Delta_{\mu}   
P  ( {\bf x}, {\bf y}, N - 1 )
\label{difflat}
\end{equation}
   
with the boundary condition $P( {\bf x}, {\bf y}, 0 ) = 
\delta_{{\bf x}, {\bf y}}$. This is solved by the Fourier
ansatz $P ({\bf K}, N) = P  ({\bf K})^N$,

\begin{eqnarray}
P  ({\bf k}) & = & 1 - { 1 \over {2D} } 
\sum_{\mu} \left( e^{ i k_{\mu} a}  - 1 \right)
\left( e^{- i k_{\mu} a}  - 1 \right) \nonumber \\
& \rightarrow & 1 - { { a^2} \over {2D} } \sum_{\mu} k^2_{\mu}
\label{retpobk}
\end{eqnarray}

For future use, notice that the gradient terms appearing in the
field theory can be traced back to this return probability; 
the gradients appear because the worldlines
have to form loops.

Using Eq. (\ref{retpobk}) and the identity $\sum_{N=1}^{\infty}
x^N / N  = - \ln ( 1 - x)$, the one loop partition function Eq. 
(\ref{oneloopZdef}) becomes,

\begin{equation}
Z_1 =  - \sum_{ {\bf k} } \ln \left( 1 - 2D P ( {\bf k} ) 
e^{-\beta \epsilon} \right)
\label{oneloopZ}
\end{equation}

The grand canonical partition function for a gas of loops is
obtained by exponentiation of the one loop partition sum,

\begin{eqnarray}
\Omega & = & e^{Z_1} \nonumber \\
       & = & \Pi_{\bf k} G_{0} ({\bf k}) \\
G_{0} ({\bf k}) & = &  { 1 \over { 1 - e^{- \beta \epsilon}
2D P ( { \bf k}) } }
\label{loopens}
\end{eqnarray}

We now observe that the product of propagators can be reproduced
by Gaussian integration over complex scalar fields $\phi ({\bf x})$,

\begin{equation}
\Omega  =  \int \; {\cal D} \phi    {\cal D} \phi^* \;
e^{ - {1 \over 2 } \int d^D x \phi^* ({\bf x}) G_0 ({\bf x})^{-1}
\phi ({\bf x}) } 
\label{fieldtheorpart}
\end{equation}

while the inverse propagator becomes in the continuum limit,

\begin{equation}
G_0 ({\bf x})^{-1} =  \partial_{\mu}^2 + m^2
\label{propfree}
\end{equation}

with a mass 

\begin{equation}
m^2 = a^{-2} \left( e^{ \beta \epsilon} - 2 D \right)  
\label{massfree}
\end{equation}

which is changing sign when the meandering entropy is overcoming
the core energy. We have recovered the famous result that Gaussian
field theory is just equivalent to a gas of random walking closed loops.

In the context of this paper, the special circumstance is that
due to the glide constraint the defect-worldlines cannot freely
meander through the embedding space-time. Instead, the loops are
oriented in the 2D planes spanned by the Burgers vectors of the
(anti)dislocations and the imaginary time axis. In the main text
(section IIIC) we follow the strategy of Landau to find that
this `reduced dimensionality' effect is not reflected in the 
dimensionality of the matter fields but only in the couplings
to the gauge fields. Can this be understood explicitly in terms
of the mapping loop gas-field theory?  

Let us consider the following loop gas problem which is clearly
representative for the problem. Consider a cubic lattice ${\bf x} =
(x_i, y_i,
\tau_i)$ (dimension $D = 3$) and assume that loops of random walkers 
occur which are exclusively meandering in either the $(x, \tau)$ or 
$(y, \tau)$ planes (dimension $D' =2$) with the same a-priori 
probability. The generalization of the one loop partition function
Eq.(\ref{oneloopZdef}) is, 
   
\begin{equation}
Z_1  =  \sum_{{\bf x}, N} { { (2 D')^N } \over N}
\left( P_x ( {\bf x}, {\bf x}, N ) + P_y ( {\bf x}, {\bf x}, N )
\right)  e^{- \beta \epsilon N} 
\label{oneloopZdim}
\end{equation}

where ${\bf x}$ refers to all points on the space-time lattice. 
The reduced dimensionality is reflected only in the factor
$(2D')^N$ and the return probabilities $P_{\alpha} ( {\bf x}, {\bf x}, 
N )$ referring now to probability of return by meandering exclusively
in the $(\alpha, \tau)$ planes, with $\alpha = x,y$). These
are given by Eq. (\ref{retpobk}) but now restricted to the meandering
planes. Explicitly ($\omega$ is Matsubara frequency), 

\begin{eqnarray}
P_{\alpha}  ({\bf k}) & = & 1 - { 1 \over {2D'} } 
\left[ \left( e^{ i k_{\alpha} a}  - 1 \right)
\left( e^{- i k_{\alpha} a}  - 1 \right) 
+ \left( e^{ i \omega a}  - 1 \right)
\left( e^{- i \omega a}  - 1 \right) \right]
\nonumber \\
& \rightarrow & 1 - { { a^2} \over {2D'} } \left[ k^2_{\alpha}
+ \omega^2 \right]
\label{rpobdir}
\end{eqnarray}

and the one loop partition function becomes (compare Eq. \ref{oneloopZ}),

\begin{equation}
Z_1 =  - \sum_{ {\bf k} } \sum_{\alpha = x, y} 
\ln \left( 1 - 2D' P_{\alpha} ( {\bf k} ) 
e^{-\beta \epsilon} \right)
\label{onedir}
\end{equation}

The grand partition function of the loop gas is as before
obtained by exponentiation,

\begin{eqnarray}
\Omega & = & e^{Z_1} \nonumber \\
       & = & \Pi_{\bf k} G_{0} ({\bf k}) \\
G_{0} ({\bf k}) & = & \Pi_{\alpha= x,y} 
{ 1 \over { 1 - e^{- \beta \epsilon} 2D' P_{\alpha} ( { \bf k}) } } 
\label{loopdir}
\end{eqnarray}

We now observe that in the continuum limit,

\begin{eqnarray}
G_{0} ({\bf k}) & \sim &  { 1 \over { (m')^2 + k_x^2 + \omega^2 } }
\; \times \; { 1 \over { (m')^2 + k_y^2 + \omega^2 } } \nonumber \\
& \sim & { 1 \over { (m')^2 } } \; \;  
{ 1 \over { (m')^2 + k^2_x + k^2_y + 2 \omega^2 + O (k^4) } }
\label{dirprop}
\end{eqnarray}

In the continuum limit factors $O (k^4)$ can be neglected and
we observe that the partition function is given in terms of
propagators living in the full embedding space time. The gradient
terms find their origin in the return probabilities and given
that the loops have the same a-priori probability to occur both
in $x$ and $y$ directions, both add gradients to the propagator. 
The effect of the loops meandering in a reduced dimensionality
just ends up in a parametric change in the mass reflecting
the reduction of the meandering entropy,

\begin{equation}
(m')^2 = { { e^{\beta \epsilon } - 2D' } \over { a^2} }
\label{massprime}
\end{equation}

and a renormalization of the velocity by a factor $\sqrt{2}$ due
to the fact that the time axis is shared. Since the propagators
are 2+1D, the fields needed to reproduce the propagators are
also 2+1D,

\begin{equation}
\Omega  \sim  \int \; {\cal D} \phi    {\cal D} \phi^* \;
e^{ - {1 \over 2 } \int d^d r d\tau \left( | \partial_x \phi |^2
+ | \partial_y \phi |^2 + { 1 \over 2} | \partial_{\tau} \phi |^2
+ (m')^2 | \phi |^2 \right) }
\label{fieldthdir}
\end{equation}
 
This demonstrates that the assertion in the main text regarding 
the nature 
of the effective field theory describing the dislocation condensate
is correct. The field $\phi$ describes the statistical distributions
of dislocation-antidislocation loops in space-time and this is a
2+1D problem. The consequence of the glide principle, that the
kinetic energy is oriented along the Burgers vectors, does
not change the dimensionality of the field but it does have 
consequences for the screening of the stress photons. 
The oriented nature of the kinetic energy becomes manifest when
screening currents are required for the shear-Meissner effect.

\section{Stress gauge fields in the classical 2D limit.}
\label{App2}

Here we rederive the stress gauge field action for the classical
two dimensional limit, corresponding with the high temperature 
limit of the 2+1D theory (see also Kleinert\cite{KleinertII}, Vol II, 
section 4.6).

In two space dimensions
the time axis is absent and one is forced to pick a Coulomb gauge,
setting the space components of the stress gauge fields to zero,
keeping only the time components. The stress gauge fields are
defined through $\sigma^a_{\mu} = \epsilon_{\mu \nu \lambda} \partial_{\nu}
B^a_{\lambda}$. Keeping only the $B^{a}_{\tau}$ fields and calling these
$A^a$ to get contact with Kleinert's notation, and ignoring
time derivatives: $\sigma^x_x = \partial_y A^x, \sigma^y_y = - \partial_x A^y,
\sigma^y_x = \partial_y A^y, \sigma^x_y = -\partial_x A^x$. 
We recognize directly Kleinert's fields and  the free energy is,

\begin{equation}
S_{2D} = { 1 \over {4\mu} } \int d^2 x \left[
(\sigma^x_x)^2 + 2(\sigma^x_y)^2 + (\sigma^y_y)^2 
- {\nu \over { 1 + \nu}} ( \sigma^x_x + \sigma^y_y)^2 \right]
\label{2Dact}
\end{equation}

after inserting the $A^a$ fields, Fourier transforming and using
the standard space longitudinal/transversal composition,

\begin{eqnarray}
A^x & = & \hat{q}_x A^L + \hat{q}_y A^T \nonumber \\
A^y & = & \hat{q}_y A^L - \hat{q}_x A^T
\label{longtrans}
\end{eqnarray}

The symmetry condition  $\sigma^x_y = \sigma^y_x$ has to be imposed 
which becomes $\partial_a A^a = 0$, implying that the $A^L$ fields
are unphysical. A simple calculation yields for the 2D action,
 
\begin{equation}
S_{2D} = { 1 \over {4\mu ( 1 + \nu) } } \int d^2 q \; q^2  |A^T|^2 
\label{2Dactproj}
\end {equation}

showing that 2D classical matter is governed by a single stress photon
mediating in fact exclusively shear stresses. Photons carrying compression
require the existence of a time axis, see the derivation of the 
compression action Eq. ({\ref{Scom1}) in section III.

\section{Relating the Goldstone propagators to the dual photons.}
\label{App3}

In this paper we use repeatedly the propagators of the stress photons
to compute the propagators and spectral functions in the Goldstone-sector.
These are related by simple general expressions, which we will derive
in this appendix. For convenience, we specialize to quantum elasticity
but the reader will recognize that the derivation is quite general. 

Consider the action of quantum-elasticity $S_0$. Define a generating functional
by adding a source ${\cal J}_{\mu}^a$ coupling to the strain fields,

\begin{equation}
G ( {\cal J} ) = { 1 \over Z} \int {\cal D} u^a 
\exp{\left( - S_0 - \int d\Omega\left [ {\cal J}_{\mu}^a \partial_{\mu} 
u^a \right] \right)} 
\label{genfunc}
\end{equation}

The elastic (or strain) propagator is defined through the functional
derivatives of the generating functional by the sources ( $\vec{r}$
is space-time coordinate),

\begin{eqnarray}
\langle \langle \partial_{\mu} u_a (\vec{r}_1)| \partial_{\nu} u_b (\vec{r}_1) 
\rangle \rangle
& \equiv & \left[ { { \delta^2   G ( {\cal J} )} \over 
{ \delta {\cal J}_{\mu}^a (\vec{r}_1) \;
\delta {\cal J}_{\nu}^b (\vec{r}_2)}} \right]_{J = 0} \nonumber \\
& = & { 1 \over Z} \int {\cal D} u^a  \; (\partial_{\mu} u_a (\vec{r}_1))\;
(\partial_{\nu} u_b (\vec{r}_2)) \; \exp{\left( - S_0 \right)}
\label{prpagatordefintion}
\end{eqnarray}

The strategy is to dualize the generating functional, to subsequently take the
functional derivatives to the sources ${\cal J}$ using the action 
expressed in the dual stress Gauge fields. Introduce auxiliary stress-fields,
so that the strain action $S_0$ turns into the stress action $S_{dual}
(\sigma^a_{\mu})$. Keeping track of the source term,

\begin{eqnarray}
G ( {\cal J} ) & = & { 1 \over Z} \int {\cal D} \sigma^a_{\mu} {\cal D} u^a
\exp{\left( -S_{dual} (\sigma^a_{\mu})  \; - \int d{\Omega} 
\left( 2 i \sigma^a_{\mu} + {\cal J}^a_{\mu} \right)  \partial_{\mu} u^a \right) }
\nonumber \\
& = & { 1 \over Z} \int {\cal D} \sigma^a_{\mu} {\cal D} u^a
\exp{\left( -S_{dual} (\sigma^a_{\mu}) \; + \int d{\Omega} \;
i \; u^a \partial_{\mu} \; \left( 2 \sigma^a_{\mu} 
- {\cal J}^a_{\mu} \right) \right) } 
\nonumber \\
& = & { 1 \over Z} \int {\cal D} \sigma^a_{\mu} {\cal D} u^a \;
\delta \left( 2 \sigma^a_{\mu} - i {\cal J}^a_{\mu} \right) 
\exp{ \left( -S_{dual} \right) }
\label{dualgenfunction}
\end{eqnarray}

The constraint can be resolved by introducing the stress gauge fields,

\begin{eqnarray}
\sigma^a_{\mu} - { i \over 2}  {\cal J}^a_{\mu} & = &  
\epsilon_{\mu \nu \lambda}
\partial_{\nu} B^a_{\lambda} \nonumber \\
\sigma^a_{\mu} & = &  \left( \epsilon_{\mu \nu \lambda}
\partial_{\nu} B^a_{\lambda} + { i \over 2} {\cal J}^a_{\mu} \right)
\label{stresssource}
\end{eqnarray}

The dual action has the form $S_{dual} = \int d\Omega \; \sigma^a_{\mu} \;
C^{-1}_{\mu \mu' a b} \; \sigma^b_{\mu'}$ and inserting Eq. (\ref{stresssource})
we obtain for the generating functional,

\begin{eqnarray}
G ( {\cal J} ) & = & { 1 \over Z} \int {\cal D} B^a_{\mu} \; \delta 
( \partial_{\mu} B^a_{\mu} ) \times \nonumber \\
& & \exp{ \left(- S_{dual} ( B^a_{\mu}) - \int d\Omega \left[ 
i {\cal J}^a_{\mu} C^{-1}_{\mu \mu' a b} \epsilon_{\mu' \nu \lambda}
\partial_{\nu} B^b_{\lambda}  - { 1 \over 4} {\cal J}^a_{\mu}
C^{-1}_{\mu \mu' a b} {\cal J}^b_{\mu'} \right] \right) }
\label{dualgenfunc}
\end{eqnarray}

where $S_{dual} ( B^a_{\mu})$ is just the `stress Maxwell' action.
Eq. (\ref{dualgenfunc}) is of the required form to deduce the
relation between strain- and stress photon propagators. Use  
Eq. (\ref{prpagatordefintion}) but now calculate the functional
derivatives using  Eq. (\ref{dualgenfunc}). Reinsert the $\sigma$'s
for notational convenience, 

\begin{equation}
\langle \langle \partial_{\mu} u_a (\vec{r}_2)| \partial_{\nu} u_b (\vec{r}_1) 
\rangle \rangle
\; =  \;  \delta (\vec{r}_1 - \vec{r}_2) \delta_{\mu, \nu} 
\delta_{a, b} C^{-1}_{\mu \mu a a} \; - \; C^{-1}_{\mu \lambda a c} 
C^{-1}_{\nu \kappa b d } \langle \sigma^c_{\lambda} (\vec{r}_1) |
\sigma^d_{\kappa} (\vec{r}_2) \rangle
\label{genpropag1}
\end{equation}

where by definition

\begin{equation}
\langle \langle \sigma^c_{\lambda} (\vec{r}_1) | 
\sigma^c_{\lambda} (\vec{r}_2) \rangle \rangle
\; \equiv  \; 
{ 1 \over Z} \int {\cal D} \sigma^a_{\mu} \delta 
( \partial_{\mu} \sigma^a_{\mu} ) \; \sigma^c_{\lambda} (\vec{r}_1) \; 
\sigma^c_{\kappa} (\vec{r}_2) \; \exp{\left( - S_{dual} ( \sigma^a_{\mu})
\right)}  
\label{ststrapr}
\end{equation}

i.e. the stress propagator which has to be averaged using the dual action.
In frequency-momentum space this becomes,

\begin{equation}
\langle \langle p_{\mu} u_a (-\vec{p})| p_{\nu} u_b (\vec{p}) \rangle \rangle
\; = \; C^{-1}_{\mu \mu a a} \; - \;
C^{-1}_{\mu \lambda a c} C^{-1}_{\nu \kappa b d } \langle \langle 
\sigma^c_{\lambda} (- \vec{p}) |
\sigma^d_{\kappa} (\vec{p}) \rangle \rangle
\label{ststrapq}
\end{equation}

which is the starting point for the computations in the main text. The
algorithm is simple: associate the strains with the stresses via the
stress-strain relations $ p_{\mu} u_a (\vec{p}) =  C^{-1}_{\mu \lambda a c}
\sigma^c_{\lambda} (- \vec{p})$ and the stress-stress propagators
$\langle \langle \sigma | \sigma \rangle \rangle$ can be computed using the stress
gauge fields. The above derivation shows that the Goldstone propagator
$\langle \langle \partial_{\mu} u^a | \partial_{\nu} u^b \rangle \rangle$ is {\em 
proportional to a constant minus} the dual photon propagators. This is
a general result, since one relates the propagator of a scalar 
Goldstone boson with the field-strength propagators $\sim \epsilon_{\mu \nu
\lambda} A_{\lambda}$ of dual gauge fields which are transversal.
We leave it as an exercise for the reader to demonstrate that in the
case of the simple XY-electromagnetism duality discussed in Appendix
A the relations are as follows: the spin wave propagator is
$\langle \langle p_{\mu} \phi | p_{\nu} \phi \rangle \rangle 
\sim p_{\mu} p_{\nu} / |p|^2$.
The field strength propagators  $\langle \langle \xi_{\mu} | \xi_{\nu} 
\rangle \rangle
\sim \delta_{\mu \nu} -  p_{\mu} p_{\nu} / |p|^2$, obeying the transversality
condition. Hence, in agreement with the above, 
$\langle \langle p_{\mu} \phi | p_{\nu} \phi \rangle \rangle = \delta_{\mu \nu} -
\langle \langle \xi_{\mu} | \xi_{\nu} \rangle \rangle$.



\begin{thebibliography}{11}
\bibitem{Stoofcom} In ref. [\cite{Stoof}] an attempt to arrive at a field
theoretical description is presented. However, this work
appears as quite incomplete.
\end{thebibliography}
\end{document}